\documentclass[preprint,12pt,authoryear]{elsarticle}

\usepackage{amssymb}
\usepackage{amsmath}
\usepackage[caption=false,font=scriptsize,labelfont=rm,textfont=rm]{subfig}
\usepackage{textcomp}
\usepackage{stfloats}
\usepackage{multirow,multicol}
\usepackage{makecell}

\usepackage{amsmath,amssymb,amsfonts}
\usepackage{algorithmic}
\usepackage{graphicx}
\usepackage{textcomp}

\usepackage[ruled,vlined]{algorithm2e}
\usepackage{amssymb}
\usepackage{amsmath}
\usepackage[caption=false,font=scriptsize,labelfont=rm,textfont=rm]{subfig}
\usepackage{textcomp}
\usepackage{stfloats}
\usepackage{multirow,multicol}
\usepackage{makecell}
\usepackage{diagbox}
\usepackage{xcolor}
\newtheorem{assumption}{Assumption}
\newtheorem{theorem}{Theorem}

\newtheorem{lemma}{Lemma}
\newtheorem{remark}{Remark}

\newenvironment{proof}{{\indent  \it Proof:}}{\hfill $\blacksquare$ \par}

\def\BibTeX{{\rm B\kern-.05em{\sc i\kern-.025em b}\kern-.08em
		T\kern-.1667em\lower.7ex\hbox{E}\kern-.125emX}}
\journal{Ocean Engineering}

\begin{document}

\begin{frontmatter}




\title{Trajectory Optimization for Unknown Maneuvering Target Tracking with Bearing-only Measurements} 

\cortext[cor1]{Corresponding author}
\author[label1,label2,label3]{Yingbo Fu} 
\author[label1,label2,label3]{Ziwen Yang\corref{cor1}}
\author[label4]{Liang Xu}
\author[label1,label2,label3]{Yi Guo}
\author[label1,label2,label3]{Shanying Zhu}
\author[label1,label2,label3]{Xinping Guan}

\affiliation[label1]{organization={State Key Laboratory of Submarine Geoscience, School of Automation and Intelligent Sensing, Shanghai Jiao Tong University},
            city={Shanghai 200240},
            country={China}}

\affiliation[label2]{organization={ Key Laboratory of System Control and Information Processing, Ministry of Education of China},
	city={Shanghai 200240},
	country={China}}

\affiliation[label3]{organization={Shanghai Key Laboratory for Perception and Control in Industrial Network Systems},
	city={Shanghai 200240},
	country={China}}
	
\affiliation[label4]{organization={School of Future Technology, Shanghai University},
	city={Shanghai 200444},
	country={China}}
\begin{abstract}
This paper studies trajectory optimization of an  autonomous underwater vehicle (AUV) to track an unknown maneuvering target both in the 2D and 3D space. Due to the restrictions on sensing capabilities in the underwater scenario, the AUV is limited to collecting only bearing measurements to the target. A framework called {\it GP-based Bearing-only Tracking (GBT)} is proposed with integration of online learning and planning. First, a Gaussian process learning  method is proposed for the  AUV to handle unknown target motion, wherein pseudo linear transformation of bearing measurements is introduced to address nonlinearity of bearings. A probabilistic bearing-data-dependent bound on tracking error is then rigorously established. Based on it, optimal desired bearings that can reduce tracking uncertainty are obtained analytically. Finally, the trajectory optimization problem is formulated and transformed into an easily solved one with parametric transformation. Numerical examples and comparison with existing methods verify the feasibility and superior performance of our proposed framework.
\end{abstract}

%
%
%
%
%
%


\end{frontmatter}



\section{Introduction}
\label{sec1}
With increasing exploitation of marine resources, autonomous underwater vehicles (AUVs) have gradually become the main role in carrying out tasks, such as underwater surveillance, feature mapping,  pollution detection, resource surveys \citep{ferri2018autonomous, su2022bearingOE}, etc. A particularly basic mission for AUVs is to track a maneuvering target of interest \citep{he2019trajectory}. Due to the rapid attenuation of underwater GPS signals and  the unknown target, underwater environments impose substantial challenges on perception and locomotion of AUVs.  The current common practice is to use onboard sensors (e.g., USBL, LBL, IMU) to obtain raw position information to achieve accurate tracking with filters. But these sensors measuring positions have a lot of disadvantages including usually enormous cost \citep{paull2013auv} and sensitivity to sound velocity errors \cite{leonard2016autonomous}. In contrast, bearing data are easy to obtain by passive bearing-only sensors (e.g. monocular cameras and hydrophones), which can serve as a supplement to position-based methods. Their advantages include low cost, low power and good concealment \citep{paull2013auv}.  Particularly, bearing sensors may be the only choice in some  scenarios requiring consistent concealment and less impacts on marine life \citep{fu2022leader}. Thus, bearing-only methods may be favored in military and marine applications where signal transmission is restricted \citep{van2022bearing}. However, bearing measurements exhibit strong nonlinear characteristics. Moreover, it is widely known that bearing-only target tracking accuracy depends not only on the filter performance but also on the relative geometry between the AUV and the target \citep{he2020trajectory}.  It is thus essential to develop trajectory optimization to improve bearing-only target tracking performance for AUVs. 

Fisher Information Matrix (FIM) is the most widely used metric in trajectory optimization. It is mainly because the inverse of FIM provides a  low bound on the estimation error covariance of an unbiased filter, known as Cramer-Rao Lower Bound (CRLB) \citep{du2022configuration}.  Following this idea,  FIM or CRLB in a finite horizon  is usually adopted  to optimize the trajectory and enhance tracking performance in many works \citep{li2023optimal, li2022three}. Besides FIM, Error Covariance Matrix (ECM) is often used, which is computationally more efficient than FIM  \citep{li2023optimal}. For example, \cite{zhou2011multirobot} and \cite{nusrat2022underwater} use the posterior covariance matrix to derive the optimal behavior to improve tracking performance. In FIM-based and ECM-based methods, the metric quantifies the informativeness of bearing data through partial derivatives so as to overcome nonlinearity of bearings naturally. However, these two methods assume a known target model. Additionally, some works directly analyze the tracking performance from a geometric perspective, mitigating  nonlinearity of bearings while optimizing the agent trajectory with geometric metrics \citep{zhang2017piecewise}. For example, in \cite{he2019trajectory}, the optimal maneuver for a point-mass aerial robot to track the piecewise-constant-maneuvering target is obtained based on a novel geometric metric. Furthermore, utilizing the orthogonal property of bearing measurements, orthogonality-based estimators are proposed to deal with nonlinearity of bearings.  To guarantee tracking performance, the persistent excitation (PE) characteristics are incorporated into the agent maneuver only for tracking a specific target, such as a static target \citep{yang2020entrapping, fu2022leader}, a constant-velocity target \citep{hu2021bearing} and an angular-rate-varying target \citep{su2022bearing}.  To deal with varying target models, the Interacting Multiple Model (IMM) algorithm has been widely used. For underwater maneuvering target tracking, the IMM-KF is exploited with TOA-DOA measurements in \citet{li2023adaptive}. In \citet{li2023adaptive}, adaptive IMM-EKF and IMM-UKF are proposed to deal with random time delay. Focusing on bearing-only  target tracking, a tracking algorithm based on IMM and PLKF is proposed in \citet{hao2021imm}.  In \citet{ebrahimi2022bearing}, higher-order Markov models are exploited to improve accuracy of the models. In \citet{qian2024maneuvering}, a turn rate identification is incorporated into IMM to alleviate the delay of the model switch. However, IMM heavily relies candidate models, which must include the actual model or represent target dynamics \citep{ali2024novel}. Therefore, IMM-based methods  can only track limited types of target motion and the candidate models must be deliberately selected upon the knowledge of the target.  When the target model is unknown, these methods cannot be specifically tailored, thereby precluding accurate target tracking. While effective for specific scenarios like constant-velocity targets, all of the above methods exhibit strong dependence on a known target model, which prove inadequate for tracking unknown maneuvering targets with unpredictable trajectory changes. Methods in the field of data prediction (such as ARIMA model, Kalman-like filters and exponential smoothing) can recover weakly nonlinear unknown target motion to some extent. But they are unable to recover greatly nonlinear maneuvers and achieve accurate tracking.  Moreover, most schemes are designed for first-order or second-order integrator systems without constraints, thereby precluding application to practical AUV models.

Recently, learning-based methods have attracted increasing attention due to their ability in complex environments and dynamic targets.  High-order polynomials are used to describe target motion with linear fitting for bearing-angle target tracking in \cite{ning2024bearing}. But  polynomial regression shows weak fitting performance when target motion is greatly nonlinear.  \cite{ali2024novel} present neural time-series models for underwater passive target tracking, enabling end-to-end learning of sonar-sequence dynamics for adaptive state estimation and maneuver identification. However, neural-network-based methods heavily rely on the richness of offline training datasets. These methods may fail if unseen maneuvering patterns of the target appear during tracking process. Recently, GP has attract much more attention in this field. Theoretically, the candidate models of GP are infinite and thus it can approximate any function, which avoids prior selection of models. Therefore,  GP has  stronger representative abilities than simple learning models (such as polynomial regression (PR), support vector regression and decision trees) and thus is more effective on tracking greatly nonlinear motion. Furthermore,  compared to some advanced learning models (such as random forest, multilayer perceptron and long short-term memory) whose structures are usually complex, GP has fewer parameters to be optimized so as to not overfit the online collected mini-batch data, and thus be more effective. Therefore, GP provides a greatly practical compromise between computation and effectiveness when considering the practical constrained computational and sensing resources of an AUV for tracking  an unknown target.   In \cite{omainska2023rigid},  GP is used to describe target motion models with the state-dependent inputs. Target tracking is achieved with visual measurements without depth. However,  GP-based schemes rely on offline batched data about target's poses and velocities which actually can not be measured online. In addition, once the target's motion does not match the offline data, the GP-based tracking error will significantly diverge. \cite{beckers2021online} and \cite{umlauft2019feedback} use batched online data modeling and control strategies to alleviate errors caused by model uncertainty.  It is noted that the above learning-based works rely on state measurements such as target positions and velocities. This is usually  infeasible for underwater scenarios as discussed previously.  Moreover, unlike position-based methods,  distinct motion patterns may generate similar bearing measurements. This ambiguity fundamentally limits the discernibility of  target motion patterns for bearing-based learning.  A key question naturally arises: what bearing data can be used to efficiently achieve unknown maneuvering target tracking? To answer this question, it is necessary to develop a performance metric to quantify informativeness of batched online bearing data for motion pattern learning, thereby optimizing the AUV's trajectory planning.

\begin{figure}[t]
	\centering
	\includegraphics[width=3.5in]{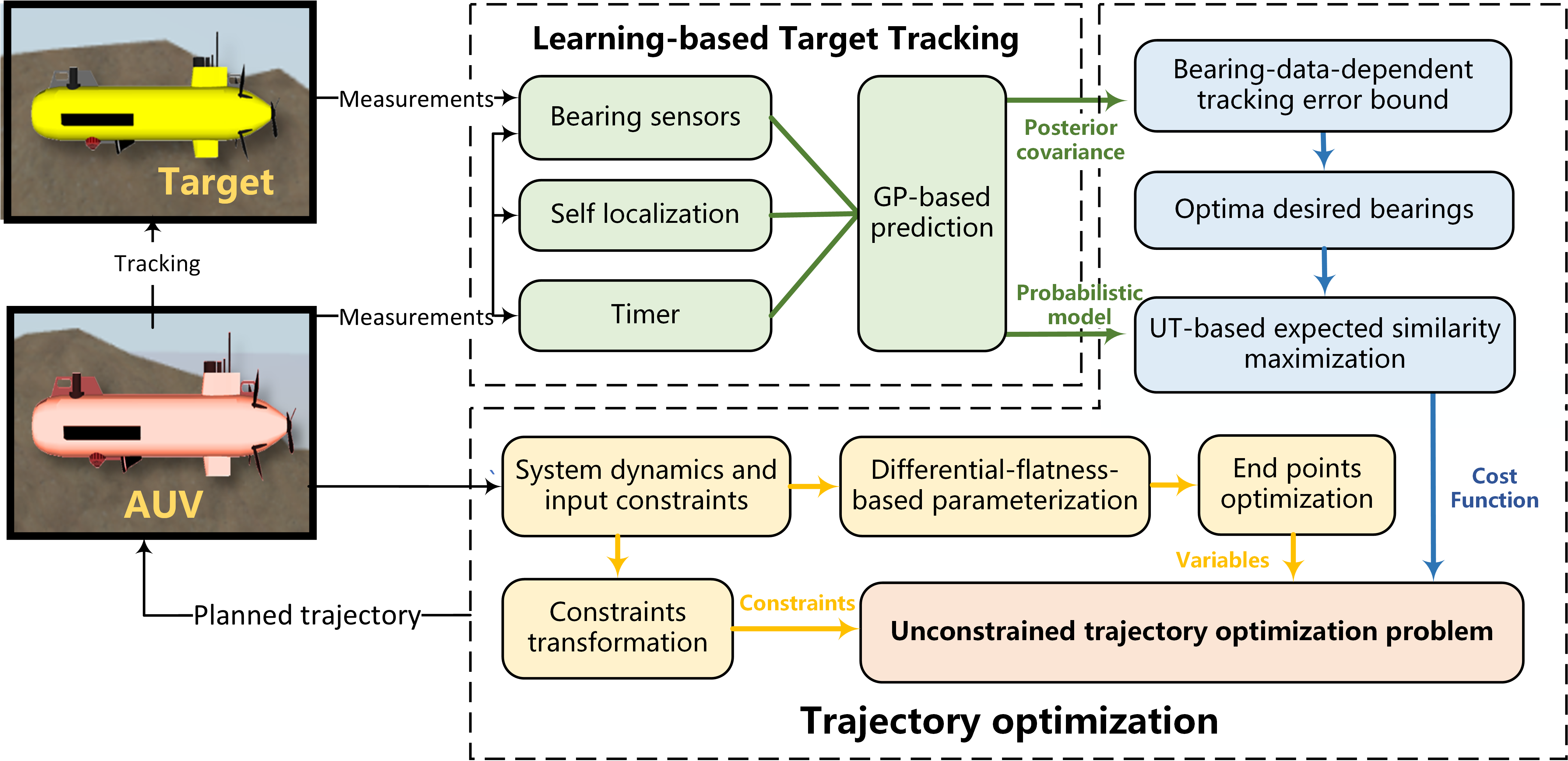}
	\caption{Schematic framework of GBT.}\label{framework}
\end{figure}

Out of the above observations, we present a novel GP-based framework of AUV trajectory optimization for unknown maneuvering target tracking using bearing-only measurements called {\it GP-based Bearing-only Tracking (GBT)}. In our framework, there are two main modules: learning-based target tracking and trajectory optimization,  depicted in Fig. \ref{framework}. Contributions of this paper are summarized as follows.

$\left.1\right)$ We propose a novel GP-based method for tracking maneuvering targets based only on bearing measurements both in the 2D and 3D space. It significantly relaxes sensing requirements compared to existing  methods that necessitate both pose and velocity measurements \citep{omainska2023rigid}. Unlike conventional approaches that rely on restrictive target motion models such as constant-velocity models \citep{li2022three, hu2021bearing}, our method does not assume any specific target dynamics. Furthermore, to effectively capture the spatio-temporal characteristics of the unknown target motion, a batch of online bearing data within a sliding time window is utilized to achieve a balance between accuracy and efficiency. At the same time, an extension to linear model of coregionalization is also given.

$\left.2\right)$ A probabilistic bound on the tracking error that explicitly depends on the acquired bearing measurements is derived. This bound characterizes the metric to qualify informative bearing data and corresponding optimal desired bearings that guarantee monotonic reduction in tracking uncertainty. The expected similarity between these optimal bearings and future measurements is formulated as a cost function to guide AUV planning. To overcome computational intractability, we develop an efficient approximation method using the Unscented Transform (UT). In contrast to prior approaches \cite{li2023optimal} and \cite{he2019trajectory}, our method leverages the GP posterior to quantify sampling efficiency with minimal computational overhead, enabling the AUV to acquire high-information-bearing data with high probability.

$\left.3\right)$ We explicitly incorporate practical AUV  models and input constraints into the trajectory optimization process for target tracking. To address infinite optimization variables arising from the continuous-time AUV model, we employ a differential-flatness-based parametric transformation technique. Furthermore, we optimize the endpoints of trajectory segment and facilitate the selection of appropriate initial guesses  to handle the non-convexity introduced by input constraints. Subsequently, constraints transformation together with the penalty function method is proposed to efficiently solve the trajectory optimization problem.

The rest parts of this paper are organized as follows. Section 2 formulates the problem. The GP-based method for bearing-only target tracking is proposed and the uniform error bound of the proposed method is also established in Section 3. In Section 4, we formulate an optimization problem for trajectory generation to improve tracking performance. Numerical simulations for validation are given in Section 5. In Section  6, the conclusion and further discussion of our work are given.

{\it Notations:} $\mathbb{R}$, $\mathbb{R}^{n}$, $\mathbb{R}^{n \times m}$ and $\mathbb{N}$ denote  sets of real numbers, real vectors with $n$ elements, real matrices with $n$ rows and $m$ columns and natural numbers, respectively. $\mathbb{S}^n$ denotes the set of unit vectors in $\mathbb{R}^{n}$. $\boldsymbol{0}_N \in \mathbb{R}^N$ is the vector containing only zeros and $\boldsymbol{I}_{N\times N} \in \mathbb{R}^{N \times N}$ is the  identity matrix. For two matrices $A, B\in \mathbb{R}^{n \times n }$, $A \leq B$ denotes that $B - A$ is positive semi-definite. $|.|$ denotes the absolute value of a real number or cardinality of a set. $\|.\|$ denotes Euclidean norm of a vector or spectral norm of a matrix.  ${\rm cond}(.)$ is  the condition number of a matrix. $\mathbb{P}(.)$ denotes probability of an event. $\mathbb{E}(.)$ denotes the expectation of a random variable. 

\section{Problem Statement}
In this section, we present the system model and bearing measurement of the AUV.

\subsection{System Models}
Consider the AUV as a rigid body in the $d$-dimension space ($d = 2$ or $3$) whose kinematics model and dynamical models follows Fossen's equations \citep{fossen2011handbook}
\begin{equation}\label{eq2}
	\left\{\begin{array}{l}
		\dot{\eta}=J(\eta) \nu,\\
		M \dot{\nu}+C(\nu) \nu+  D(\nu)\nu=\tau,
	\end{array}\right.
\end{equation}
where $\eta= \left[P_A, \psi\right]^{\mathrm{T}} \in \mathbb{R}^{d+1}$ denotes the position in the $d$-dimension space and yaw angle of the AUV in earth-fixed frame $O_e$, $\nu= \left[v_A^{\mathrm{T}} , r\right]^{\mathrm{T}}  \in \mathbb{R}^{d+1}$ denotes linear and angular velocities of yaw angle in the body-fixed frame $O_b$, and $\tau = [F_v^{\mathrm{T}} , F_r]^{\top} \in \mathbb{R}^{d+1}$ describes control forces and moments satisfying $ -\tau_{\max} \leq \tau \leq \tau_{\max}$. Here, $J(\eta)$ describes the transformation matrix from body-fixed frame to the earth-fixed frame, $M$ and $C(\nu)$ describe the inertia matrix and the Coriolis and centripetal matrix with the added mass, $D(\nu)$ describes the damping matrix,  and $\tau$ describes control forces and moments.  For simplicity, we assume that the mass distribution of the  AUV is homogeneous. Correspondingly, for the 2-D space and 3-D space, a 3-DOF AUV and a 4-DOF AUV are considered, respectively.


The problem of interest is to track an  underwater maneuvering  target
\begin{align}\label{eq1}
	\dot{P}_T(t) = \boldsymbol{h}(t, P_T(t)), 
\end{align}
where $P_T(t) \in \mathbb{R}^{d}$ is the position of the moving target in the $d$-dimensional space and each element of $ \boldsymbol{h}_x(t, P_T)$ is {\it unknown}, piecewise continuous in $t$, and locally Lipschitz in $P_T(t)$ for all $t\geq 0$.  The target model is quite different from previous works such as \cite{li2022three} and \cite{hu2021bearing}, in which a known constant-velocity target is considered. In comparison, the target model  we consider is {\it unknown} and {\it nonlinear}, which greatly increase challenges on tracking.

\begin{figure}[t]
	\centering
	\includegraphics[width=4in]{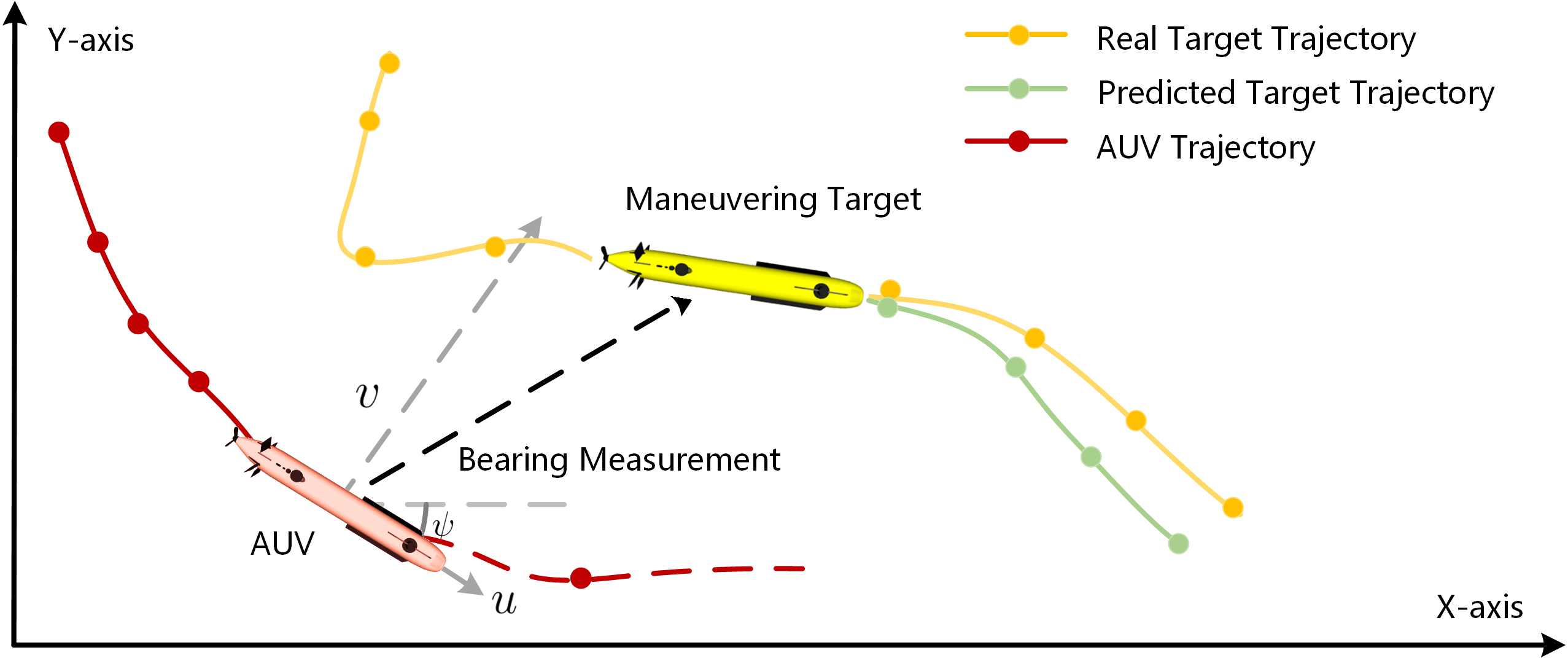}%
	\label{variables}
	\caption{Illustrations of the  problem statement.}
\end{figure}

\subsection{Bearing Measurement Model}

The  AUV can measure its own state $\boldsymbol{x}$ and the bearing to the target at each discrete time instant $t_{k} = kT$ for $k \in \mathbb{N}$ and $T$ is a positive constant. Let $q(t_k) = P_T(t_{k}) - P_A(t_{k})$ be the relative position from the AUV to the target. The bearing measurement $\lambda(t_k)$ $\in$ $\mathbb{S}^d$ is
\begin{align}\label{eq4}
	\lambda(t_k) =\frac{q(t_{k}}{\|q(t_{k})\|}.
\end{align}
 Focusing on solving unknown target models, which is the main contribution of the paper, we assume that the bearing measurement is noise-free, which is often seen as in \cite{sui2024adaptive} and \cite{hu2021bearing}. Even though noise-free measurements are considered, it is still challenging because bearings are incomplete and may lead to indistinguishable motion patterns when tracking an unknown maneuvering target, which can follow any motion mode. At the same time, it is clear that bearing measurements exhibit strong nonlinear characteristics, leading to the non-Gaussian uncertainty propagation.
%

From beginning, the collected measurements are all stored in the AUV, which is denoted as $\mathcal{D}_{0:k} = \{t_i, \lambda(t_i),P_A(t_i)\}_{i=0}^{k}$. A diagram of key variables are given in Figure. 2.


\subsection{Objectives}

The overall goal of this work is to track an unknown maneuvering target  by  trajectory optimization for the  AUV based on bearing measurements. No information transmitted from the target to the AUV can be used. The AUV stores all collected data and processes them locally online onboard so that our method is communication-free and communication delay can be avoided. 

To track a target with an unknown model, we first solve the {\it learning-based target tracking} problem to achieve simultaneous motion learning and tracking with online collected nonlinear bearing data.  Sequentially, we solve the {\it trajectory optimization} problem to overcome the limitation on discernibility of target motion  patterns in bearing-only target tracking, thereby achieving accurate tracking.

1) (Learning-based target tracking) For each time instant $t_{k}$, using the online collected data set $\mathcal{D}_{0:k}$, design a learning-based predictor $\mu_k(\cdot)$ to predict positions of the target for  all time instants in $\mathcal{T}_{k} = \{t_{k}, \dots, t_{k+n}\}$, where $n$ is the length of prediction horizon.

2) (Trajectory optimization) To decrease the tracking error, at each time instant $t_k$ and with the predicted target motion, optimize the AUV's trajectory $\big(\boldsymbol{x}(t), \tau(t)\big)$ of the AUV in the future time interval $[t_{k}, t_{k+p}]$ with $p \leq n$ as the length of planning horizon under input constraints.

Our proposed framework achieves accurate target tracking through tight integration of online learning-based target
tracking and trajectory optimization of the AUV.

\section{Learning-based Bearing-only Target Tracking}
In this section, we propose a GP-based method using a batch of bearing measurements to track the target and provide the upper bound on the learning-based tracking error rigorously. 

\subsection{GP Learning for Unknown Target Tracking}
In using the GP learning to track the target $P_T(t)$, traditional procedure consisting of specifying a prior distribution on $P_T(t)$ with a {\it positive-definite} kernel $k(t,t')$, i.e., $P_T(t) \sim \mathcal{G} \mathcal{P}\left(\boldsymbol{0}_d, \boldsymbol{k}(t, t')\right)$, where $\boldsymbol{k}(t, t') = k(t, t') \boldsymbol{I}_{d \times d}$. GP learning requires that the target positions at previous time instants are sampled. However, in bearing-only target tracking, target position measurements $P_T(t)$ are not available but only bearing measurements $\lambda(t)$ in (\ref{eq4}), which are nonlinear. This means that traditional GP can not be directly applied.

To address the above issue, we  exploit  the orthogonal property of bearing measurements and define a new variable as
\begin{align}\label{eq5}
	\bar{\lambda}(t_k) = {\rm orth}(\lambda(t_k))\in \mathbb{R}^{d \times (d-1)}, 
\end{align}
which is the orthonormal basis matrix of the tangent plane at $\lambda(t_k)$ satisfying that $\bar{\lambda}(t_k)^{\top} \lambda(t_k) =  \boldsymbol{0}_{d-1}$ and $\bar{\lambda}(t_k)^{\top}\bar{\lambda}(t_k) = \boldsymbol{I}_{(d-1)\times (d-1)}$.
Multiplying $\bar{\lambda}(t_k)^{\top}\|q(t_k)\| $ on both sides of (\ref{eq4}) yields that
\begin{align}\label{eq9}
		\bar{\lambda}(t_k)^{\top}\lambda(t_k)\|q(t_k)\|
		= \bar{\lambda}(t_k)^{\top}(P_T(k) - P_A(t_k)),
		= \boldsymbol{0}_{d-1}.
\end{align}
This motivates us to define the pseudo linear measurement  $y(t_k) = \bar{\lambda}(t_k)^{\top}P_A(t_k)$. Therefore, the nonlinear bearing measurement equation (\ref{eq4}) can be transformed into a pseudo linear equation as
\begin{align}\label{eq10}
	y(t_k)  = \bar{\lambda}(t_k)^{\top}P_T(t_k). 
\end{align}

Due to computational complexity of GP learning, a mini-batch data set $\mathcal{D}_{c:k} = \{t_i, \lambda(t_i),P_A(t_i)\}_{i=c}^{k}$ where $c = \max(0, k - N_c + 1)$, including the latest $N=\min\{N_c, k - 1\}$ data,  is used where $N_c$ is the maximum capacity. With this data set, target positions in the current time instant and future $n$ time instants are predicted.

\begin{remark}
The learning effectiveness of GP will improve as $N_c$ increases, but the required computation time will also increase. When the target motion is greatly nonlinear, more data are often needed for training. A trade-off between learning  effectiveness and computational efficiency needs to be considered when setting $N_c$. The length of the forecast data $n$ usually depends on the task requirements. Usually, a task for long-term forecast requires a large $n$ and the short-term forecast requires a small $n$. However, predictive uncertainty grows as predictive time instants get far from the time instants in the data set  \citep{snelson2008flexible}. The learned GP model is based on historical $N_c$ data, so tracking performance may deteriorate as $n$ increases, especially when the target maneuvers severely and  motion modes during historical horizon and forecast horizon are greatly different.
\end{remark}

Considering all time instants in the data set $\mathcal{D}_{c:k}$, pseudo linear measurements $\boldsymbol{{y}} = [{y}(t_c), {y}(t_{c+1}), \dots, {y}(t_{k})]^{\top}$ can be written as $\boldsymbol{y} = \boldsymbol{G}\boldsymbol{P}_T$,
where $\boldsymbol{P}_{T} = [P_T(t_c)^{\top}, \dots, P_T(t_{k})^{\top}]^{\top}$ and $\boldsymbol{G} = {\rm diagblk}\{\bar{\lambda}(t_c)^{\top}, \dots, \bar{\lambda}(t_{k})^{\top}\} $.

Therefore, given the prior GP on $P_T(t)$, i.e., $P_T(t) \sim \mathcal{G} \mathcal{P}\left(\boldsymbol{0}_{d}, \boldsymbol{k}(t, t')\right)$, for pseudo linear measurements $\boldsymbol{{y}}$  at sampling time instants $\boldsymbol{t}$ and the unknown ${P}_T(t^{\star})$ for  $t^{\star} \in \mathcal{T}_k$, the {\it prior}  mean values are
\begin{align} \label{eq13}
\mathbb{E}(\boldsymbol{y} | \boldsymbol{t}) = &\boldsymbol{0}_{(d-1)N}, \nonumber\\
\mathbb{E}(P_T(t^{\star}) | t^{\star} ) =&\boldsymbol{0}_d,
\end{align}
and the {\it prior} covariance matrices are
\begin{align}  \label{eq14}
	\Omega_{\boldsymbol{y} \boldsymbol{y}} =& {\rm Cov}(\boldsymbol{y}, \boldsymbol{y} | \boldsymbol{t}) =  \boldsymbol{G}\boldsymbol{K}\boldsymbol{G}^{\top}, \nonumber\\
	\Omega_{\boldsymbol{y} P}^{\star} =& {\rm Cov}(\boldsymbol{y}, P_T(t^{\star}) | \boldsymbol{t},  t^{\star}) =\boldsymbol{G}\boldsymbol{K}^{\star}, \nonumber\\
	\Omega_{PP}^{\star}=& {\rm Cov}(P_T(t^{\star}), P_T(t^{\star}) |  t^{\star}) =\boldsymbol{k}(t^{\star}, t^{\star}),
\end{align}
where $\boldsymbol{K}^{\star} = [\boldsymbol{k}(t, t^{\star})]_{t \in \boldsymbol{t}} \in \mathbb{R}^{dN \times d}$ and $\boldsymbol{K} = [\boldsymbol{k}(t, t')]_{t, t' \in \boldsymbol{t}} \in \mathbb{R}^{dN \times dN}$.
By (\ref{eq13}) and (\ref{eq14}), the {\it prior}  Gaussian distribution where $\boldsymbol{y}$ and $P_T(t^{\star})$ jointly follows is given as
\begin{align}\label{eq15}
	\left[\begin{array}{c}
		\boldsymbol{{y}}\\
		{P}_T(t^{\star})
	\end{array}\right]
	\sim \mathcal{N}\left(\left[\begin{array}{c}
		\boldsymbol{0}_{(d-1)N} \\
		\boldsymbol{0}_d
	\end{array}\right],\left[\begin{array}{ccccc}
		\Omega_{\boldsymbol{y}\boldsymbol{y}} & 	\Omega_{\boldsymbol{y}P}^{\star}\\
		\Omega_{\boldsymbol{y}P}^{\star\top}  & \Omega_{PP}^{\star}
	\end{array}\right]\right). &
\end{align}
By applying \cite{williams2006gaussian}, we obtain the conditional distribution for multivariate Gaussian distribution as$	P_T(t^{\star}) |  \boldsymbol{y}  \sim \mathcal{N}\big(\mu_k(t^{\star}),  	\Sigma_k(t^{\star})\big)$, where the {\it posterior} mean function is
\begin{align}\label{eq16}
	\mu_k(t^{\star}) \triangleq \mathbb{E}({P}_T(t^{\star}) \mid  \boldsymbol{{y}})= \Omega_{\boldsymbol{y}P}^{\star\top}(\Omega_{\boldsymbol{y}\boldsymbol{y}})^{-1}\boldsymbol{y}, \!\!\!\!\!\!
\end{align}
and the {\it posterior} covariance is
\begin{align}\label{eq17}
	\Sigma_k(t^{\star})
	\triangleq {\rm Cov}({P}_T(t^{\star}), {P}_T(t^{\star}) | \boldsymbol{G}, \boldsymbol{{y}}) 
	= \Omega_{PP}^{\star} - 
	\Omega_{\boldsymbol{y}P}^{\star\top}(\Omega_{\boldsymbol{y}\boldsymbol{y}})^{-1}\Omega_{\boldsymbol{y}P}^{\star}.
\end{align}
Then,  we can use the posterior mean $\mu_k(t^{\star})$ as the predictor of $P_T(t^{\star})$ and the posterior covariance $\Sigma_k(t^{\star})$ as the uncertainty measure.

It is noted that $\mu_k(t^{\star})$ and $\Sigma_k(t^{\star})$ are related to the kernel matrices and thus rely on hyperparameters of kernels, such as the length scale $l$ in the Squared Exponential (SE)  kernel $k(t, t') = \exp(\frac{-|t-t'|^2}{2l^2})$ \citep{lederer2023gaussian}, which are denoted as $\boldsymbol{\theta}$. To infer the hyperparameters, log-likelihood maximization is adopted
\citep{williams2006gaussian}. However, in most existing GP-based methods, log-likelihood maximization is designed for state measurements, which can not be directly used for bearing measurements. 
Base on pseudo linear transformation, it can be extended as
\begin{align} \label{like1}
	\max\limits_{\boldsymbol{\theta}} \,  \log(p(\boldsymbol{y}|\boldsymbol{t})) = -\frac{1}{2}\big(\boldsymbol{y}^{\top}\Omega_{\boldsymbol{y}\boldsymbol{y}}^{-1}\boldsymbol{y} - \log(\det(\Omega_{\boldsymbol{y}\boldsymbol{y}})) - (d-1)N\log(2\pi)\big),
\end{align}
where $\boldsymbol{\theta}$ are hyperparameters of the kernel and $\Omega_{\boldsymbol{y}\boldsymbol{y}}$ in (\ref{eq14}) is related to  the kernel $\boldsymbol{k}(.,.)$ with specific hyperparameters.
It is noted that (\ref{like1}) is an unconstrained optimization problem so that it can be solved by gradient-based methods \citep{lederer2023gaussian}. The algorithm of GP-learning-based  target tracking is given in Algorithm \ref{GP}.

\begin{algorithm}\label{GP}
	\caption{GP-learning-based  Target Tracking}
	\SetKwProg{Fn}{Initialization}{:}{\KwRet}
	\Fn{}{
		Initialise the data set $D_{-1:0} = \emptyset$\;
		Initialise iteration $k = 0$;
	}
	\textbf{For} $k = k + 1$ \Repeat{maneuvering target tarcking}{
		Collect data $\{t_{k}, \lambda(t_k), P_A(t_k)\}$ and update $\mathcal{D}_{c:k}$\;
		Compute  $\boldsymbol{G}$, $\boldsymbol{y}$  by (\ref{eq5}) and (\ref{eq10}) \;
		Solve (\ref{like1}) to tune hyperparameters $\boldsymbol{\theta}$\;
		\ForAll{$t^{\star} \in \mathcal{T}_k$}{
			Compute $\Omega_{\boldsymbol{y}\boldsymbol{y}}$ and $\Omega_{\boldsymbol{y}P}^{\star}$ by (\ref{eq15})\; 
			Compute the expectation $\mu_{k}(t^{\star})$ by (\ref{eq16})\;
			Output $\mu_{k}(t^{\star})$ as the tracking result of $P_T(t^{\star})$;
		}
	}
\end{algorithm}

\subsection{Probabilistic Bearing-data-dependent Tracking Error Bound}

To analyze performance of the proposed GP-Learning-based  target tracking, a key assumption to indicate the relationship between the target position $P_T(t)$ and the prior Gaussian process $\mathcal{G} \mathcal{P}(\boldsymbol{0}_d, \boldsymbol{k}(t, t'))$ is given as follows. 
\begin{assumption}\label{ass2} 
	The unknown function $P_T(t)$ is a sample function from a multioutput Gaussian process $\mathcal{G} \mathcal{P}(\boldsymbol{0}_d, \boldsymbol{k}(., .))$, where  $\boldsymbol{k}(t, t') = k(t, t')\boldsymbol{I}_{d \times d}$ with a  positive-definite bounded kernel $k(t, t')$.
\end{assumption}

We note that Assumption \ref{ass2} imposes  a rather mild assumption on the manuvering target motion compared with the constant-velocity, constant-turn and constant-acceleration models, which supports severely nonlinear trajectory of the target. By {\it Karhunen-Loeve Expansion} \citep{kanagawa2018gaussian}, Gaussian process ${P}_T(t) \sim \mathcal{G} \mathcal{P}(\boldsymbol{0}_d, \boldsymbol{k}(t, t'))$ can be expressed using the orthonormal basis $\{\upsilon^{1/2}\phi_i(t) \in \mathbb{R}^d | i\in \mathcal{I}\}$ of $\mathcal{H}_{k}$ and standard Gaussian random variable vectors $z_i \sim \mathcal{N}(\boldsymbol{0}_d, \boldsymbol{I}_{d \times d})$, where $\mathcal{I} $ is a set of indices. Therefore, Assumption \ref{ass2} actually specifies the admissible functions for regression via the space of sample functions. The hyperparameters of the kernel typically specify the shape of the corresponding probability distribution \citep{lederer2022cooperative}.

\begin{remark}
 The support of the process for SE kernel is equal to the space of all continuous functions, so that  a sample function that approximates any continuous function infinitely  theoretically can be found \citep{van2011information}. In practice, the sample paths of the SE kernel are analytic functions, which may much smoother than real trajectories.   If the true target motion is rougher than what the default kernel would produce,  the error caused by this distinction will appear. To address this distinction, the hyperparameters are tuned online to fit the less smooth trajectory and samples are drawn from a quasi-uniform sampling, which has been verified effective in practice to overcome this distinction \citep{wynne2021convergence}.
\end{remark}


Under this assumption, we have connected target motion with the prior GP.  To improve the tracking accuracy of $\mu_{k}(\cdot)$ with collected bearing data, we next give the following probabilistic bearing-data-dependent tracking error bound.
\begin{theorem} \label{the1} Consider the unknown target position $P_T(t)$ in the $d$-dimensional space ($d=2$ or $d=3$)  and a prior Gaussian process $\mathcal{G} \mathcal{P}(\boldsymbol{0}_d, \boldsymbol{k}(t, t'))$ satisfying Assumption \ref{ass2}.  For a finite set $\mathcal{T}_{k}$ to achieve target tracking, the posterior mean function (\ref{eq16}) conditioned on the data set $\mathcal{D}_{c:k}$ admits a probabilistic tracking error bound, i.e.,
	\begin{align}\label{eq19}
		\mathbb{P}(\|P_T(t_i)-\mu_k(t_i)\| \leq \beta(\delta)\bar{\sigma}_{k}(t_i), \forall t_i \in  \mathcal{T}_{k})  \geq 1-\delta, 
	\end{align}
	holds for $\delta \in (0, 1]$, where  $\beta(\delta) = \sqrt{\chi^2_{d, 1-\frac{\delta}{|\mathcal{T}_k|}}}$ and
	\begin{align}\label{eq20}
		\bar{\sigma}_{k}(t_i) =& \sqrt{
			k(t_i, t_i) -\frac{\min\limits_{t \in \mathcal{D}_{c:k}}k^2(t_i,t)}{\sqrt{\frac{d}{d-1}}\max\limits_{t\in \mathcal{D}_{c:k}}k(t, t) {\rm cond}\Big({P}_{c:k}\Big) } }, \nonumber\\
		P_{c:k} = &\sum_{i=c}^{k}\boldsymbol{I}_{d\times d} - \lambda(t_{i})\lambda(t_{i})^{\top}.
	\end{align}
\end{theorem}
\begin{proof} 
	See Appendix A.
\end{proof}

\begin{remark}
		To reveal a nominal instruction about optimal geometry of the measurement sequence, a noise-free design matrix is assumed and leads to an optimistic error bound, which is analogous to the common practice in most relevant works, such as FIM-based methods \citep{hu2021bearing} and geometry-based methods \citep{he2019trajectory}, and mainly works for guiding sampling. In the process of derivation, all our bounding steps consider the worst case so as to reveal the qualitative link.
\end{remark}

With Theorem \ref{the1}, a probabilistic performance guarantee is provided for bearing-only target tracking. From (\ref{eq19}), we can find that it is related to the condition number of the cumulative bearing matrix ${P}_{c:k}$, which is decided by the bearing data $\{\lambda(t_i)\}_{i=c}^{k}$. Obviously, the collected bearing data is greatly related to the relative trajectory of the  AUV and the target.

\subsection{Extension to Linear Model of Coregionalization}

Considering the cross-axis correlation of target motion, we extend  to employ the {\it linear model of coregionalization } (LMC) \citep{liu2018remarks}.  Target motion $P_T(t)$ is assumed to be a combination of latent functions $u(t)$ with a weight matrix $A$, i.e., $P_T(t) = AP_u(t)$, where $P_u(t) = [P_{u, 1}(t), \dots, P_{u, Q}(t)]^{\top} \in \mathbb{R}^Q$ is the latent function vector including $Q \geq d$ latent functions and $A \in \mathbb{R}^{d\times Q}$ is a weight matrix realizing the transformation from latent functions to target motion. Here, each function is assumed be sampled from a zero-mean Gaussian Process, i.e., ${\rm Cov}\big(P_{u, q}(t), P_{u, q}(t')\big) = k_q(t, t')$ for all $q$ with a positive-definite kernel $k_q(t, t')$ and $A$ is assumed to be row full rank. The prior Gaussian Process that each function samples from is assumed to be independent, i.e., ${\rm Cov}\big(P_{u, q}(t), P_{u, q'}(t')\big) = 0$ for all $q$ and $q'$.

Under the above modeling, the resulting covariance for $P_T(t)$ and $P_T(t')$ can be given by
\begin{align}
	\boldsymbol{k}_f(t, t') \triangleq {\rm Cov}\big(P_T(t), P_T(t')\big)  =& A\boldsymbol{k}_Q(t, t')A^T,\nonumber
\end{align}  
where $\boldsymbol{k}_Q(t, t') = {\rm diagblk}(k_1(t, t'), \dots, k_Q(t, t')) \in \mathbb{R}^{Q \times Q}$ is the covariance matrix of $Q$ latent functions. Extended from the original GP-based learning method in {\it Section 3.1}, the kernel matrix $\boldsymbol{k}(t, t')$ is replaced by $\boldsymbol{k}_f(t, t')$ and others remain unchanged. More hyperparameters need to be optimized, which includes hyperparameters of kernels for latent functions and the weight matrix $A$.

To extend the theoretical probabilistic tracking error bound with LMC,  the following new assumption is given.
\begin{assumption}\label{ass3} 
	The unknown function $P_T(t)$ is a combination of a latent function vector $u(t)$ with a row full rank weight matrix $A$, i.e., $P_T(t) = AP_u(t)$,  where $P_u(t)$ is a sample function vector from a multioutput Gaussian process $\mathcal{G} \mathcal{P}(\boldsymbol{0}_{q}, \boldsymbol{k}_Q(., .))$.
\end{assumption}
Assumption 1 without considering cross-axis correlation can be treated as a special case of Assumption 2 when $A = \boldsymbol{I}_{d \times d}$, and only considers that each axis of $P_T(t)$ is of the same smoothness. In contrast, the matrix $A$ is used to capture correlation of motion in each axis and  latent function with different smoothness are used.  Under the cross-axis correlation, the new theorem about probabilistic tracking error bound is given as follows. 

\begin{theorem} \label{the2} Consider the unknown target position $P_T(t)$ in the $d$-dimensional space ($d=2$ or $d=3$) and $Q$ prior Gaussian processes $\mathcal{G} \mathcal{P}(0, k_q(., .))$ with $q = 1,\dots, Q$ satisfying Assumption \ref{ass3}.  For a finite set $\mathcal{T}_{k}$ to achieve target tracking, the posterior mean function conditioned on the data set $\mathcal{D}_{c:k}$ admits a probabilistic tracking error bound, i.e.,
\begin{align}\label{eqnew_bound}
	\mathbb{P}(\|P_T(t_i)-\mu_k(t_i)\| \leq \beta(\delta)\bar{\sigma}_{k}(t_i), \forall t_i \in  \mathcal{T}_{k})  \geq 1-\delta, 
\end{align}
holds for $\delta \in (0, 1]$, where $\beta(\delta) = \sqrt{\chi^2_{d, 1-\frac{\delta}{|\mathcal{T}_k|}}}$  and 
	\begin{align}\label{eqnew_sigma}
		\bar{\sigma}_{k}(t_i) =& \sqrt{
			\|\boldsymbol{k}_f(t_i,t_i)\| -\frac{\min\limits_{t \in \mathcal{D}_{c:k}}\lambda_{\min}(\boldsymbol{k}_f(t_i,t))^2}{(\frac{d}{d-1})^{\frac{3}{4}}\max\limits_{t\in \mathcal{D}_{c:k}}\|\boldsymbol{k}_f(t,t)\| {\rm cond}\Big({P}_{c:k}\Big) } }.
	\end{align}
\end{theorem}
\begin{proof}
	See Appendix B.
\end{proof}

With Theorem \ref{the2}, a similar result is obtained and we can also find that it is related to the condition number of the cumulative bearing matrix ${P}_{c:k}$.  It motivates us to optimize the AUV's trajectory to decrease the tracking error bound.

\section{Trajectory Optimization for Unknown Target Tracking}

To improve the target tracking performance, we focus on how to collect bearing data by the AUV that can reduce tracking error bound  in (\ref{eq19}).

\subsection{Trajectory Optimization Problem}

As the condition number of $P_{c:k}$ approaches to its minimum value $1$, the upper bound of the tracking error gets smaller. Therefore, the condition number of bearing data is used as the metric to measure the quality of bearing data on reflecting target motion patterns.

It is noted that the number of bearing data $N$ is changing until it reaches the storage capacity  $N_c$. Considering $d < m \leq N_c$ bearing data, the desired bearings for the AUV are given as follows
\begin{align} \label{eq22}
	\{\lambda^{\star}_i\}_{i=0}^{m-1} = \underset{\left\{\lambda_i\right\}_{i=0}^{m-1} \subset \mathbb{S}^{d}}{\arg \min }{\rm cond} (P_{m}),
\end{align}
where $P_{m}$ is the cumulative bearing matrix for $m$ bearing data as $P_{m} = \sum_{i=0}^{m-1} \big(\boldsymbol{I}_{d\times d} - \lambda_i \lambda^{\top}_i\big)$.

\begin{figure*}[!t]
	\centering
	\subfloat[2D optimal  bearings.]{\includegraphics[width=0.25\linewidth]{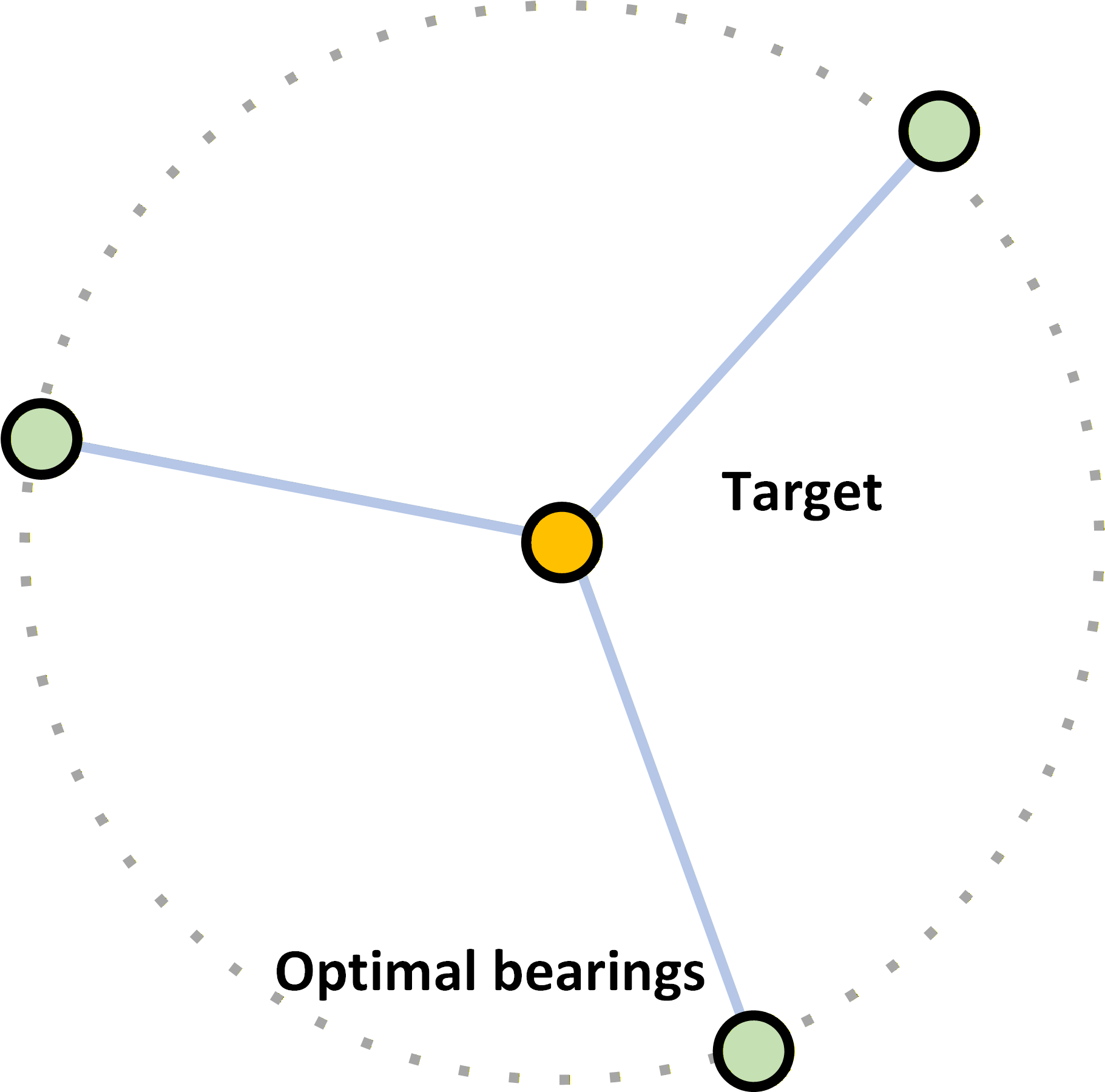}} \qquad \qquad
	\subfloat[3D optimal  bearings.]{\includegraphics[width=0.25\linewidth]{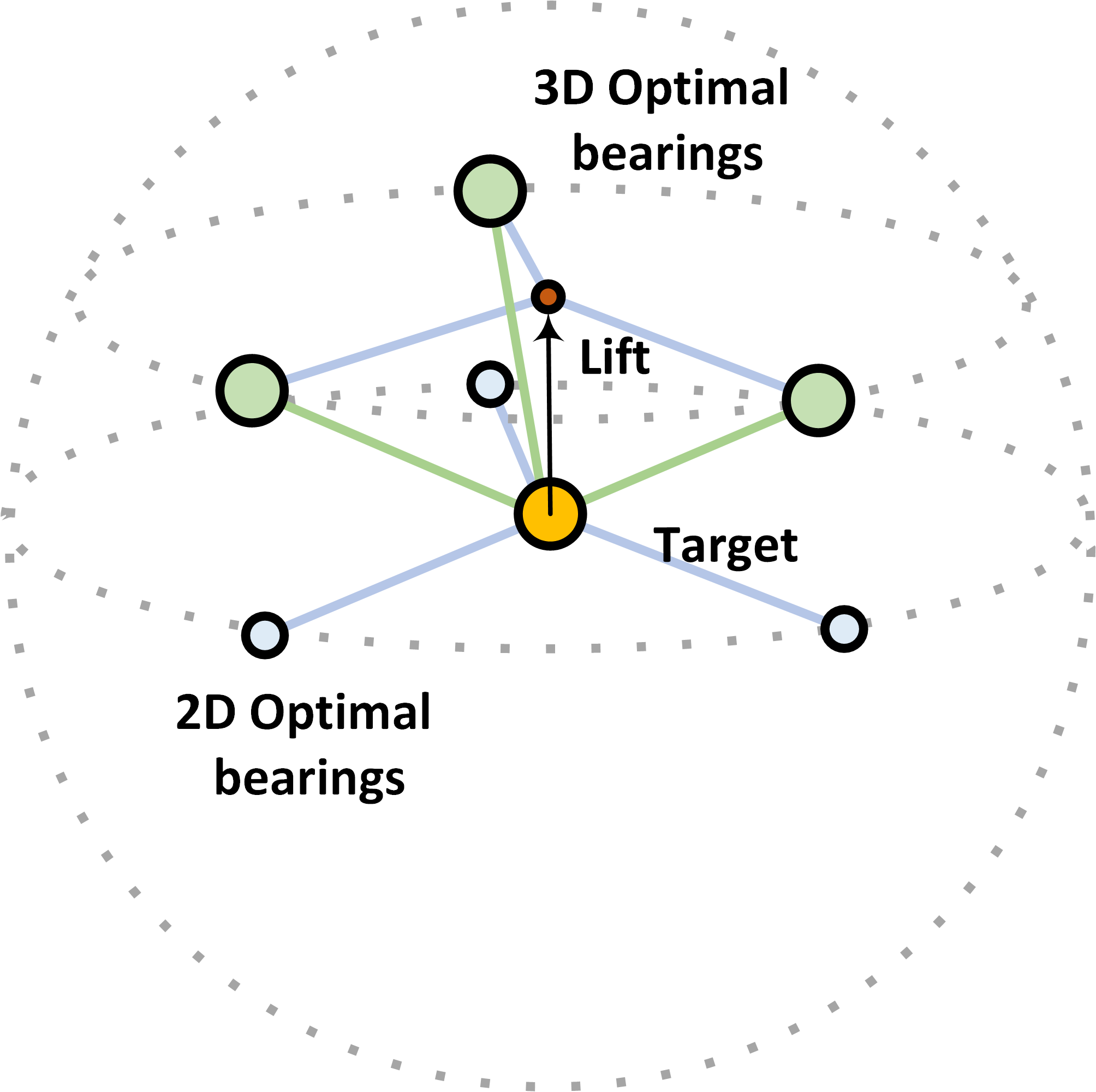}}
	\caption{Optimal desired bearings.}
	\label{bearings}
\end{figure*}

From \cite{zhao2013optimal}, we can know that in the 2D space, the distribution of bearings in a circular orbit is one of the optimal solutions. According to  \citep{benedetto2003finite}, 3D optimal desired bearings can be gotten by lifting 2D ones to the latitude $\frac{\sqrt{3}}{3}$ so as to be evenly spaced in a circle with a radius of $\frac{\sqrt{6}}{3}$ on an unit ball.  An illustration is given in Figure.\ref{bearings}. In summary, the optimal desired bearings for 2D and 3D space are given as
\begin{equation}
	\lambda^{\star}_i = \left\{
	\begin{aligned}
		&\cos(\frac{2\pi i }{m} )\lambda_1+\sin(\frac{2\pi i }{m})\lambda_2,& d=2, \\
		&\frac{\sqrt{3}}{3}\lambda_0 + \frac{\sqrt{6}}{3}(\cos(\frac{2\pi i }{m}  )\lambda_1 + \sin(\frac{2\pi i }{m} )\lambda_2),&  d=3,
	\end{aligned}\right.
\end{equation}
where for $d=2$, $\lambda_1$ and $\lambda_2$ are the unit basis vectors of $\mathbb{R}^2$, and for $d=3$, $\lambda_0$ is the preset unit vector and $\lambda_1$ and $\lambda_2$ are the unit basis vectors of the tangent plane at $\lambda_0$. Therefore, in order to improve the tracking performance in the 2D and 3D space, the AUV needs to sample {\it the optimal desired bearing} as
\begin{equation}\label{op_bearings}
	\lambda^{\star}(t) = \left\{
	\begin{aligned}
		&\cos(\omega t )\lambda_1+\sin(\omega t)\lambda_2,& d=2, \\
		&\frac{\sqrt{3}}{3}\lambda_0 + \frac{\sqrt{6}}{3}(\cos(\omega t  )\lambda_1 + \sin(\omega t  )\lambda_2),&  d=3,
	\end{aligned}\right.\
\end{equation}
where $\omega = \frac{2\pi}{Tm}$ is a preset angular rate with a positive integer $m$.

\begin{remark}
	Collecting bearings data in a circular orbit is  a widely used strategy in many model-based methods, such as FIM-based methods in \cite{li2022three} and \cite{zhao2013optimal}, and PE-based methods in \cite{yang2020entrapping}. Although similar in form, the tracking metric for GP-based target tracking based on data-dependence analysis differs significantly from traditional metrics in terms of mechanism.
\end{remark}

With {\it the optimal desired bearing} $\lambda^{\star}(t)$, the next question is how to sample them without knowing the true position of the target under motion constraints. A common choice is to use the predicted target states in the trajectory optimization problem \citep{he2020trajectory}. But when the initial tracking error is large, this setting may lead to unfavorable measurements being collected.  Alternatively, with the posterior distribution (\ref{eq16}) and (\ref{eq17}) about the unknown target position $P_T(t_{k+i})$, the {\it bearing to be sampled} $\lambda(t_{k+i})$ in the future can be determined with $P_T(t_{k+i}) \sim \mathcal{N}\big(\mu_k(t_{k+i}), \Sigma_{k}(t_{k+i}) \big)$ and the AUV position $P_A(t_{k+i})$. Therefore, the posterior distribution about future bearings to be sampled ${\lambda}(t_{k+i})$  can be obtained. We thus adopt the expected similarity function
\begin{align}\label{eq26}
	\mathcal{F}(\boldsymbol{x}_{k+1: k+p}) =  \sum_{i=1}^{p} \mathbb{E}_{{P}_T(t_{k+i}) }
	\Big(\langle\lambda^{\star}(t_{k+i}), 
	{\lambda}(t_{k+i})\rangle\Big), 
\end{align}
as the cost function,  where $\langle\lambda^{\star}(t_{k+i}), 
{\lambda}(t_{k+i})\rangle$ is the inner product of the desired bearing $\lambda^{\star}(t_{k+i})$ and the bearing to be sampled ${\lambda}(t_{k+i})$.

Due to nonlinearity of bearing measurements, the analytical form of $\mathcal{F}(\boldsymbol{x}_{k+1: k+p})$ in (\ref{eq26}) is not available.  We exploit the Unscented Transformation (UT) \citep{haykin2004kalman} to approximate it with little computation.  To calculate the expectation of this nonlinear function, $2d+1$ sigma points are given by
\begin{align} \label{eq27}
	\chi_0(t_{k+i}) &= \mu_{k}(t_{k+i}),\nonumber\\
	\chi_{j}(t_{k+i}) &= \mu_{k}(t_{k+i}) - \sqrt{(d + \kappa) {\Sigma}_{k}(t_{k+i})}_j, \nonumber\\
	\chi_{d + j}(t_{k+i}) &= \mu_{k}(t_{k+i}) + \sqrt{(d + \kappa) {\Sigma}_{k}(t_{k+i})}_j, \forall j = 1, \dots, d,
\end{align} 
where $\sqrt{(d + \kappa) {\Sigma}_{k}(t_{k+i})}_j$ is the $j$-th column of the matrix square root of $(d + \kappa) {\Sigma}_{k}(t_{k+i})$, $\kappa$ is an adjusting factor. With these sigma points, the bearing from the  AUV to sigma points $\chi_{j}(t_{k+i})$ can be given as
	$\hat{\lambda}_{j}(t_{k+i}) = \frac{\chi_{j}(t_{k+i}) - P_A(t_{k+i})}{\|\chi_{j}(t_{k+i}) - P_A(t_{k+i})\|}, \forall j = 0, \dots, 2d.
$
Then,  the cost function (\ref{eq26}) can be approximated by
\begin{align}\label{eq28}
	\hat{\mathcal{F}}(\boldsymbol{z}_{k+1:k+p}) =  \sum_{i=1}^{p} \sum_{j=0}^{2d}w_j\langle\lambda^{\star}(t_{k+i}), 
	\hat{\lambda}_{j}(t_{k+i})\rangle,
\end{align}
where  $w_0 = \frac{\kappa}{d + \kappa}$ and $w_j = \frac{1}{2d + 2\kappa}$ for $j=1,\dots, d$ are weights. The approximated trajectory optimization problem is given as follows
\begin{subequations}\label{eq29}
	\begin{align}
		\max\limits_{\boldsymbol{x}(t), \tau(t)} \quad & \hat{\mathcal{F}}(\boldsymbol{x}_{k+1: k+p}) \label{eq29a}\\
		\text { s.t. }
		&\dot{\boldsymbol{x}}(t) = f(\boldsymbol{x}(t)) + M^{-1}\tau(t), \quad \forall t \in [t_{k}, t_{k+p}] \label{eq29b}\\
		& -\tau_{\max} \leq \tau(t) \leq \tau_{\max}, \quad \forall t \in [t_{k}, t_{k+p}] \label{eq29c}\\
		& \boldsymbol{x}(t_k) = \boldsymbol{x}_0, \label{eq29d}
	\end{align}
\end{subequations}
where (\ref{eq29b}) is the system dynamics of the  AUV, the constraint (\ref{eq29c}) is control input limits of  $\tau$ and (\ref{eq29d}) is about initial states $\boldsymbol{x}_0 = [\eta(t_k)^{\top}, \nu(t_k)^{\top}]^{\top}$.


\subsection{Differential-flatness-based Parameterization}

Obviously, directly solving the optimization problem (\ref{eq29}) is very hard due to the complex differential system. To avoid it, we exploit the property of {\it differential flatness}  of the AUV system \citep{wang2022geometrically}, and thus the states and control inputs of the AUV system can be algebraically expressed by a set of flat outputs and their derivatives. Choosing $z = [P_A^{\top}, \psi]$ as the flat output of the ($d+1$)-DOF AUV in the $d$-dimensional space,  the states $\boldsymbol{x} = [P_A^{\top}, \psi, v_A^{\top}, r]^{\top}$ and inputs $\tau = [F_v^{\top}, F_r] $ can be decided by
\begin{align}\label{eq30}
	\boldsymbol{x} =& \Psi_{\boldsymbol{x}}(z, \dot{z}) = \left[\begin{array}{c}
		z \\
		J(z)^{\top}\dot{z}
	\end{array}\right], \nonumber\\
	\tau =& \Psi_{\tau}(z, \dot{z}, \ddot{z})
	=
	 M(J(z)^{\top}\ddot{z} + \dot{J}(z)^{\top}\dot{z}) + \Big(C(z,\dot{z}) + D(z,\dot{z})J(z)^{\top}\Big)\dot{z}.
\end{align}

By differential flatness, we can decide the whole trajectory by finding the optimal flat outputs $z(t)$. However,  there are actually infinite flat outputs in the continuous-time domain. To address this issue,  we split the trajectory into $p$ pieces and parameterize the flat output $z_i(t)$ for the $i$-th piece  using $d$-order  polynomials as
\begin{align} \label{eq31}
	z(t)    =& z_i(t), \forall t \in \left[t_{k+i-1}, t_{k+i}\right], \nonumber\\
	z_i(t)	=& \boldsymbol{c}_i^{\top}\beta(t - t_{k+i-1}), \forall t \in \left[t_{k+i-1}, t_{k+i}\right], 
\end{align}
where $\beta(x) = [1, x, \dots, x^{L}]^{\top} \in  \mathbb{R}^{(L+1)}$ is the polynomial basis, and $\boldsymbol{c}_i  \in \mathbb{R}^{(L+1) \times (d+1)}$ the coefficient matrix with $L+1$ coefficients for each state. Therefore, the whole trajectory  can be described by the coefficient matrix $\boldsymbol{c} = [\boldsymbol{c}^{\top}_1, \dots, \boldsymbol{c}^{\top}_{p}]^{\top} \in \mathbb{R}^{ p(L+1) \times (d+1)}$.

It is noted that the cost function (\ref{eq29a}) is  about the flat output, i.e., $\bar{\boldsymbol{z}} =  [z(t_{k+1})^{\top}, \dots, z(t_{k+p})^{\top}]^{\top} \in \mathbb{R}^{(d+1)p}$, which are endpoints of each piece of the trajectory.  Therefore, we choose $L=2$ and the coefficients $\boldsymbol{c}$ can be obtained by solving initialization constraints and continuity constraints, i.e., 
\begin{equation}\label{eq32}
	\left\{
	\begin{array}{lr}
		\boldsymbol{c}_1^{\top}\beta(0)= \eta(t_{k}), & \\
		\boldsymbol{c}_1^{\top}\dot{\beta}(0) =  J(\eta(t_k))\nu(t_{k}), &  \\
		\boldsymbol{c}_i^{\top}{\beta}(T) =  \bar{z}(t_{k+i}), i = 1, 2, \dots, p, &  \\
		\boldsymbol{c}_{i}^{\top}\beta(T)=  \boldsymbol{c}_{i + 1}^{\top}\beta(0), i = 1, 2, \dots, p - 1,& \\
		\boldsymbol{c}_{i}^{\top}\dot{\beta}(T)= \boldsymbol{c}_{i + 1}^{\top}\dot{\beta}(0), i = 1, 2, \dots, p -1,& 
	\end{array}
	\right.
\end{equation}
which are linear equations and can be solved efficiently.

Compared with directly optimizing the coefficients, fewer variables need to be optimized when optimizing intermediate points $\bar{\boldsymbol{z}}$ and continuity of piece-wise trajectories can be guaranteed. It is noted that (\ref{eq29c}) is non-convex by (\ref{eq2}), so the optimization problem (\ref{eq29}) is sensitive to the selection of initial solutions for numerical solvers. When endpoints are optimized, initial solutions about optimal variables can be set by analyzing their physical meanings, which can help improve solving efficiency. Then the trajectory optimization problem (\ref{eq29}) can be rewritten as
\begin{subequations} \label{eq33}
	\begin{align}
		\max\limits_{\bar{\boldsymbol{z}}} \quad  &\hat{\mathcal{F}}(\bar{\boldsymbol{z}}) \label{eq33a}\\
		\text { s.t. } 	
		&  -\tau_{\max} \leq \mathcal{G}_i \leq \tau_{\max}, \quad \forall t \in [t_{k + i- 1}, t_{k+i}],  \forall i \in\{1,\dots, p\} \label{eq33b}
	\end{align}
\end{subequations}
where $\mathcal{G}_i  = \mathcal{G}_i(z_0, \dot{z}_0, \bar{\boldsymbol{z}}, t) \in \mathbb{R}^{d+1}$ is  the expression of $\tau(t)$ calculated with (\ref{eq30}) and (\ref{eq31}) using intermediate points $\bar{\boldsymbol{z}}$ given $z_0$ and $\dot{z}_0$.

Note that $z_0$ and $\dot{z}_0$ are embedded in the constraint (\ref{eq33b}), leading to infinite constraints on control inputs. Motivated by \cite{li2021constrained}, we transform continuous-time infinite constraints into finite constraints by the integral of constraint violations. For the piece-wise trajectory, we define
\begin{align} \label{eq34}
	\mathcal{I}(\bar{\boldsymbol{z}}) =& \sum\limits_{i=1}^{p} \sum\limits_{\rho = 1}^{d+1} \int_{t_{k +i -1}}^{t_{k+i}}g\big(\mathcal{G}^{\rho}_{i}(z_0, \dot{z}_0, \bar{\boldsymbol{z}}, t) - \tau^{\rho}_{\max}\big) + \nonumber\\
	& g\big(-\mathcal{G}^{\rho}_i(z_0, \dot{z}_0, \bar{\boldsymbol{z}}, t) -\tau_{\max}^{\rho}\big)dt,
\end{align}
where $\mathcal{G}^{\rho}_{i}(z_0, \dot{z}_0, \bar{\boldsymbol{z}}, t)$ and $\tau^{\rho}_{\max}$ are the $\rho$-the element of $\mathcal{G}^{\rho}(z_0, \dot{z}_0, \bar{\boldsymbol{z}}, t)$, $\tau_{\min}$ and $\tau_{\max}$, respectively, and
\begin{align} 
	g(x) = \left\{\begin{array}{ll}
		x, &x > \varpi, \nonumber\\
		\frac{(x -  \varpi)^2}{4\varpi}, & -\varpi \leq x \leq \varpi, \nonumber\\
		0, &x < -\varpi,\nonumber
	\end{array}\right.
\end{align}
and $\varpi$ is a positive number. Once $	\mathcal{I}(\bar{\boldsymbol{z}}) = 0$, the constraints (\ref{eq33b}) hold for all time instants in $t \in [t_{k}, t_{k+p}]$. Moreover, we exploit the Trapezoidal Rule to approximate the integral in (\ref{eq34}) as
\begin{align}\label{eq35}
	\hat{\mathcal{I}}(\bar{\boldsymbol{z}}) = & \sum\limits_{i=1}^{p} \sum\limits_{\rho = 1}^{d+1}\frac{T}{n_c} \sum_{j=0}^{n_c}\omega_{j}g\big(\mathcal{G}^{\rho}_{i}(z_0, \dot{z}_0, \boldsymbol{z}, t_{k+i-1} + \frac{jT}{n_c}) - \nonumber\\
	&\gamma\tau^{\rho}_{\max}\big) + 
	\omega_{j}g\big( -\mathcal{G}^{\rho}_{i}(z_0, \dot{z}_0, \boldsymbol{z}, t_{k+i-1} + \frac{jT}{n_c}) -\gamma\tau^{\rho}_{\max}\big), 
\end{align}
where $n_c$ is the resolution rate, $(\omega_0, \omega_1, \dots, \omega_{n_c - 1}, \omega_{n_c}) = (1/2, 1, \dots, 1, 1/2)$ are weights in the trapezoidal rule and $\gamma \in (0, 1]$ is a scaling factor for avoiding large constraint violation in the penalty method.

Combining these two approximated functions (\ref{eq28}) and (\ref{eq35}), with the penalty method, we can rewrite the constrained optimization problem (\ref{eq33}) into the following unconstrained one
\begin{align} \label{eq36}
	\min\limits_{\bar{\boldsymbol{z}}} \quad \mathcal{J}(\bar{\boldsymbol{z}}) = - \hat{\mathcal{F}}(\bar{\boldsymbol{z}})+ w_p \hat{\mathcal{I}}(\bar{\boldsymbol{z}}),
\end{align}
where $w_p > 0$ is the penalty weight. The overall GBT framework is summarized in Algorithm \ref{Overall}, which includes sampling, target tracking, trajectory optimization and trajectory following.  Through the GBT framework with learning-based target tracking and trajectory optimization, the  AUV can achieve efficient tracking for  an unknown maneuvering target.

\begin{remark}
		In the prediction step, the complexity of hyperparameters optimization in each iteration is $\mathcal{O}(N^3)$ where $N \leq N_c$, and the complexity of the posterior prediction is $\mathcal{O}(nN^3)$ \citep{williams2006gaussian, alvarez2008sparse}. In the optimization step, the unconstrained NLP can be solved using the gradient descent method, whose complexity is $\mathcal{O}(Lp)$ in each iteration and $L = \Theta(pn_c)$ is the exact asymptotic order. With the bounded complexity in each iteration, the maximum solving time can be decreased by setting a lower maximum solving iteration and the average solving efficiency can be improved by tuning tolerance \citep{mattingley2012cvxgen}.  According to computing abilities, we can tune these parameters to ensure that the overall computing time does not exceed the minimum real-time requirement.
\end{remark}

\begin{algorithm}[!h]\label{Overall}
	\caption{GP-based Bearing-only Tracking (GBT)}
	\SetKwProg{Fn}{Initialization}{:}{\KwRet}
	\Fn{}{
		Initialise the data set $D_{-1:0} = \emptyset$\;
		Initialise optimal desired bearings $\lambda^{\star}(t)$ by (\ref{op_bearings})\;
		Initialise iteration $k = 0$;
	}
	\textbf{For} $k = k + 1$ \Repeat{Maneuvering target tarcking}{
		\SetKwProg{Fn}{Sampling}{:}{\KwRet}
		\Fn{}{
			Collect data $\{t_{k}, \lambda(t_k), P_A(t_k)\}$ and Update $\mathcal{D}_{c:k}$ ;
		}
		\SetKwProg{Fn}{Target tracking}{:}{\KwRet}
		\Fn{}{
			Compute $\boldsymbol{G}$, $\boldsymbol{y}$  by  (\ref{eq5}) and (\ref{eq10}) \;
			Solve (\ref{like1}) to tune hyperparameters $\boldsymbol{\theta}$\;
			\ForAll{$t^{\star} \in \mathcal{T}_k$}{
				Compute $\Omega_{\boldsymbol{y}\boldsymbol{y}}$ and $\Omega_{\boldsymbol{y}P}^{\star}$ by (\ref{eq15})\; 
				Compute the expectation $\mu_{k}(t^{\star})$ by (\ref{eq16}) and output it;
			}
		}
		\SetKwProg{Fn}{Trajectory Optimization}{:}{\KwRet}
		\Fn{}{
			\ForAll{$t^{\star} \in \mathcal{T}_k$}{
				Compute sigma points $\chi_{j}(t^{\star})$ by (\ref{eq27})
			}
			Compute optimal endpoints $\bar{\boldsymbol{z}}$ by solving (\ref{eq36})\;
			Compute optimal coefficients $\boldsymbol{c}$  by (\ref{eq32})\;
			Compute  trajectory $(\boldsymbol{x}(t), \tau(t))$ for $ t \in [t_{k}, t_{k+p}]$ by (\ref{eq30});
		}
		\SetKwProg{Fn}{Trajectory Following}{:}{\KwRet}
		\Fn{}{
			AUV follows planned trajectory $(\boldsymbol{x}(t), \tau(t))$ for  $t \in [t_{k}, t_{k+1}]$;
		}
	}
\end{algorithm}
\section{Numerical Validation}
In this section, we conduct simulation to verify practicability and effectiveness of our proposed GBT framework.

\subsection{Simulation Setup}
We consider the  AUV model  with rigid body parameters and hydrodynamic coefficients, which is from the identified dynamic model of {\it Falcon}. \cite{shen2016integrated}. The mass $m=116(\mathrm{kg})$; the moment of inertia with respect to $z$-axis $I=13.1\left(\mathrm{kg} \cdot \mathrm{m}^2\right)$. The values of added mass in surge, sway, heave  and yaw are $X_{\dot{u}}=-167.6, Y_{\dot{v}}=-477.2$, $Z_{\dot{w}} = -235.7$ $N_{\dot{r}}=-15.9$. And  the values of linear drag in surge, sway, heave  and yaw are $X_u=26.9, Y_v=35.8, Z_{w} = 6.19, N_r=3.5$. And And  the values of quadratic drag in surge, sway, heave  and yaw are $D_u=241.3, D_v=503.8, D_w = 119.1$, and $D_r=76.9$. Here, the maximum and minimum control forces $F_{\nu, \max}= 5000 \cdot \boldsymbol{1}_{d}(\mathrm{N})$, and $ F_{r,\max} =1500(\mathrm{N \cdot m })$. On implement of target tracking, we consider the following points about simulation and algorithm setting.

(1) The AUV samples the bearing to target in $10$ Hz,i.e., $T = 0.1(s)$, which is a common setting as a baseline. In robustness analysis, performance with different periods will be tested.

(2) Consider the limited computation capacity of the AUV,  $N_c = 20$ for 2D space and $N_c = 12$ for 3D space are used.  On the length of prediction horizon and planning horizon, we select $n = 11$ and $p = 5$. Different lengths  will be compared with simulations.

(3) On choice of kernels, the Squared Exponential  kernel is used to fit target motion, which is the most commonly used kernel. The nugget regularization is exploited to improve computational stability. 

(4) For trajectory optimization, we select the desired bearings with $m=10$ for 2D space and $m=4$ for 3D space and parameters on optimization problem transformation with $\kappa = 1$, $n_c = 20$, $\gamma = 0.98$ and $w_p = 1000$.

\subsection{Effectiveness of the Proposed GBT} 

The following six different types of target motion are considered, in which Case 1-3 are designed in 2D space and Cases 4-6 are designed in 3D space.

{\it Case 1. Constant-velocity target}:  The target moves with a constant velocity $[1, 1](m/s)$ from  $P_T(t_0)=[-1,-1]^{\top}(m)$, that is, $x_T(t) = -1 + t, y_T(t) = -1 + t.$

{\it Case 2.  Target on an 8-shaped orbit}: The target moves on an 8-shaped orbit centered at $[0, 0](m)$ from $P_T(t_0)=[3,0](m)$, that is, $x_T(t) = \frac{3\cos(\frac{\pi}{8} t)}{(1+\sin^2(\frac{\pi}{8} t))^2}$, $y_T(t) = \frac{1.5\sin(\frac{\pi}{4} t)}{(1+\sin^2(\frac{\pi}{8} t))^2}$.

{\it  Case 3. Target on a circular orbit with time-varying velocity}: The target follows the 2D nonholonomic model and  moves from $P_T(t_0)=[4, 0](m)$ on a circular orbit centered at $[0, 0](m)$ and a time-varying angular velocity with $\psi_T(0) = 0 ({\rm rad})$, i.e.,$x_T(t) = 4\cos(\psi_T(t))$, $y_T(t) = 4\sin(\psi_T(t))$, $\dot{\psi}_T(t) = 0.3 + 0.2 \sin(0.4t)$.

{\it Case 4. Constant-velocity target}:  The target motion in the $x-y$ plane is the same as the one in Case 1. In the $z$-axis, the target moves with a constant velocity $0.1(m/s)$ from $0(m)$, i.e, $z_T(t) = 0.1t$.

{\it Case 5.  Target on an 8-shaped orbit in the $x-y$ plane and with periodic motion in $z$-axis}:  The target motion in the $x-y$ plane is the same as the one in Case 2. In the $z$-axis, the target moves periodically from $0(m)$, i.e, $z_T(t) =  3\sin(\frac{\pi}{8} t)$.

{\it  Case 6. Target on a circular orbit with time-varying velocity in the $x-y$ plane and periodic motion in $z$-axis }: The target motion in the $x-y$ plane is the same as the one in Case 3. In the $z$-axis, the target moves periodically from $0(m)$, i.e, $z_T(t) = 5\sin(0.2t)^2$.

\begin{figure*}[!t]
	\centering
	\subfloat[Case 1.]{\includegraphics[width=1.8in]{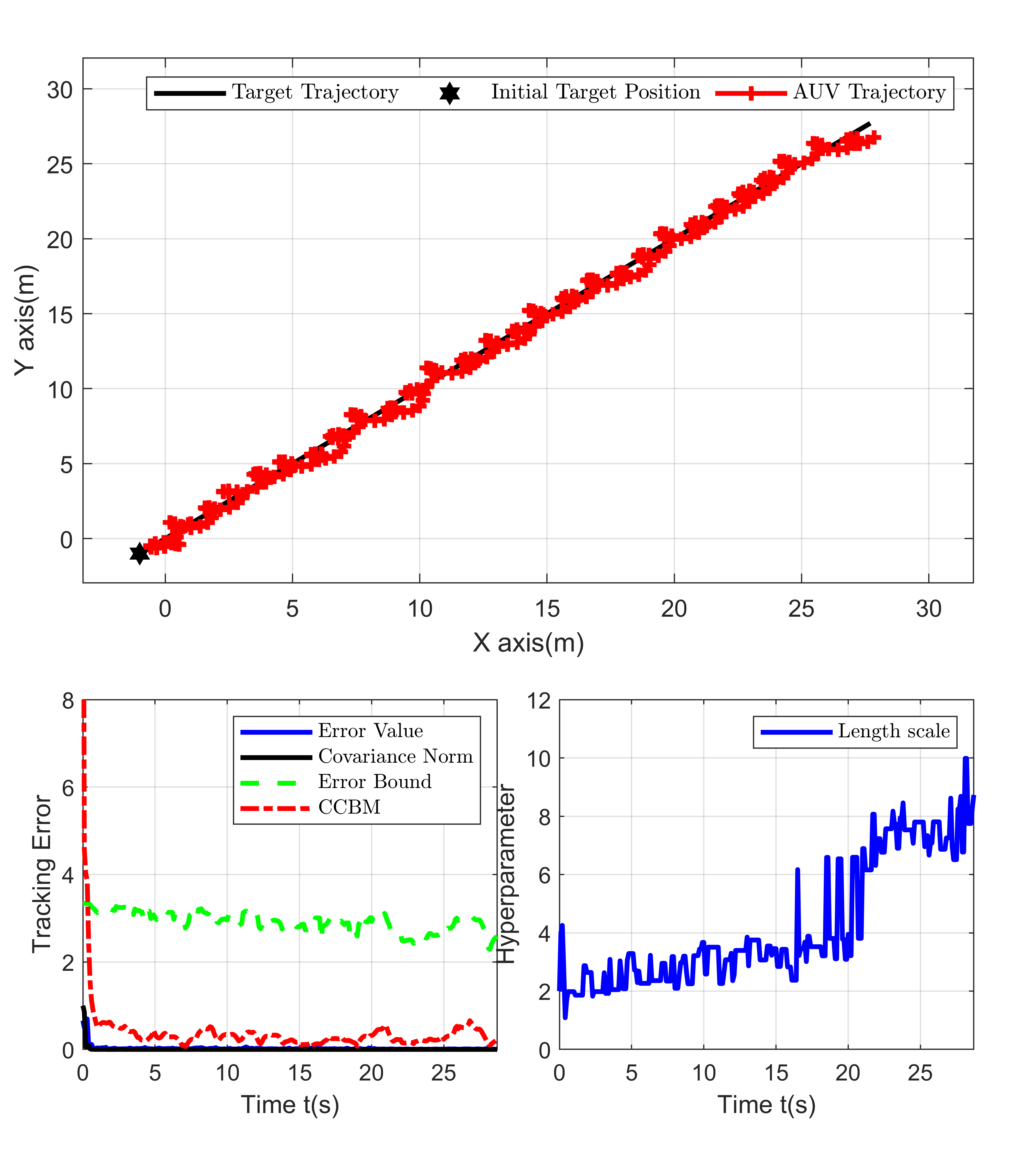}%
		\label{2D_Case1}}
	\subfloat[Case 2.]{\includegraphics[width=1.8in]{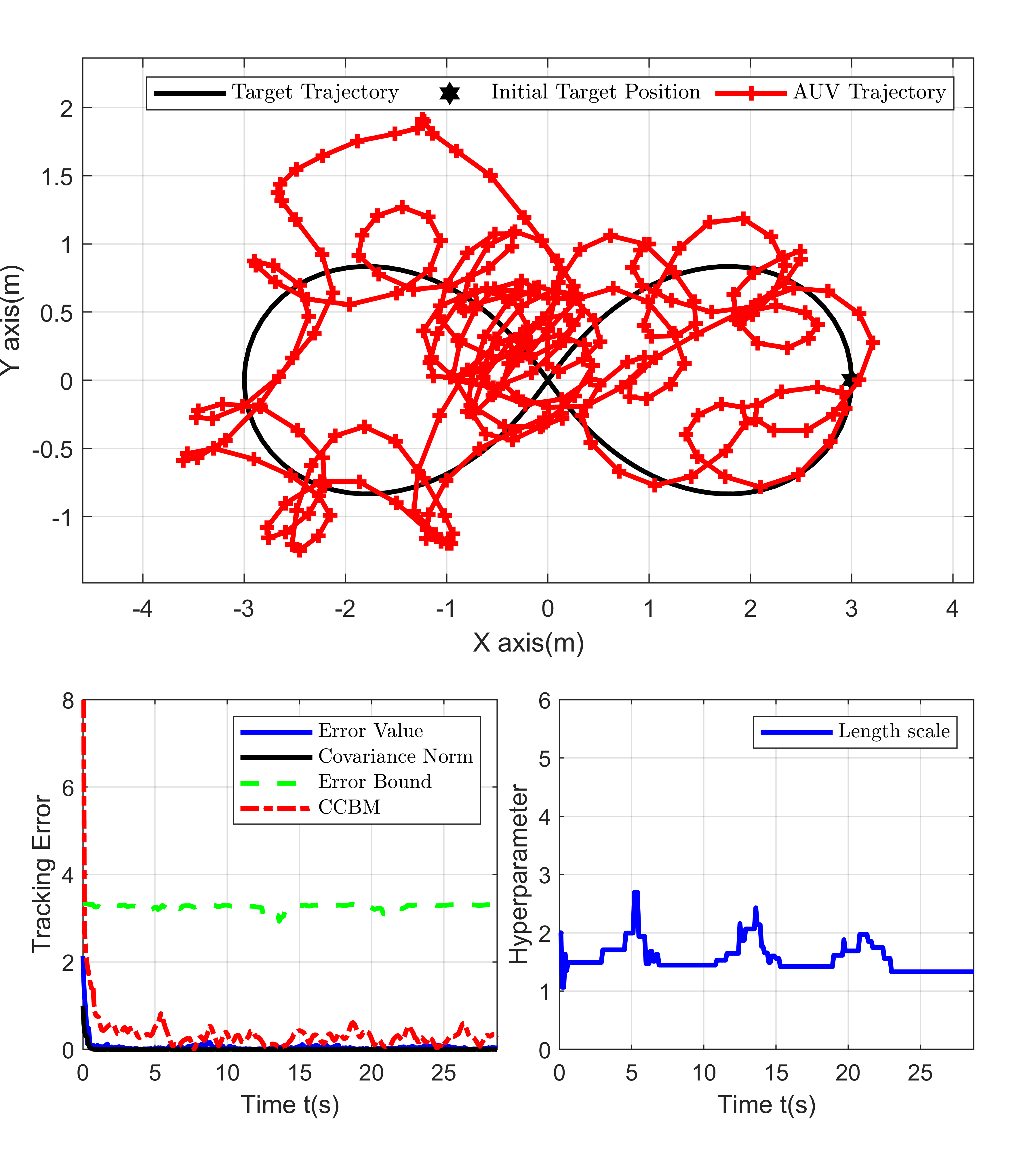}%
		\label{2D_Case2}}
	\subfloat[Case 3. ]{\includegraphics[width=1.8in]{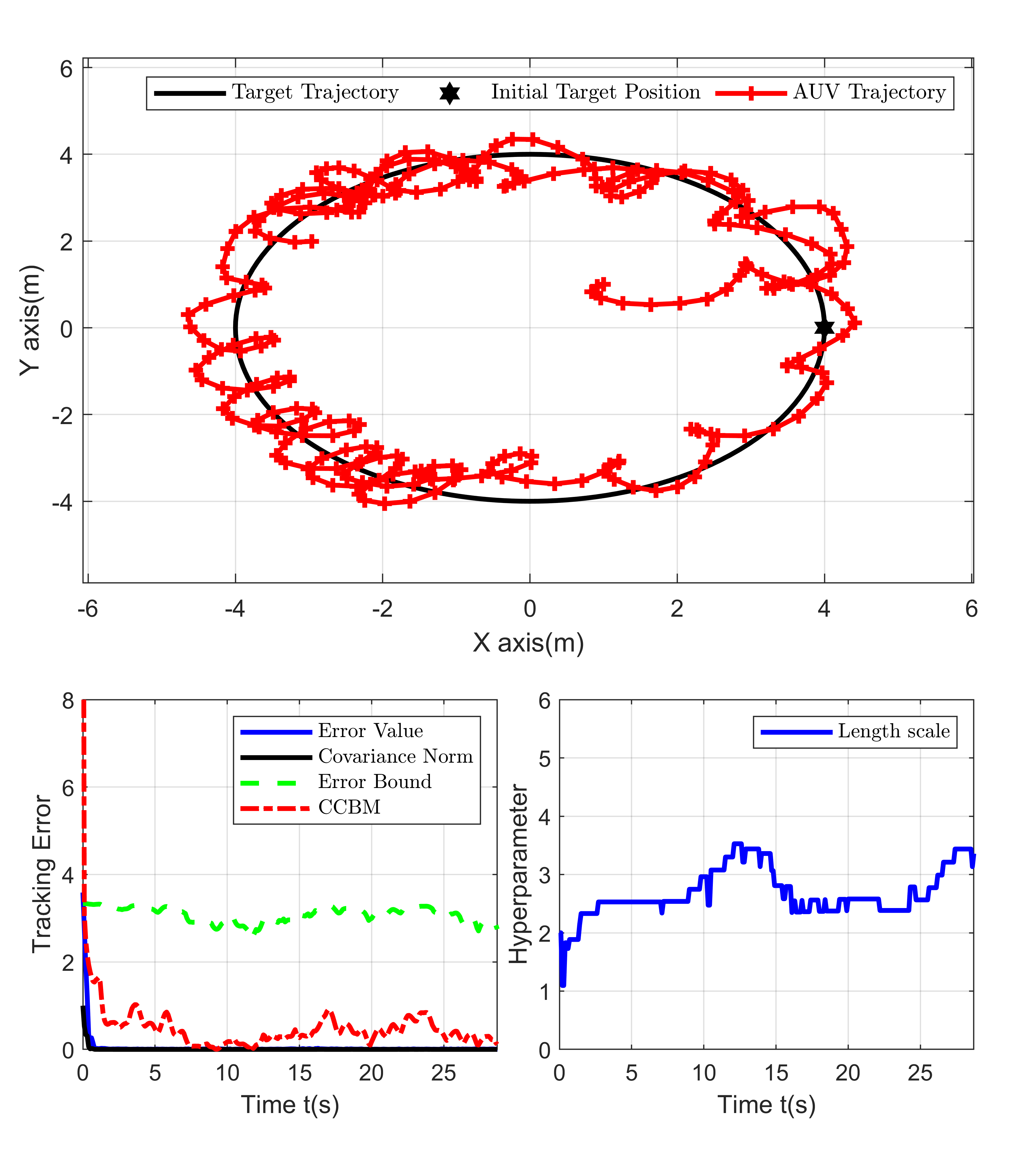}%
		\label{2D_Case3}}\\
	\subfloat[Case 4.]{\includegraphics[width=1.8in]{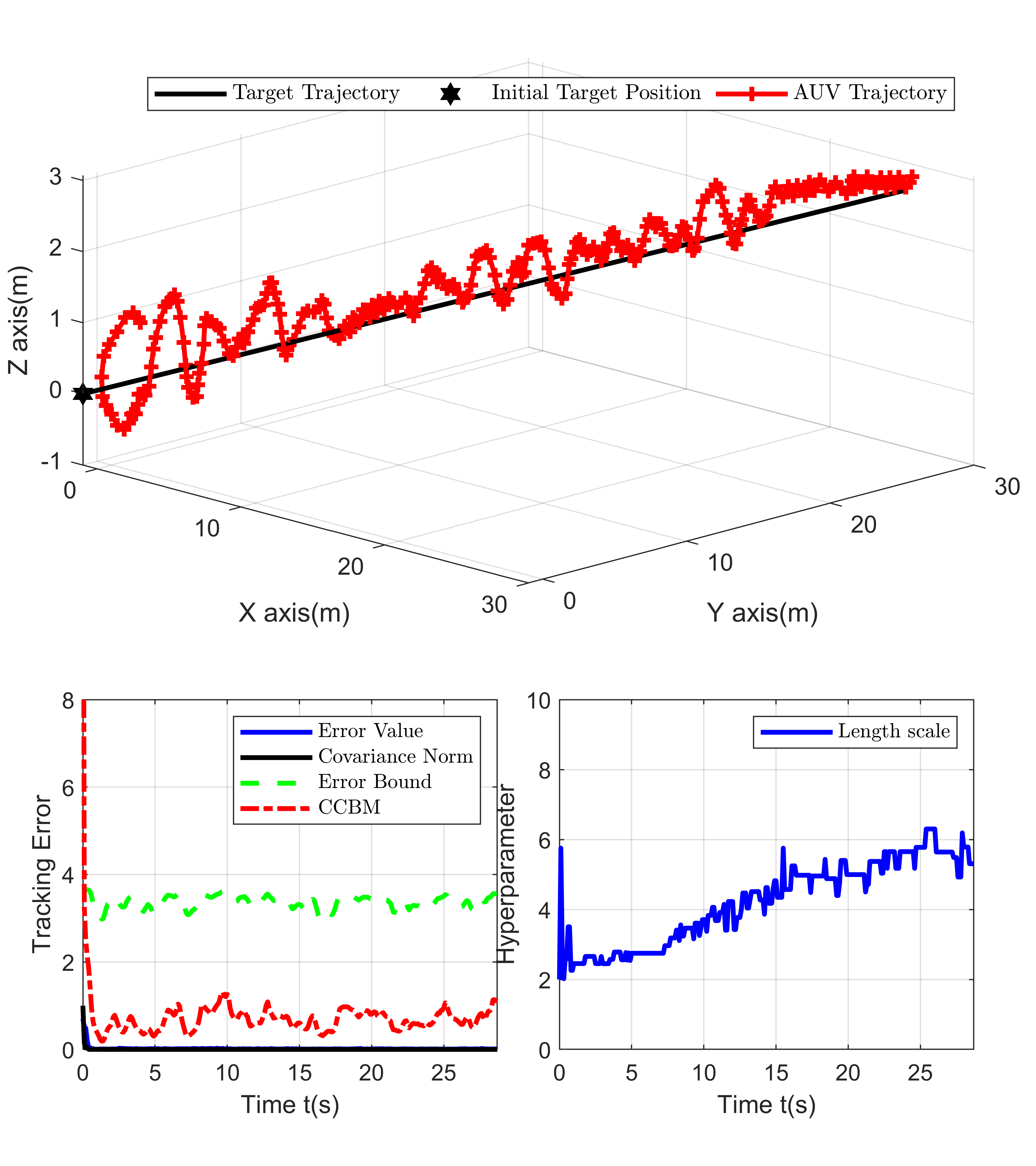}%
		\label{3D_Case4}}
	\subfloat[Case 5.]{\includegraphics[width=1.8in]{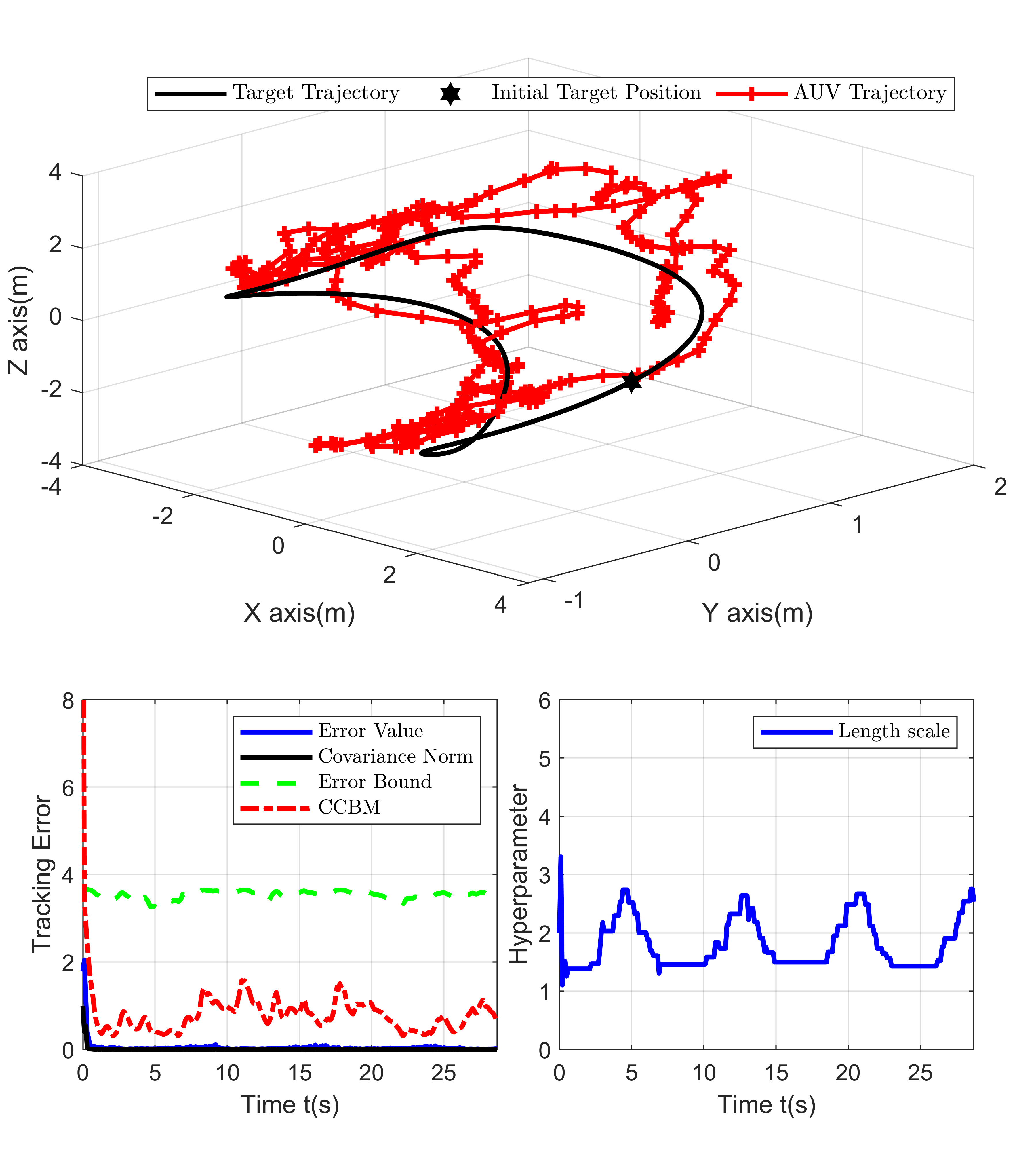}%
		\label{3D_Case5}}
	\subfloat[Case 3. ]{\includegraphics[width=1.8in]{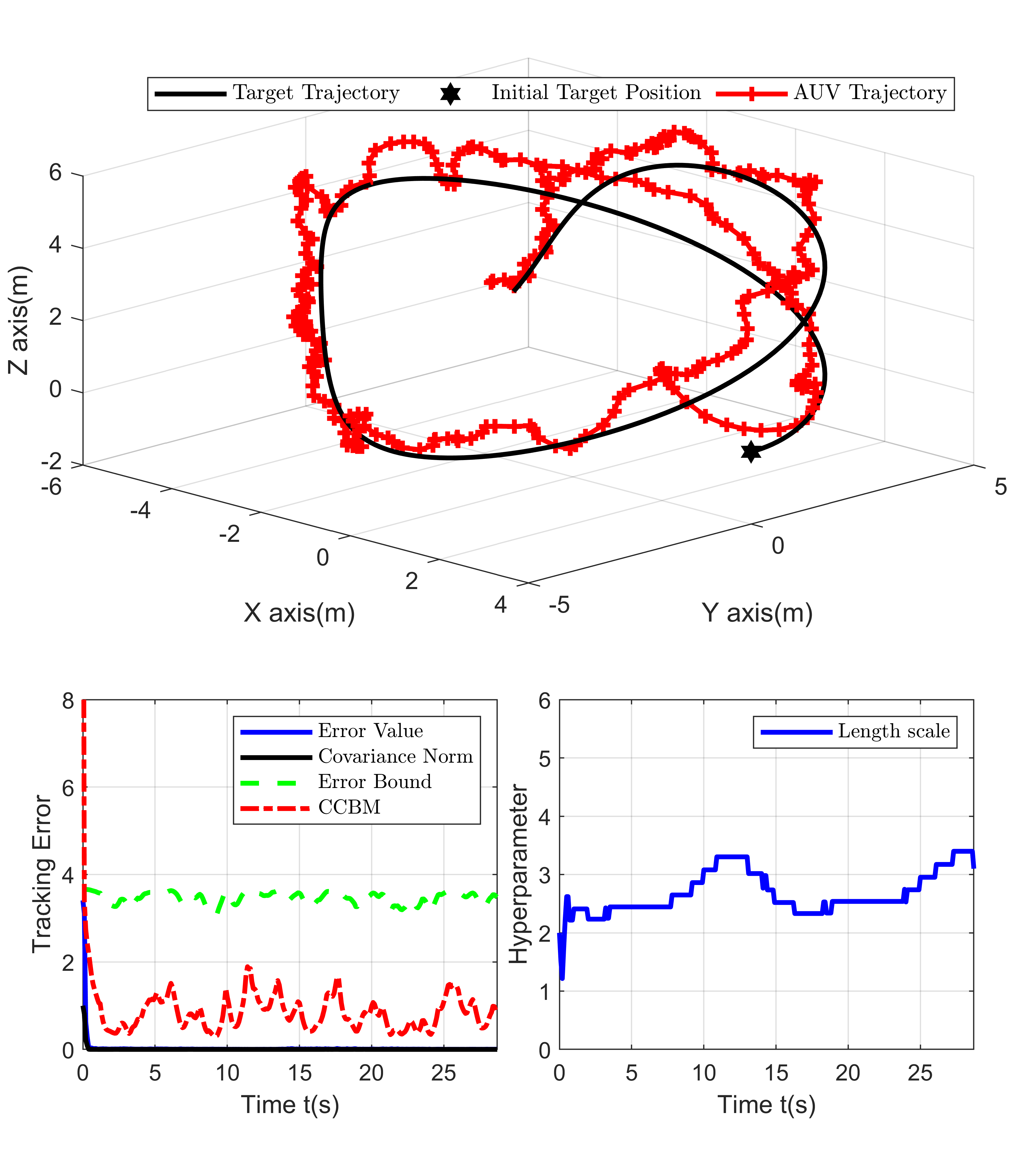}%
		\label{3D_Case6}}
	\caption{Simulation results.}
	\label{Figure4}
\end{figure*}

Simulation results for various different types of target motion in the 2D space and 3D space are given in Figure.\ref{Figure4}\subref{2D_Case1}-\subref{2D_Case3} and Figure.\ref{Figure4}\subref{3D_Case4}-\subref{3D_Case6}, respectively. Here, motion  motion trajectories, tracking error and the hyperparameter of the kernel  are given in each figure.

First, from each figure of motion trajectories, in the 2D space, the AUV moves surrounding the moving target to collect bearing measurements in a circular orbit, which are optimal desired bearings by our analysis. On the other hand, in the 3D space, AUV follows the target along the helical line to collect optimal bearing bearing measurements. The behaviour of the AUV is generated by our proposed method of trajectory optimization to improve tracking performance.

Second, with the actively collected bearing measurements, the mean tracking error and covariance norm of all time instants in $\mathcal{T}_k$ decreases rapidly within $0.1(m)$, showing effectiveness of our framework both in the 2D and 3D space. It indicates that the posterior covariance generated by GP is a useful metric. It provide solid foundations for analysis about the error bound. The corresponding bearing-data-dependent error bound  and the logarithm of the condition number of the cumulative bearing matrix (CCBM) are also shown.   And it clearly indicates validity of the probabilistic tracking error bound and strong correlation between tracking error and CCBM, statistically validating CCBM as an effective  quality metric of bearing data in target tracking applications.

Third, in both 2D and 3D space, the SE kernel is used in Gaussian process learning and prediction. The corresponding length scale is also shown. From them, we can see that the hyperparameter is adjusted online to fit target motion locally so that highly nonlinear unknown motion can be captured.

\subsection{Effects of AUV Motion for Sampling Bearing Data}

To quantitatively validate the necessity of trajectory optimization for bearing-only target tracking, we conduct a comparative tracking performance analysis under different AUV motion modes: static motion, constant-force motion, and randomly moving motion as follows, labeled as {\it Static}, {\it Constant-force} and {\it Random}, respectively.

{\it  Mode 1. Static}: The AUV is static in the initial position without control force, i.e., $\tau = \boldsymbol{0}_{d+1}$.

{\it  Mode 2. Constant-force}: The AUV works with a constant control force, i.e., $F_{\nu} = 500 \cdot \boldsymbol{1}_d$ and $F_{r} = 0$.

{\it  Mode 3. Random}: The AUV works with the random control force and torque.

\begin{figure*}[!t]
	\centering
	\subfloat[Tracking error(Case 2).]{\includegraphics[width=2.5in]{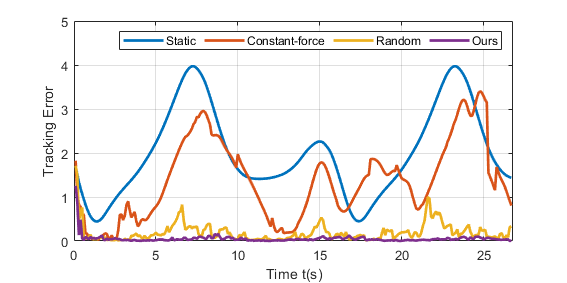}%
		\label{Figure_4(a)}}
	\subfloat[Tracking error(Case 5).]{\includegraphics[width=2.5in]{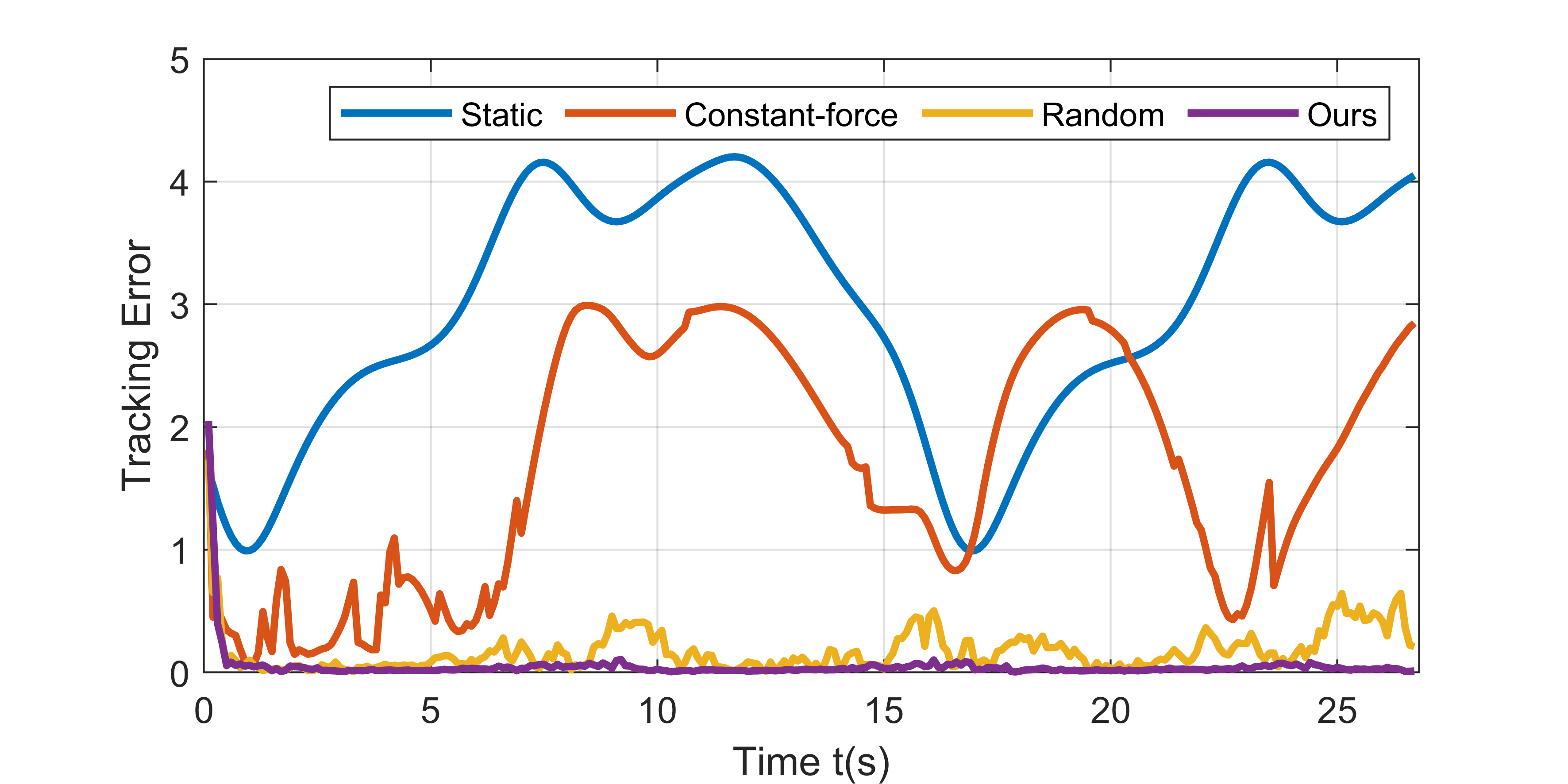}%
		\label{Input_case2}}
	\caption{Comparison of results with different AUV motion modes.}
	\label{Figure_5}
\end{figure*}

Here, the Case 2 and Case 5 are considered as typical scenes in 2D and 3D space, respectively. The simulation results  are shown in  Fig. \ref{Figure_5}. It can be observed that the tracking error for a static AUV  diverges, which shows that motion of the AUV is necessary for learning unknown target motion  with bearing-only measurements.  On the other hand, the tracking error for a constant-force AUV and randomly moving AUV both decrease, but there are still significant residuals about $0.3-1(m)$ at some instants. As mentioned before, bearing-based learning heavily requires that the collected bearing data should fully reflect the motion patterns of the target. Although the same number of bearing data are collected for these models, uncertainty about target motion has not been eliminated by a sufficiently rich amount of information. In comparison, the AUV with trajectory optimization can accurately track the moving target.  The proposed method enables improved learning of target motion patterns from the bearing data, which can not be realized by other motion modes.

\subsection{Comparison with Other Tracking Methods}

To show the superior performance,  we compare our GBT framework with the methods in \cite{yang2020entrapping}, \cite{hu2021bearing} and \cite{li2022three},  which are very representative methods in bearing-only target tracking. In \cite{yang2020entrapping}, a  static target in entrapped with an orthogonality-based estimator, denoted by OE-S. In \cite{hu2021bearing}, a constant-velocity target is tracked with its proposed orthogonality-based estimator, denoted as OE-CV. In \cite{li2022three}, a pseudo linear Kalman filter (PLKF) is proposed to track a constant-velocity target. While OE-S, OE-CV, and PLKF are not designed for varying-velocity targets, they are included as representative examples of classical bearing-only tracking algorithms. For most cases, when the motion of the target is not particularly complex,  modeling as a constant-velocity target is effective in practice. Their inclusion serves to illustrate the performance gap that arises when using methods that assume restrictive motion models in a maneuvering target scenario. As typical methods in underwater and bearing-only target tracking, IMM combined with KF using full measurements in \cite{li2023adaptive} and IMM combined with PLKF \citep{hao2021imm} are compared, and denoted by IMM-KF-F and IMM-PLKF, respectively. In both IMM-based methods, CV, CT and CA models are used. At the same time, in \cite{ning2024bearing}, a $n$-th order polynomial is used to approximate target motion, which is also compared and denoted by PR. 

\begin{figure*}[!t]
	\centering
	\subfloat[Tracking error(Case 1).]{\includegraphics[width=2.5in]{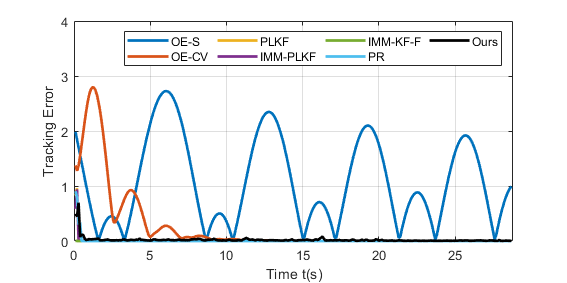}%
		\label{Comparison_Case1}}
	\subfloat[Tracking error(Case 2).]{\includegraphics[width=2.5in]{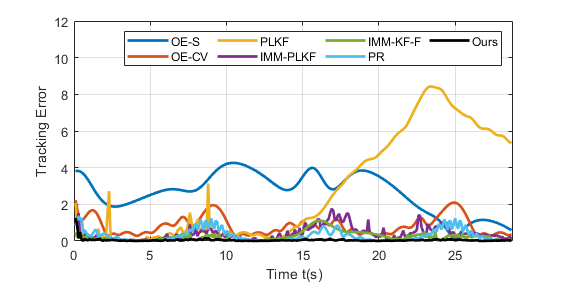}%
		\label{Comparison_Case2}}\\
	\subfloat[Tracking error(Case 3). ]{\includegraphics[width=2.5in]{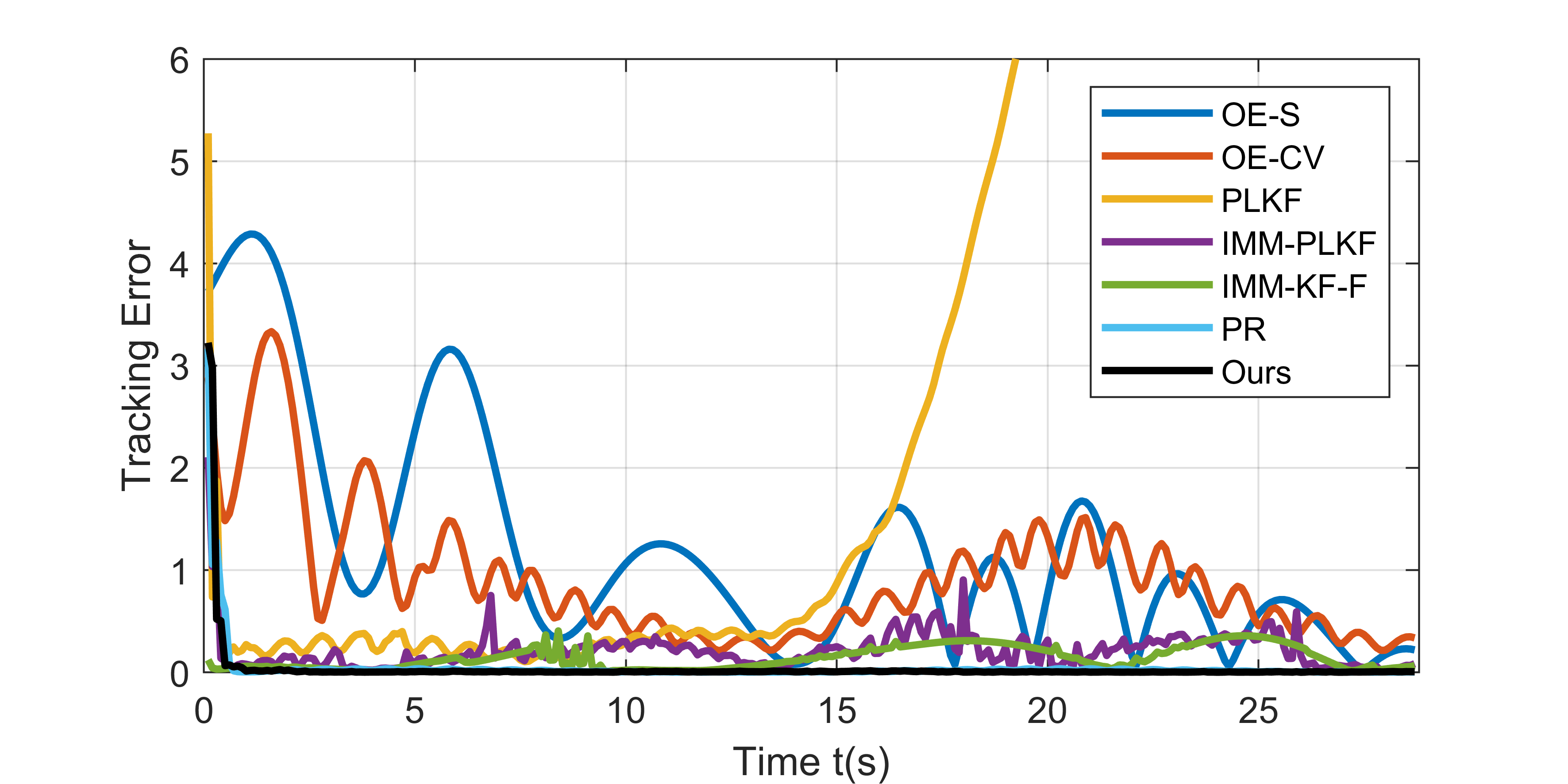}%
		\label{Comparison_Case3}}
	\subfloat[Tracking error(Case 4). ]{\includegraphics[width=2.5in]{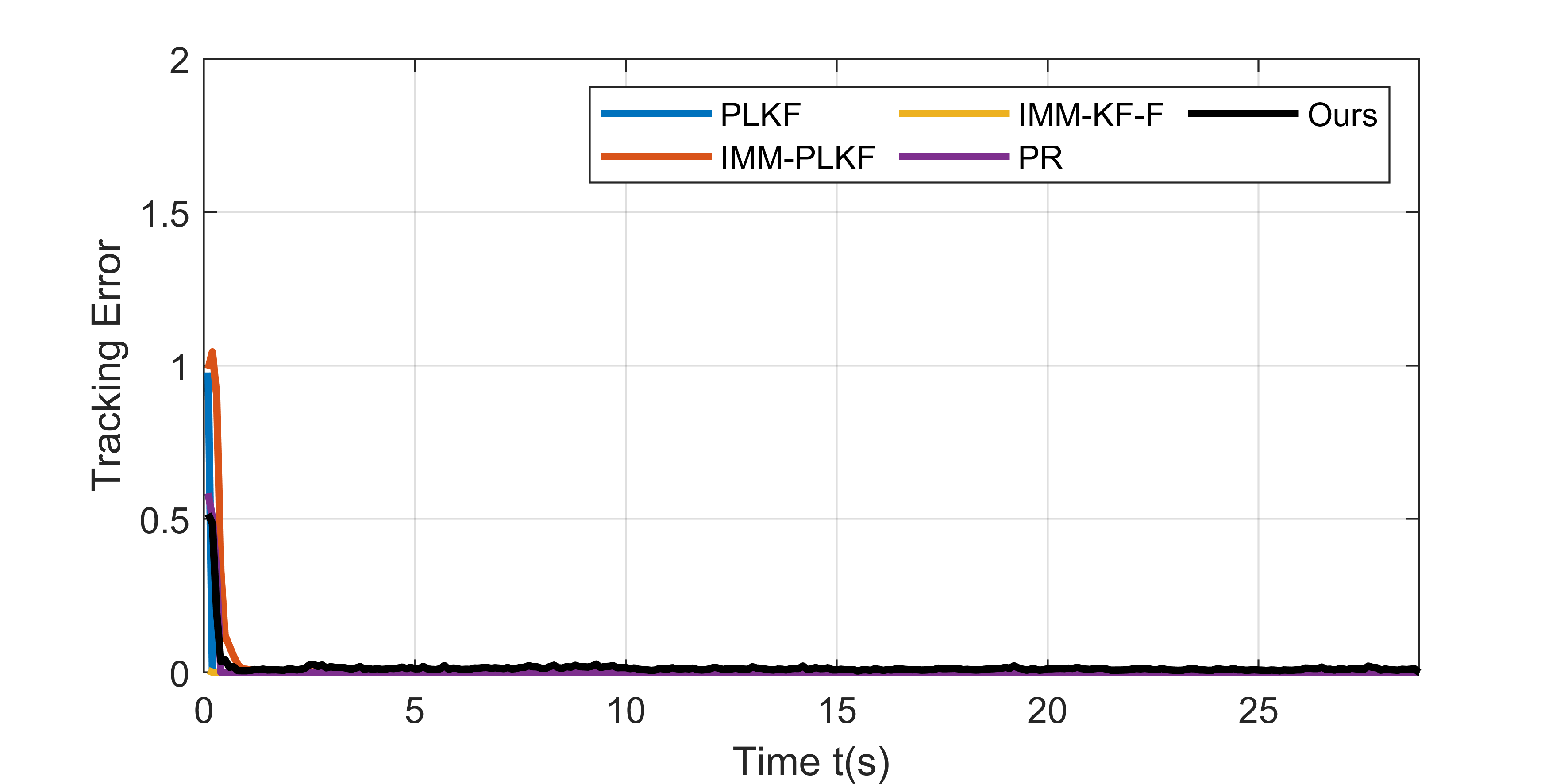}%
		\label{Comparison_Case4}}\\
	\subfloat[Tracking error(Case 5). ]{\includegraphics[width=2.5in]{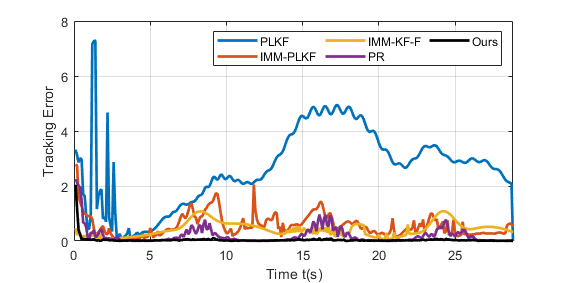}%
		\label{Comparison_Case5}}
	\subfloat[Tracking error(Case 6). ]{\includegraphics[width=2.5in]{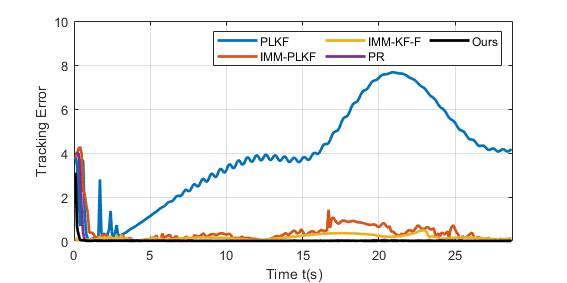}%
		\label{Comparison_Case6}}
	\caption{Comparison with other methods.}
	\label{Figure6}
\end{figure*}
	
To tune parameters of compared methods, grid search is exploited to optimize the parameters of compared methods.  The parameter grid of compared methods in grid search are given as follows. For OE-S, the parameter grid includes gains $k_1 = \{0.5, 1, 5, 10\}$. For OE-CV, the parameter grid includes gains $c_1 = \{0.5, 1, 5, 10\}$ and $c_2 = \{0.1, 0.5, 1, 2\}$. For PLKF and IMM-PLKF, the parameter grids include initial covariance $P_0 = \{0.5, 1, 5, 10\}$. For PR, the parameter grid includes polynomial orders $\{3, 5, 8, 10\}$. To avoid unfairness caused by AUV mobility strategies, PE-based and FIM-based methods have been included for a comparison for tracking targets. For OE-S and OE-CV,  the method in \cite{wang2024target} is used to trigger their PE conditions and the parameter grid includes gains $\beta = \{1, 2, 5, 10\}$, which has been proven to be effective for bounded-velocity targets. For 2D space, the tracking performance of PLKF, IMM-PLKF and PR are improved by directly placing AUVs in the optimal configuration positions based on FIM analysis proposed in \cite{zhao2013optimal} and the parameter grid includes configuration numbers $n = \{10, 15, 20\}$. For 3D space, the navigation law of maximizing the determinant of FIMs in \cite{li2022three} is adopted to improve the performance of PLKF, IMM-PLKF, and PR, whose parameter grid includes gains $N = \{ 0.5 , 1.5, 3, 5\}$, $c_z = \{1.5, 3.5, 5.5\}$, $c_y = \{1.5, 3.5, 5.5\}$ and $k = \{0.3, 0.5, 0.7\}$.

Average tracking errors in these six cases  are given in Figure. \ref{Figure6}. $50$ times of simulations with randomly initialized AUV positions have been conducted. The average tracking error and standard  deviation are summarized in Table.\ref{table2} and Table.\ref{table3}. 

\begin{table}[!h]
	\vspace{-5mm}
	\begin{center}
		\caption{Average Error (2D space)}
		\begin{tabular}{c|c|c|c} 
			\hline
			\quad	& Case 1  & Case 2 & Case 3   \\
			\hline
			OE-S &  $1.069 \pm 0.128$  &  $1.114 \pm 0.117 $ & $1.119\pm 0.266$ \\
			\hline
			OE-CV & $0.264 \pm 0.081$  &  $0.862\pm 0.049 $ & $1.060\pm 0.057 $  \\
			\hline
			PLKF & $0.010 \pm 0.003 $  &  $2.993 \pm 0.321 $ & $4.225\pm 0.078$  \\
			\hline
			IMM-PLKF & $0.010 \pm 0.001 $  &  $0.412 \pm 0.063 $ & $0.226\pm 0.036$\\
			\hline
			IMM-KF-F&  $\boldsymbol{0.003 \pm 0.000}$ &  $0.243 \pm 0.000$ &  $0.129 \pm 0.000$ \\
			\hline
			PR & $0.011\pm 0.001 $  &  $0.311 \pm 0.002 $ & $0.053\pm 0.001$ \\
			\hline
			Ours & $0.026 \pm 0.002 $  &  $\boldsymbol{0.059\pm 0.004} $ & $\boldsymbol{0.052 \pm 0.005}$ \\
			\hline
		\end{tabular}			\label{table2}
	\end{center}
	\vspace{-9mm}
	\begin{center}
		\caption{Average Error (3D space)}
		\begin{tabular}{c|c|c|c} 
			\hline
			\quad & Case 4  & Case 5 & Case 6 \\
			\hline
			PLKF & $0.010 \pm 0.004 $ &  $2.866 \pm 0.492$ &  $4.1233 \pm 0.318$ \\
			\hline
			IMM-PLKF&  $0.010 \pm 0.002$ &  $0.589 \pm 0.107$ &  $0.399 \pm 0.079$ \\
			\hline
			IMM-KF-F&  $\boldsymbol{0.003 \pm 0.000}$ &  $0.355 \pm 0.000$ &  $0.167 \pm 0.000$ \\
			\hline
			PR & $0.008\pm 0.001$  &  $0.214 \pm 0.012 $ & $0.084\pm 0.008$ \\
			\hline
			Ours & $0.014 \pm 0.002$  &  $\boldsymbol{0.053 \pm 0.004} $ & $\boldsymbol{0.035 \pm 0.003}$\\
			\hline
		\end{tabular}			\label{table3}
	\end{center}
	\vspace{-7mm}
\end{table}

From Figure.\ref{Figure6}\subref{Comparison_Case1} and Figure.\ref{Figure6}\subref{Comparison_Case4}, we can see that  the error with OE-S does not decrease and the error with  other methods converge to the value near zero because the OE-S assumes a static target which does not match the real one. In Table \ref{table2} and Table \ref{table3}, for Case 1 and Case 4, IMM-KF-F works best because most informative full measurements are used compared to others and target motion conforms to the candidate models. At the same time, the errors of other methods including ours are still within an acceptable range.

However,  from Figure.\ref{Figure6}\subref{Comparison_Case2}-\subref{Comparison_Case3} and Figure.\ref{Figure6}\subref{Comparison_Case5}-\subref{Comparison_Case6}, OE-S, OE-CV and PLKF lead to large tracking errors.  In Table \ref{table2} and Table \ref{table3},  for Case 2-3 and Case 5-6, the average errors of OE-S, OE-CV and PLKF are all large. It is because the target does not move with a constant or slowly varying velocity.  But for a greatly nonlinear maneuvering target, a constant-velocity model is not an effective approximation.  At the same time, we can see that IMM-PLKF, IMM-KF-F and PR significantly reduce tracking error although target motion does not match their assumptions.  It is because  IMM-PLKF, IMM-KF-F and PR can approximate nonlinear unknown target motion with more diverse motion patterns. However,  there are still significant residuals and they fail to achieve accurate target tracking.  On IMM-based methods, its tracking ability heavily relies richness of candidate models, which may fail when target motion is complex to describe. On PR, it is because the fitting effect based on polynomials is very limited, and it will fail when encountering severely  nonlinear motion modes. By contrast, our proposed GP-based method works effectively on the same amount of mini-batch bearing data across all six cases and is greatly suitable for online tracking complex target maneuvers.  Above simulations demonstrate  the superior performance of our GP-based method compared with other methods.

\begin{figure*}[!t]
	\centering
	\subfloat[Tracking error (Case 1).]{\includegraphics[width=2.4in]{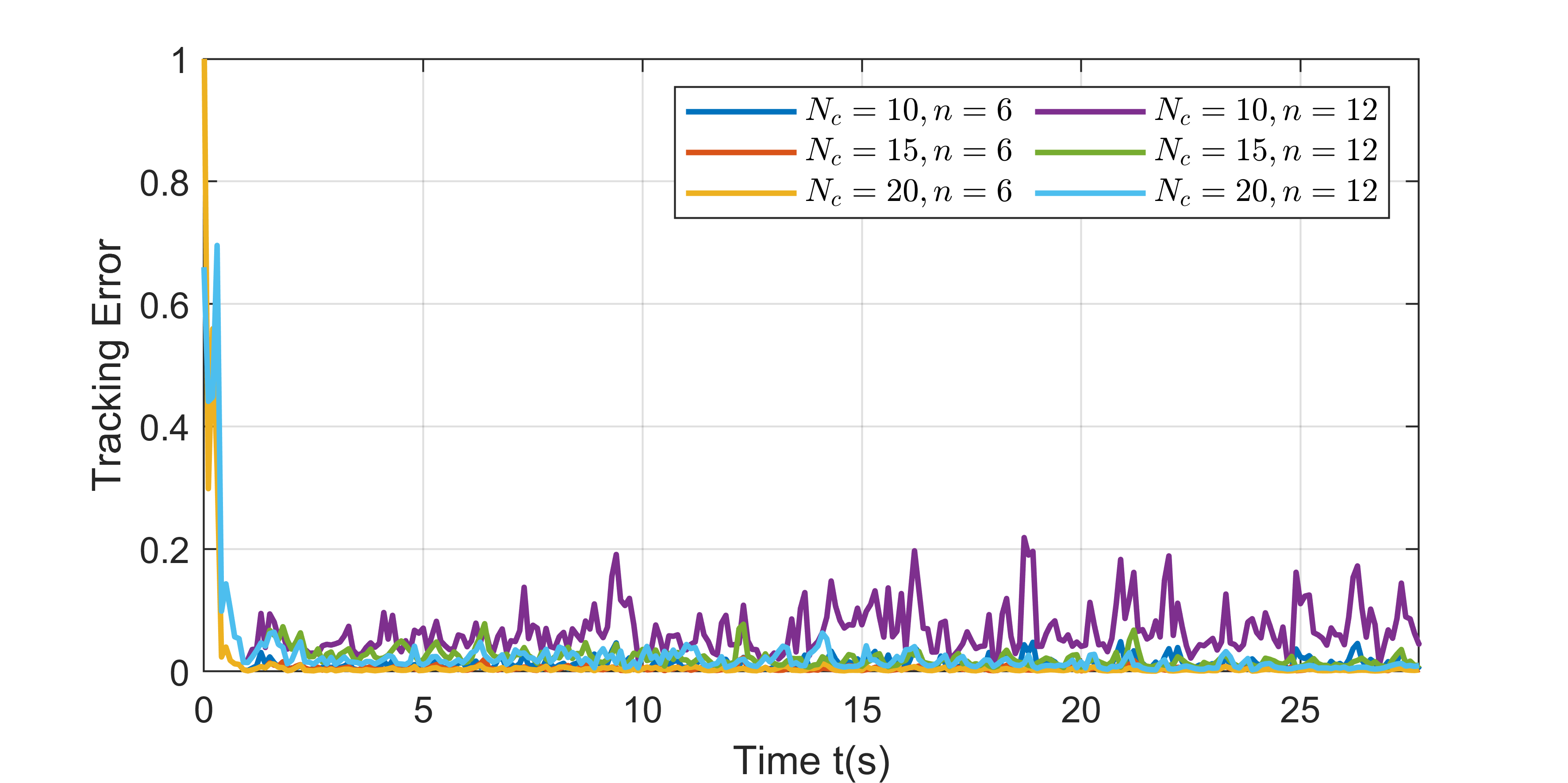}%
		\label{Tuning1_2D}}
	\subfloat[Tracking error  (Case 5).]{\includegraphics[width=2.4in]{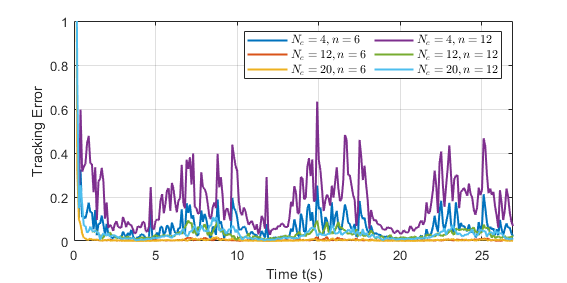}%
		\label{Tuning1_3D}}\\
	\subfloat[Variance norm  (Case 1).]{\includegraphics[width=2.4in]{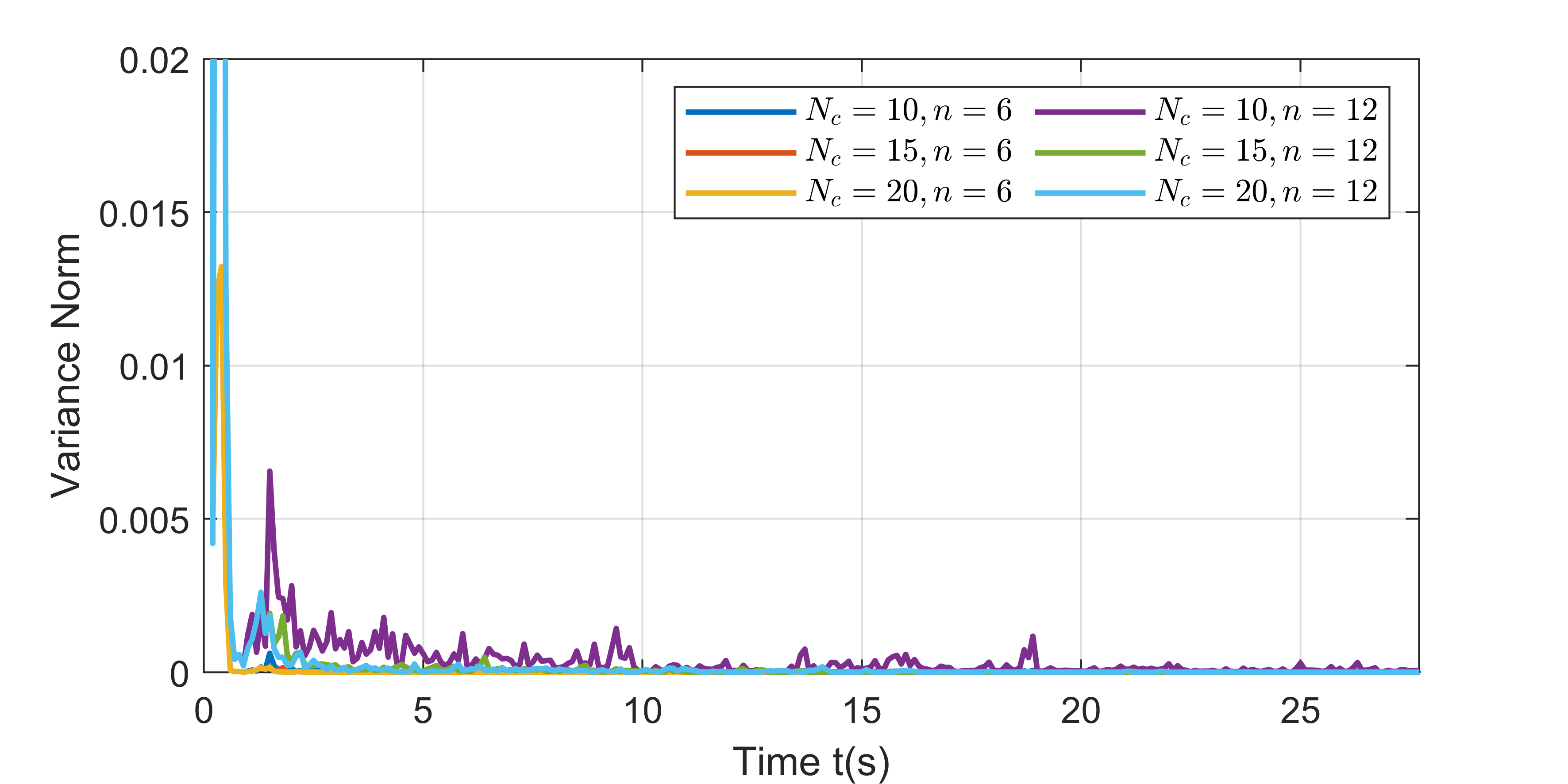}%
		\label{Tuning2_2D}}
	\subfloat[Variance norm  (Case 5).]{\includegraphics[width=2.4in]{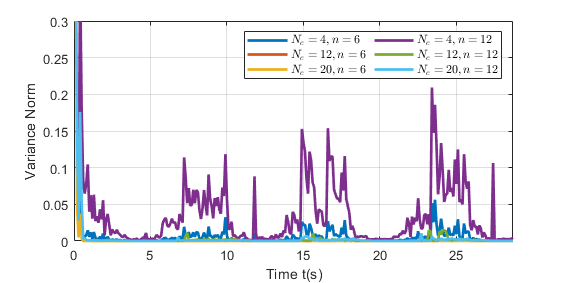}%
		\label{Tuning2_3D}}
	\caption{Comparison of different lengths.}
	\label{Figure7}
\end{figure*}

\subsection{Tracking Performance with different Lengths}

Next, we study the impacts of the lengths of GBT on the
target tracking performance. In our proposed GBT, there are two key lengths: the length of target historical data $N_c$ and the length of the forecast data $n$. Here, the Case 1 and Case 5 are considered as typical scenes in 2D and 3D space, respectively.

The simulation result is shown in Figure.\ref{Figure7}. It can be observed that the average tracking error and variance norm increase as decreasing $N_c$ and increasing $N$. For a larger $n$, more data and thus a larger $N_c$ are needed to achieve similar performance because near extrapolation involves only local function behavior and can be captured with few samples, whereas far extrapolation spans more complex global behavior, requiring more data to constrain the model. Therefore, for case 1 where target motion not severely varies, low errors can be achieved with a small $N_c$ even for a large $n$ and it fails for the other. In practice, to obtain suboptimal parameters, given a specific $n$, we can increase the amount of data until the tracking error or the variance norm no longer decays significantly and chose the current amount of data as $N_c$. For case 1, $N_c = 15, n = 12$ and $N_c = 10, n=6$ are  suboptimal parameters, and for case 5, $N_c = 12$ are suboptimal both for $n = 12$ and $n=6$.

\subsection{Robustness Test}

\begin{figure*}[!t]
	\centering
	\subfloat[Tracking error (Case 3, Scenario 1).]{\includegraphics[width=2.5in]{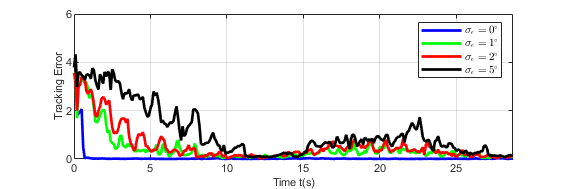}%
	\label{Figure_31}}
	\subfloat[Tracking error (Case 6, Scenario 1).]{\includegraphics[width=2.5in]{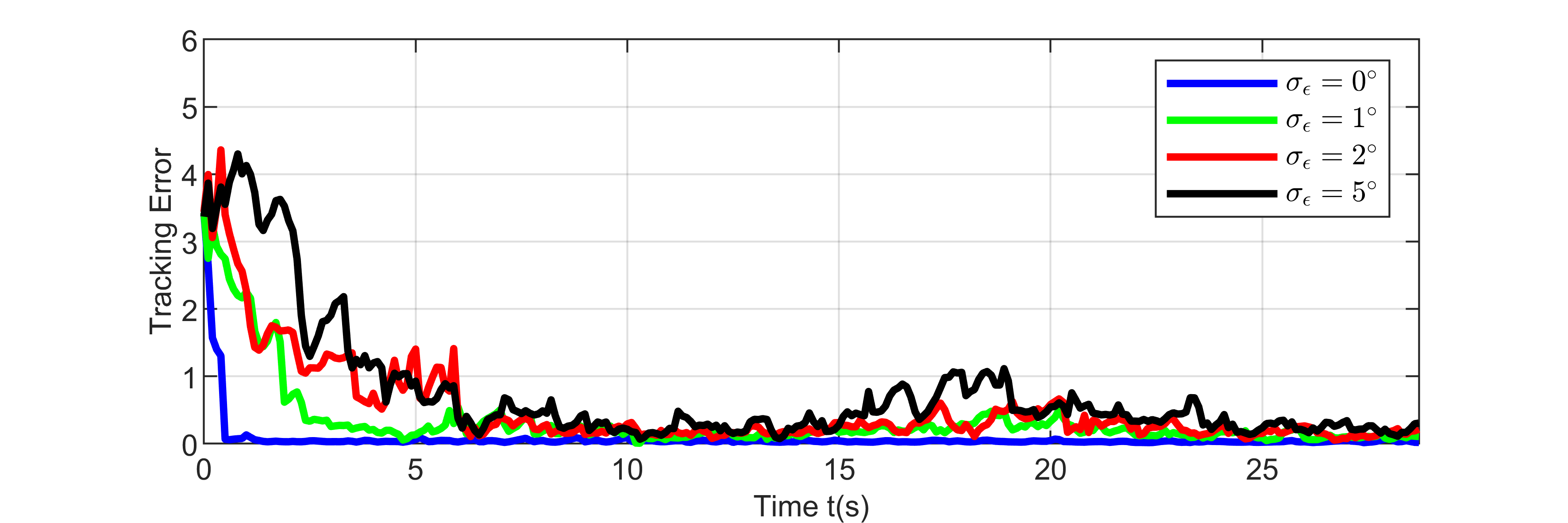}%
	\label{Figure_61}}\\
	\subfloat[Tracking error (Case 3, Scenario 2).]{\includegraphics[width=2.5in]{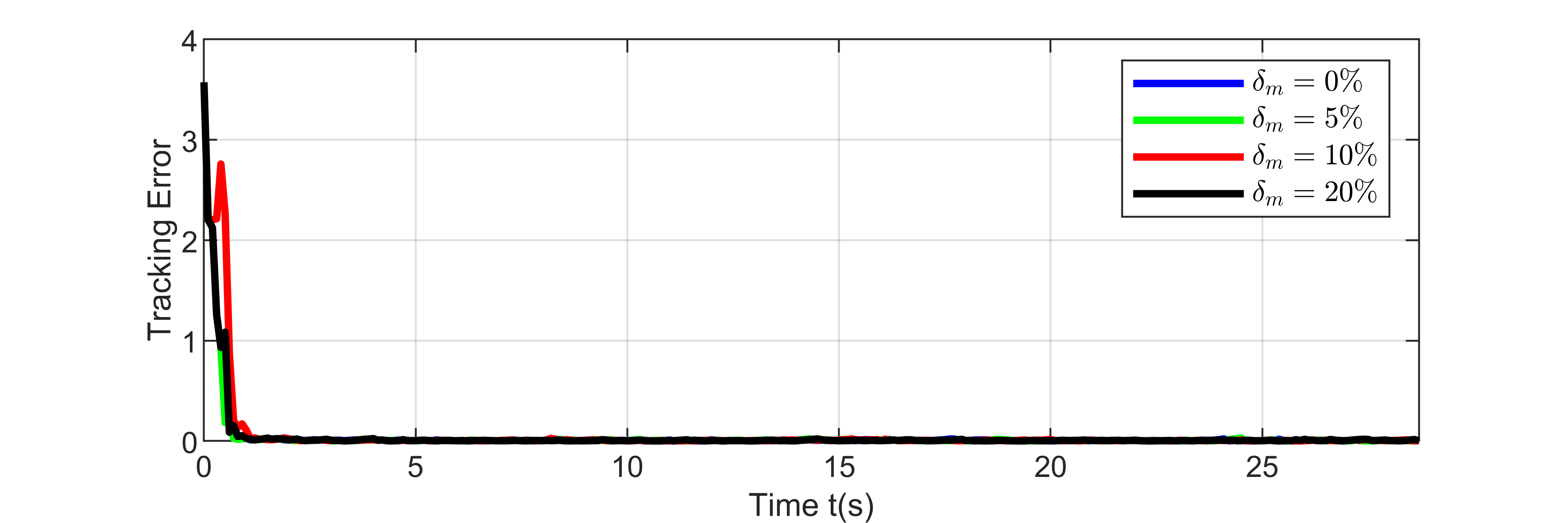}%
\label{Figure_32}	}
	\subfloat[Tracking error (Case 6, Scenario 2).]{\includegraphics[width=2.5in]{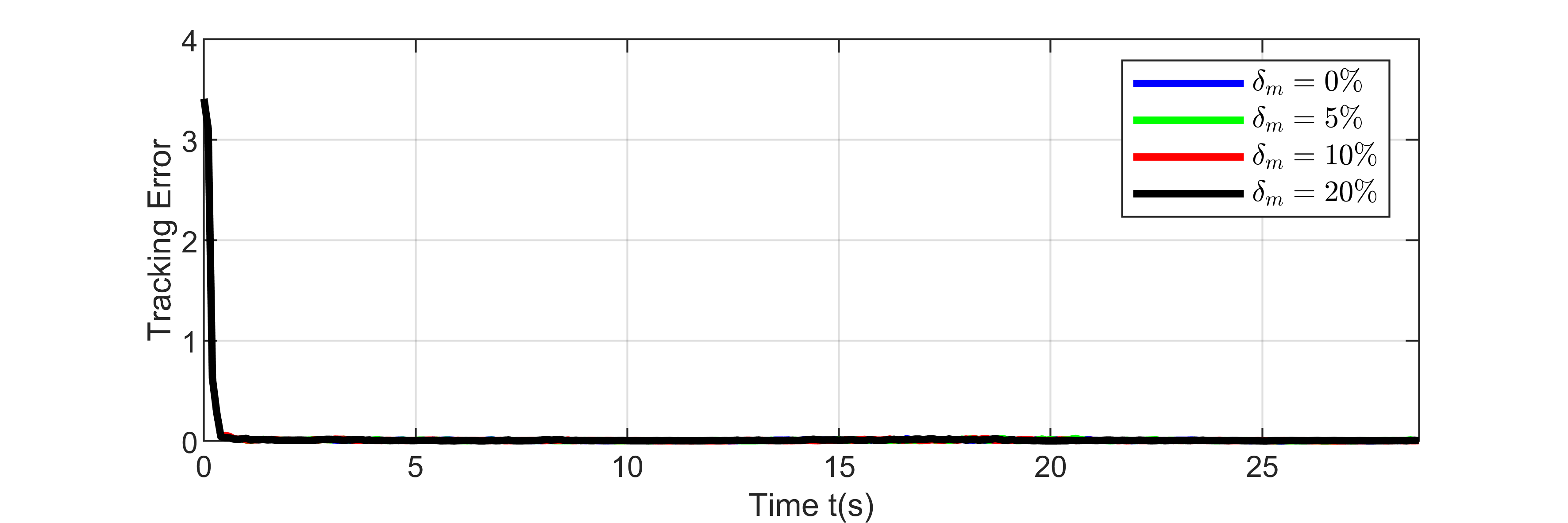}%
	\label{Figure_62}}\\
	\subfloat[Tracking error (Case 3, Scenario 3).]{\includegraphics[width=2.5in]{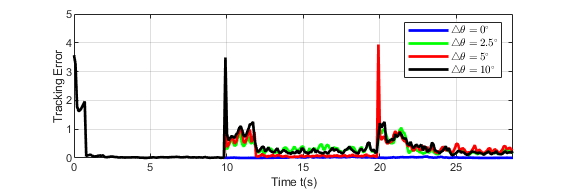}%
\label{Figure_33}	}
	\subfloat[Tracking error (Case 6, Scenario 3).]{\includegraphics[width=2.5in]{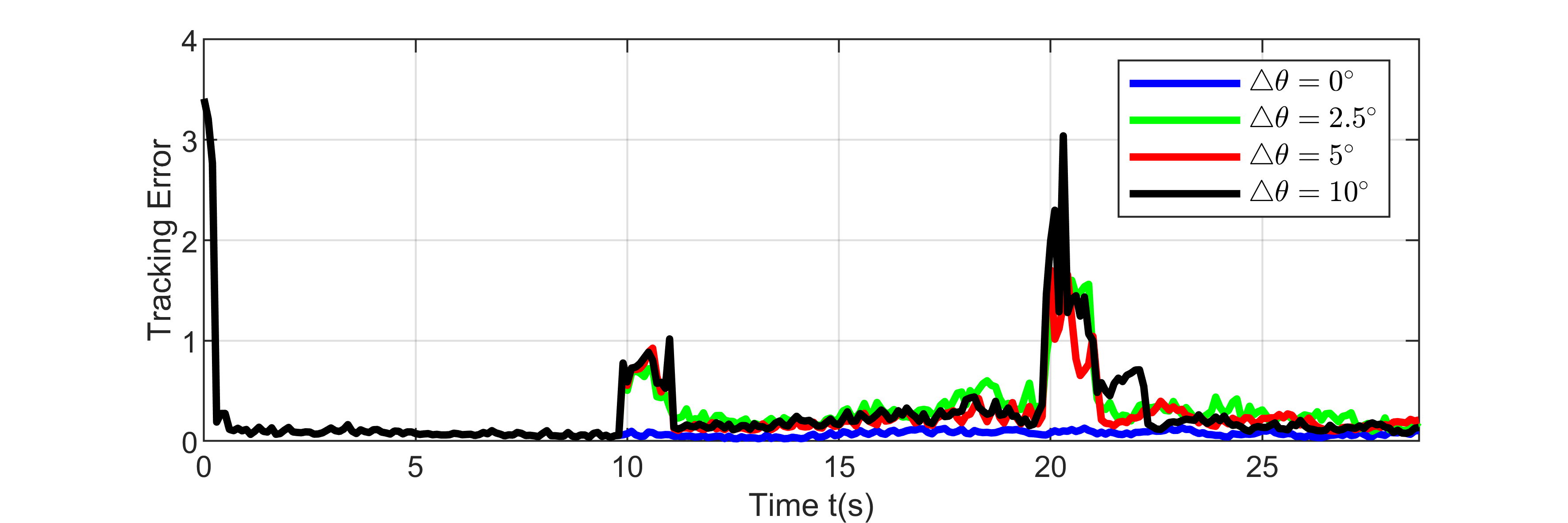}%
	\label{Figure_63}}\\
	\subfloat[Tracking error (Case 3, Scenario 4).]{\includegraphics[width=2.5in]{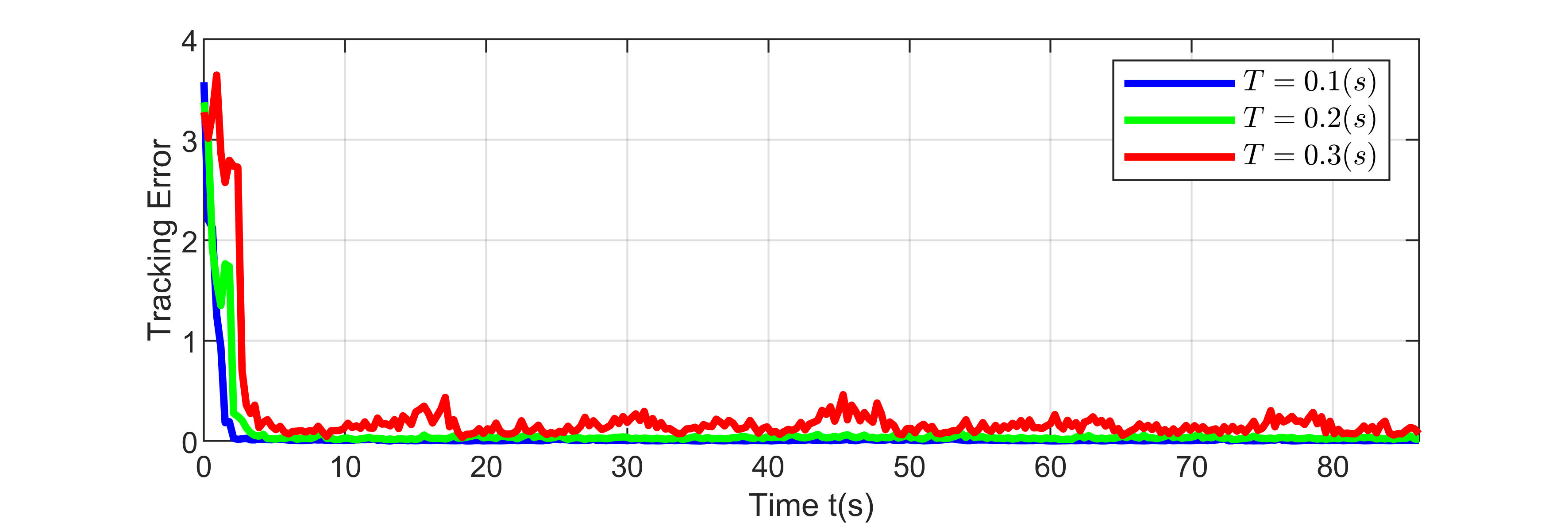}%
	\label{Figure_34}}
	\subfloat[Tracking error (Case 6, Scenario 4).]{\includegraphics[width=2.5in]{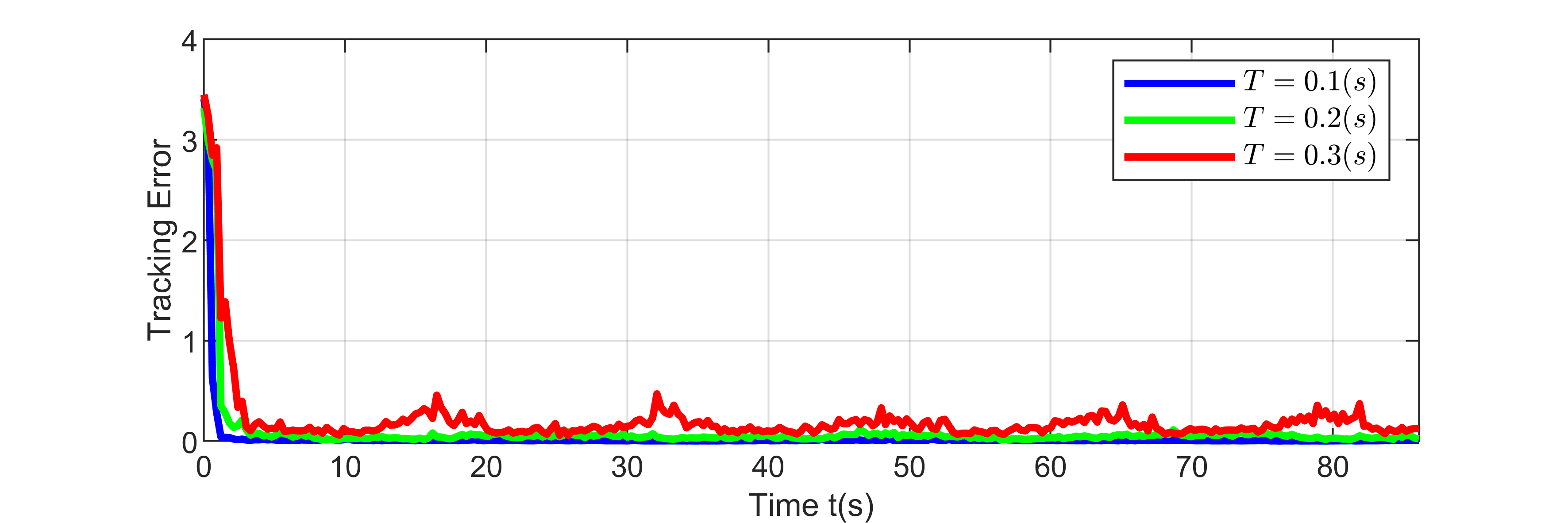}%
	\label{Figure_64}}\\	
	\subfloat[Tracking error (Case 3, Scenario 5).]{\includegraphics[width=2.5in]{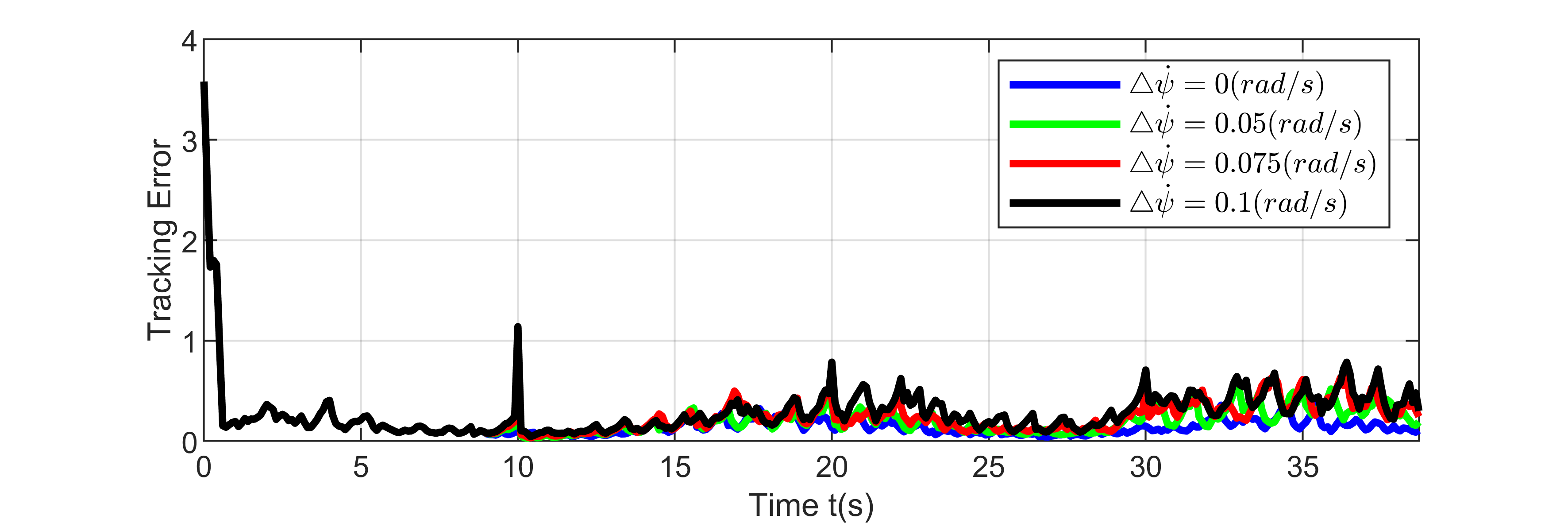}%
	\label{Figure_35}}
	\subfloat[Tracking error (Case 6, Scenario 5).]{\includegraphics[width=2.5in]{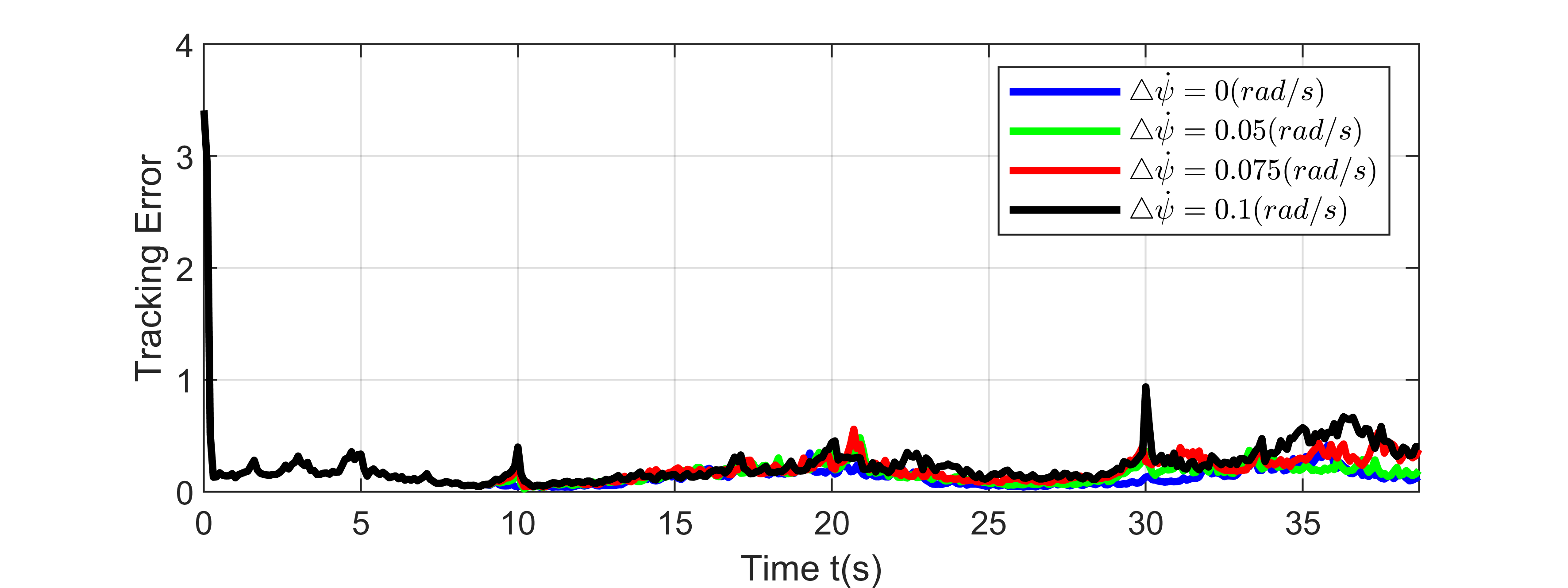}%
	\label{Figure_65}}\\
	\caption{Robustness test.}
\label{Figure8}
\end{figure*}

In this section, we check the robustness of our proposed framework in different scenarios. The following challenging scenarios are considered in our simulation. Here, the Case 3 and Case 6 are considered as typical scenes in 2D and 3D space, respectively.

{\it Scenario 1. High sampling noise}: Bearing measurement noises  in the angular domain are considered.  Usually, most angle domain noises do not exceed $1^{\circ}$ \citep{clapp2006single, kleeman1999fast}. Noises with different levels of standard deviation, i.e., $\sigma_{\epsilon} = \{0^{\circ}, 1^{\circ}, 2^{\circ}, 5^{\circ}\}$ are considered. As shown in Figure.\ref{Figure8}\subref{Figure_31}-\subref{Figure_61}. It can be observed that as the deviation of noise increases, the tracking error grows larger. But ultimate errors are always within $1(m)$, which indicate that although our proposed framework simplifies the bearing measurement model by considering a noise-free case, it still shows great robustness.

{\it Scenario 2. Missed detection}:  A realistic scenario where the bearing measurement is missed at different probabilities $\delta_{m} = \{0\%, 5\%, 10\%, 20\%\}$ at each time step is considered. The simulation result is shown in Figure.\ref{Figure8}\subref{Figure_32}-\subref{Figure_62}. In the initial stage, missed detection may lead to increase of time required to achieve accurate learning. The impact of a single outlier is temporary and corrected as new data arrive. As the amount of data gradually increases, the impact of outliers is less significant, because target motion has been well fitted with historic data in lack of latest data. The target can be tracked accurately even when the probability of  missed detection is still large at each time step.

{\it Scenario 3. Outliers}: Large, sporadic errors $\triangle \delta$ are injected into the bearing measurements at each $10(s)$ to test the resilience of the GP learning process against spurious data for $\triangle \delta = \{0^{\circ}, 2.5^{\circ}, 5^{\circ}, 10^{\circ}\}$. The simulation result under outliers is shown in Figure.\ref{Figure8}\subref{Figure_33}-\subref{Figure_63}. It can be clearly seen that sporadic errors on bearing measurements cause significant tracking error within $1(m)$ for a certain period of time. Therefore, as long as the outliers are not particularly frequent, which do not occur frequently in practice, our method has strong robustness to this.

{\it Scenario 4. Slower sampling rates}:  The framework under lower sampling frequencies with periods of $T = \{0.1, 0.2, 0.3 \}(s)$ is evaluated. The simulation result with slower sampling rates is shown in Figure.\ref{Figure8}\subref{Figure_34}-\subref{Figure_64}, from which it can be seen that tracking error increases as the sampling periods $T$ increases. However, tracking errors for these three periods are all within $0.3(m)$, showing robustness for slower sampling rates.

{\it Scenario 5. Abrupt maneuvers}:  A new target motion based on the Case 3 and 6 considering the sudden change on $\dot{\psi}_T$ for each $10(s)$ with $\triangle \dot{\psi}_T = \{0,   0.05, 0.075,  0.1\}({\rm rad/s})$ is designed. The simulation result with abrupt maneuvers is shown in Figure.\ref{Figure8}\subref{Figure_35}-\subref{Figure_65}. It can be seen that our method can effectively handle it and the tracking error reduces rapidly  after each abrupt maneuver by rapidly adapting the GP hyperparameters.

\subsection{Effectiveness of LMC-based Methods}

\begin{figure*}[!t]
	\centering
	\subfloat[Case 2.]{\includegraphics[width=2.4in]{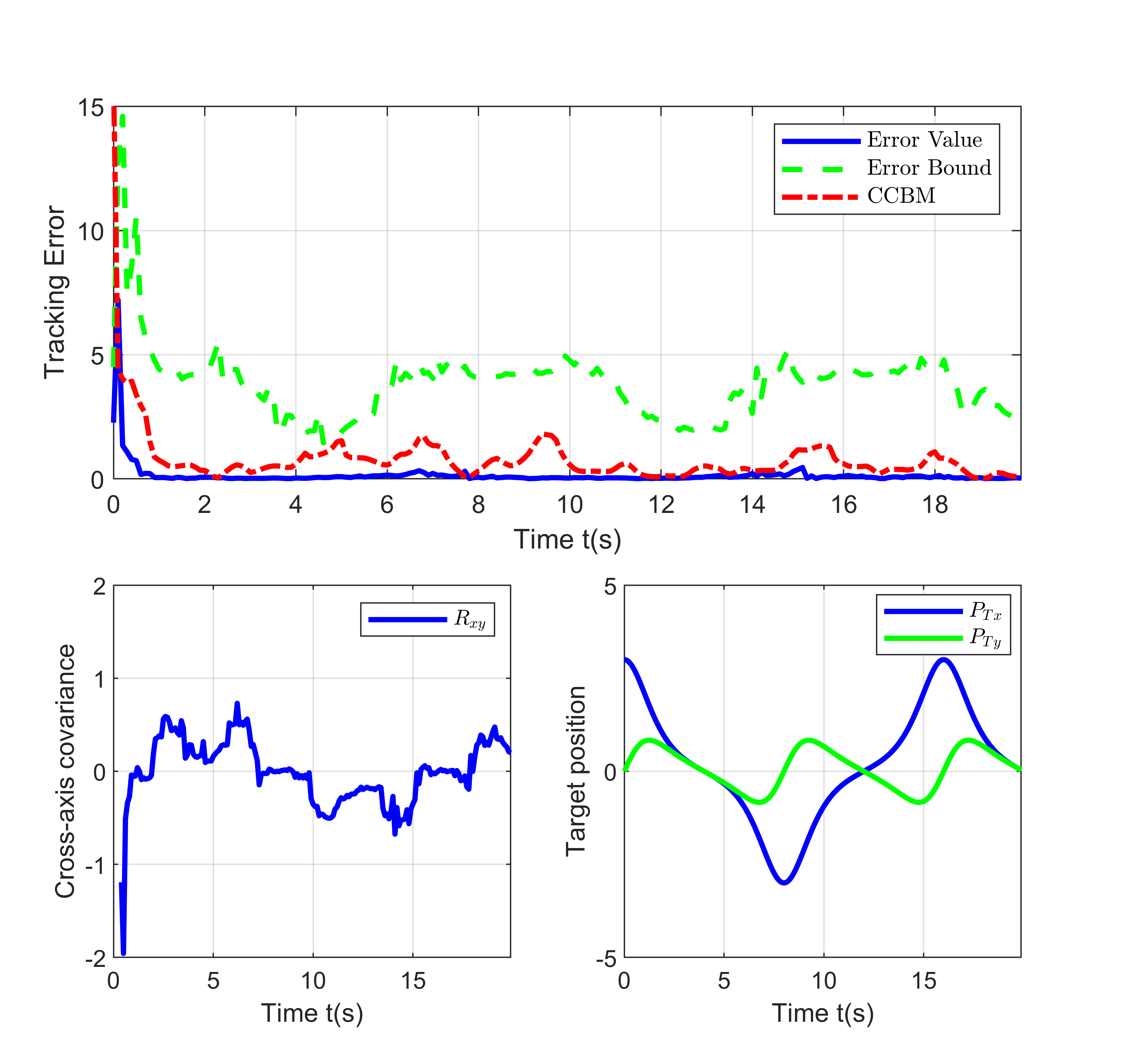}%
		\label{Ex_2D_Case2}}
	\subfloat[Case 6.]{\includegraphics[width=2.4in]{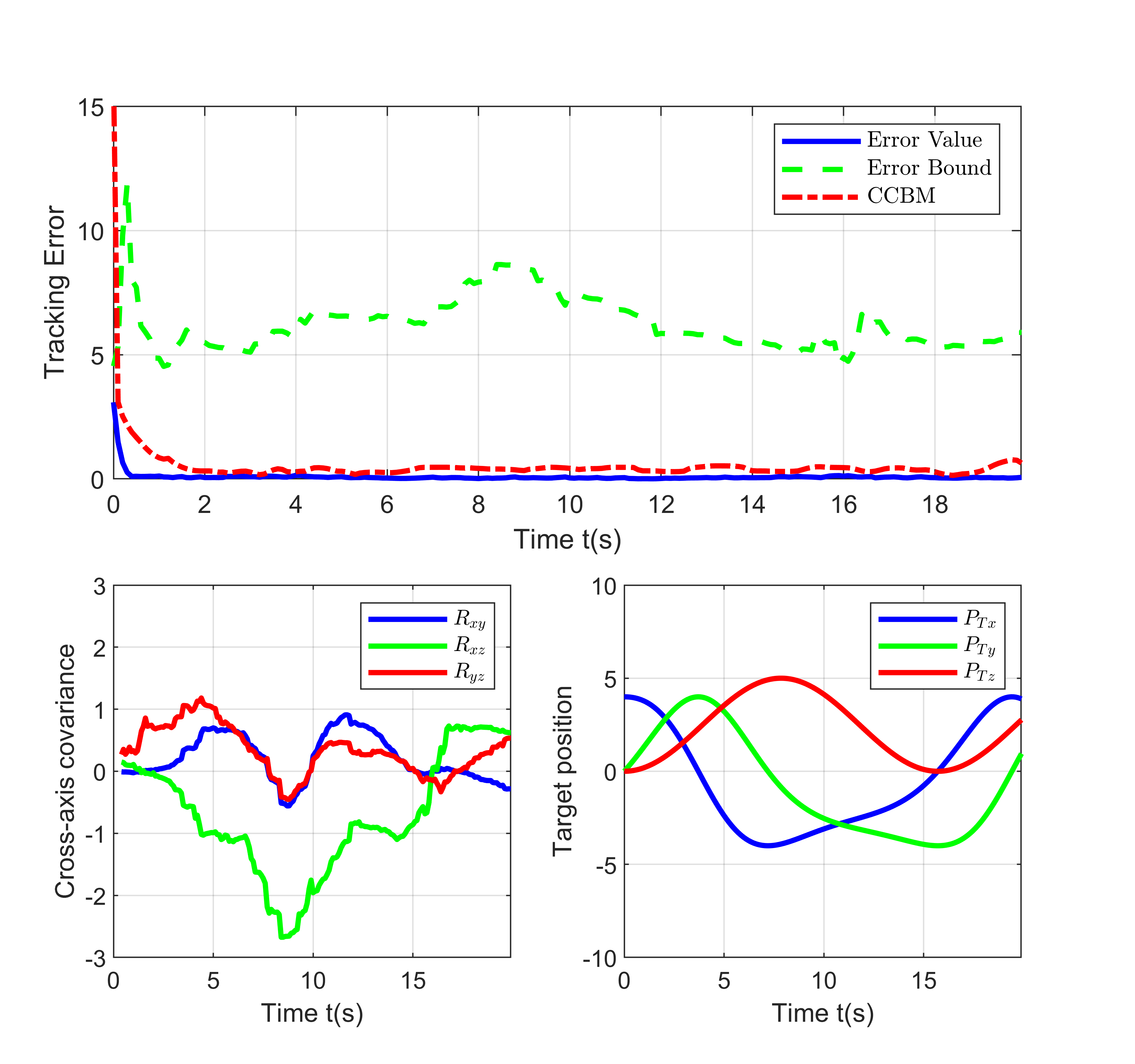}%
		\label{Ex_3D_Case6}}
	\caption{Simulation results of LMC.}
	\label{Figure9}
\end{figure*}

In this section, we testify the tracking performance of the extended LMC-based method in Section 3.3.  Here, the Case 2 and Case 6 are considered as typical scenes in 2D and 3D space, respectively. For both cases, $d$ latent functions are considered and the SE kernel is used for all latent functions.
The simulation result is given in Figure.\ref{Figure9}, where the tracking error and the cross-axis correlation is shown. 

First, as shown in the above subfigure, the mean tracking error of all time instants in $\mathcal{T}_k$ decreases rapidly within $0.1(m)$ showing the effectiveness of the extension. At the same time, the tracking error is within the theoretical bound, showing the correctness of Theorem 2. And The positive correlation between tracking error and CCBM can be clearly seen.

Next, to analyze the cross-axis correlation, we show the prior covariance between each axis in $\boldsymbol{k}_f(t, t)$ at each time instant in the bottom subfigure, denoted by $R_{xy}$, $R_{xz}$ and $R_{yz}$. At the same time, the value of each axis at each time instant is also given. Taking Case 2 from 0 to 5 seconds as an example, X and Y show opposite changes at first, so $R_ {xy}$ is initially negative. And then both show the same trend of change, at which point $R_ {xy}$ becomes positive.  By observing the covariance between coordinates and the true position changes throughout the entire process, we can clearly see that the coupling relationship between coordinates is accurately captured.

\section{Conclusions and Future Work}
This article has addressed the problem of trajectory optimization for bearing-only target tracking by an AUV.  A novel GP-based method with online bearing data was designed to simultaneously learn target motion patterns and track target positions. The proposed method can efficiently track the  target from online data with minimal computational cost. Furthermore, a theoretical bearing-data-dependent tracking error bound was explicitly proven, providing the formal relationship between tracking performance and data collection, based on which the high-value bearing data is obtained. And then a trajectory optimization problem for collecting high-value bearing data has been proposed, which was transformed into an easily solvable one, enabling feasible and efficient trajectory planning. Numerical simulation verification and comparison have demonstrated its powerful tracking capability for unknown maneuvering targets. It can be clearly seen that the proposed framework eliminates the strong dependence on the target model and can achieve accurate and efficient target tracking with bearing-only data. In this paper, it is assumed that the measurement is accurate and the target can always be detected at any time, which raise high requirements on the onboard sensors and bring great challenges due to nonlinearity of bearings. 	A framework for noisy bearing measurements needs further investigation.

\appendix
\section{Proof of Theorem 1}
To prove Theorem {\ref{the1}}, we first prove the following lemma.

\begin{lemma} \label{lem2} Consider the unknown target position $P_T(t)$ in the $d$-dimensional space ($d=2$ or $d=3$) and a prior Gaussian process $\mathcal{G} \mathcal{P}(\boldsymbol{0}_d, \boldsymbol{k}(t, t'))$ satisfying Assumption \ref{ass2}.  For a finite set $\mathcal{T}_{k}$ to achieve target tracking and a  probability $\delta \in (0, 1]$, it holds that
	\begin{align}\label{eqA1}
		\mathbb{P}(\|P_T(t_{i}) -\mu_{k}(t_i)\| \leq \beta(\delta)\sqrt{\|\Sigma_k(t_i)\|}, \forall t_i \in  \mathcal{T}_{k})  \geq 1-\delta, 
	\end{align}
	where  $\mu_{k}(t_i)$ and $\Sigma_{k}(t_i)$ are the posterior expectation and covariance function, respectively, and
	$\beta(\delta) = \sqrt{\chi^2_{d, 1-\frac{\delta}{|\mathcal{T}_k|}}}$ is the $(1-\frac{\delta}{|\mathcal{T}_k|})$-quantile of the $\chi^2_{d}$ distribution.
\end{lemma}
\begin{proof} Under Assumption \ref{ass2}, for any time instant $t_i \in \mathcal{T}_{k}$, the unknown ${P}_T(t_i)$ follows the posterior Gaussian distribution
${P}_T(t_i) \sim \mathcal{N}\left(\mu_k(t_i), \Sigma_k(t_i)\right).$ The posterior covariance can be represented by eigendecomposition as $\Sigma_k(t_i) = Q_i\Lambda_iQ_i^{\top}$. Considering $Z_i = Q_i^{\top}(P_T(t_i)-\mu_k(t_i)) = [Z_{i, 1}, \dots, Z_{i, d}] \in \mathbb{R}^d$, we define $\bar{Z}_{i, j} =\frac{Z_{i,j}^2}{\lambda_{i,j}} \cdot 1_{\lambda_{i,j} \neq 0}$ for all $j = 1, \dots, d$. 
Obviously, for a fixed $c > 0$, it holds that
\begin{align}\label{eqA5}
	\mathbb{P}(\|P_T(t_i)-\mu_k(t_i)\|^2 > \|\Sigma_{k}(t_i)\|c) \leq \mathbb{P}(\sum\limits_{j=1}^{d} \bar{Z}_{i, j} > c).
\end{align}
Furthermore, $\sum\limits_{j=1}^{d} \bar{Z}_{i, j}$ follows the chi-square distribution with $r = {\rm rank}(\Sigma_k(t_i))$ degrees of freedom, i.e., $\sum\limits_{j=1}^{d} \bar{Z}_{i, j} \sim \chi_{r}^2$. With the property of the chi-square distribution, for $ \varphi > 0$
\begin{align}\label{eqA6}
	\mathbb{P}(\sum\limits_{j=1}^{d} \bar{Z}_{i, j} > \chi^2_{d, 1-\varphi}) \leq \mathbb{P}(\sum\limits_{j=1}^{d} \bar{Z}_{i, j} > \chi^2_{r, 1-\varphi}) = \varphi,
\end{align}
where $\chi^2_{r, 1-\varphi}$ is the the $(1-\varphi)$-quantile of the $\chi^2_{r}$ distribution because $\chi^2_{d, 1-\varphi} \geq \chi^2_{r, 1-\varphi}$ for all possible $r$. Combining (\ref{eqA5}) and (\ref{eqA6}) and let $c = \chi^2_{d, 1-\varphi}$ we can obtain that
\begin{align} \label{eqA7}
	\mathbb{P}(\|P_T(t_i)-\mu_k(t_i)\|^2 > \|\Sigma_{k}(t_i)\|\chi^2_{d, 1-\varphi}) \leq \varphi.
\end{align}
Applying the union bound, which states that the probability of a union of events is at most the sum of their individual probabilities, we get \ref{eqA1} and completes the proof.
\end{proof}

Based on Lemma 1, we give the proof of Theorem \ref{the1}.

\underline{\it Proof of Theorem 1}. For any time instant $t_i \in \mathcal{T}_{k}$, it holds that
\begin{align} \label{eqA8}
	\Sigma_k(t_i) \leq  \boldsymbol{k}(t_i, t_i) -\frac{\sum\limits_{j=c}^{k} \boldsymbol{k}(t_i,t_j)P(t_j)\boldsymbol{k}(t_i,t_j)}{\|\boldsymbol{G}\boldsymbol{K}\boldsymbol{G}^{\top}\|}. 
\end{align}
At the same time,   $\|\boldsymbol{G}\boldsymbol{K}\boldsymbol{G}^{\top}\| = \|\boldsymbol{G}^{\top}\boldsymbol{G}\boldsymbol{K}\|$ holds.  For a vector $v = [v_1^{\top}, \dots, v_N^{\top}] \in \mathbb{R}^{dN}$ with $v_i \in \mathbb{R}^d$ satisfying $\|v\| = 1$, it holds that
\begin{align}\label{eqA9}
	\|v\boldsymbol{G}^{\top}\boldsymbol{G}\boldsymbol{K}\| 
	\leq&  \max\limits_{j = c,\dots, k} \sqrt{N \|\sum_{i=1}^{N}k(t_{c+i-1}, t_{j})P(t_{c+i-1})v_i\|^2}, \nonumber\\
	\leq & \sqrt{N\max\limits_{t\in \mathcal{D}_{c:k}}k(t, t)^2 \|\sum_{i=1}^{N}{P}(t_{c+i-1})\| \big( \sum_{i=1}^{N}\|{v}_i\|_2^2\big)},
\end{align}
with the Cauchy-Schwarz inequality. 
It is noted that $\frac{(d-1)N}{d} \leq \lambda_{\max}\Big(P_{c:k}\Big) \leq N$. Combining it with  (\ref{eqA9}), one can further derive that
\begin{align}\label{eqA14}	\|\boldsymbol{G}^{\top}\boldsymbol{G}\boldsymbol{K}\| \leq \sqrt{\frac{d}{d-1}} \max\limits_{t\in \mathcal{D}_{c:k}}k(t, t)
	\lambda_{\max} \Big({P}_{c:k}\Big). 
\end{align}
In addition, it holds that
\begin{align}\label{eqA15}
	\sum\limits_{j=c}^{k} k^2(t_i,t_j)P(t_j) \geq  \min\limits_{t \in \mathcal{D}_{c:k}}k^2(t_i,t)P_{c:k}. \!\!\!\!\!\!\!\!
\end{align}
Therefore, with (\ref{eqA14}) and (\ref{eqA15}), we can obtain that
\begin{align}
	\|\Sigma_k(t_i)\| 
	\leq k(t_i,t_i) - \frac{\min\limits_{t \in \mathcal{D}_{c:k}}k^2(t_i,t)\lambda_{\min}\Big(P_{c:k}\Big)}{\sqrt{\frac{d}{d-1}}\max\limits_{t
			\in \mathcal{D}_{c:k}}k(t, t)
		\lambda_{\max} \Big(P_{c:k}\Big)}.\nonumber
\end{align}
Under Assumption \ref{ass2}, combining the result of Lemma \ref{lem2}, we can complete the proof.

\section{Proof of Theorem 2}

\underline{\it Proof of Theorem 2}. 
For a vector $v = [v_1^{\top}, \dots, v_N^{\top}] \in \mathbb{R}^{dN}$ with $v_i \in \mathbb{R}^d$ satisfying $\|v\| = 1$, it holds that
\begin{align}\label{eqB2}
	\|v\boldsymbol{G}^{\top}\boldsymbol{G}\boldsymbol{K}\| 
		\leq &  \max\limits_{j = c,\dots, k} \sqrt{N \|\sum_{i=1}^{N}\boldsymbol{k}_f(t_{c+i-1}, t_{j})P(t_{c+i-1})v_i\|^2}.
\end{align}
Let $P_f(t_{c+i-1}) = \boldsymbol{k}_f(t_{c+i-1}, t_{j})P(t_{c+i-1})$ and we can infer that
\begin{align}\label{eqB3}
	\|\sum_{i=1}^{N}P_f(t_{c+i-1})v_i\|^2 \leq&  \|\sum_{i=1}^{N}{P}_f(t_{c+i-1})\| \sum_{i=1}^{N}(v_i^{\top}P_f(t_{c+i-1})v_i),\nonumber\\
	\leq &  \max\limits_{t \in \mathcal{D}_{c:k}}\|\boldsymbol{k}_f(t, t)\|^2 \sqrt{N\|\sum_{i=1}^{N}{P}(t_{c+i-1})\|}
\end{align}
In addition, it holds that
\begin{align}\label{eqB4}
	\lambda_{\min}\big(\sum\limits_{j=c}^{k} \boldsymbol{k}_f(t_i,t_j)P(t_j)\boldsymbol{k}_f(t_i,t_j)^{\top}\big) \geq  \lambda_{\min}\big(\min\limits_{t \in \mathcal{D}_{c:k}}\lambda_{\min}(\boldsymbol{k}_f(t_i,t))^2 P_{c:k}\big). \!\!\!\!\!\!\!\!
\end{align}
It is noted that $\frac{(d-1)N}{d} \leq \lambda_{\max}\Big(P_{c:k}\Big) \leq N$. With (\ref{eqB3}) and (\ref{eqB4}), we can conclude that
\begin{align}
	\|\Sigma_k(t_i)\| 
	\leq \|\boldsymbol{k}_f(t_i,t_i)\| - \frac{\min\limits_{t \in \mathcal{D}_{c:k}}\lambda_{\min}(\boldsymbol{k}_f(t_i,t))^2\lambda_{\min}\Big(P_{c:k}\Big)}{(\frac{d}{d-1})^{\frac{3}{4}}\max\limits_{t
			\in \mathcal{D}_{c:k}}\|\boldsymbol{k}_f(t, t)\|
		\lambda_{\max} \Big(P_{c:k}\Big)}.\nonumber
\end{align}
Under Assumption \ref{ass3}, combining the result of Lemma \ref{lem2}, we complete the proof of Theorem \ref{the2}.

\section*{Acknowledgment}
This work was supported in part by Natural Science Foundation of China
(under grant No. 62473251, 62373239, 62333011) and Shanghai Municipal Natural Science Foundation (under grant No. 25ZR1402270).

\bibliographystyle{elsarticle-harv}
\bibliography{reference}

\begin{thebibliography}{47}
\expandafter\ifx\csname natexlab\endcsname\relax\def\natexlab#1{#1}\fi
\providecommand{\url}[1]{\texttt{#1}}
\providecommand{\href}[2]{#2}
\providecommand{\path}[1]{#1}
\providecommand{\DOIprefix}{doi:}
\providecommand{\ArXivprefix}{arXiv:}
\providecommand{\URLprefix}{URL: }
\providecommand{\Pubmedprefix}{pmid:}
\providecommand{\doi}[1]{\href{http://dx.doi.org/#1}{\path{#1}}}
\providecommand{\Pubmed}[1]{\href{pmid:#1}{\path{#1}}}
\providecommand{\bibinfo}[2]{#2}
\ifx\xfnm\relax \def\xfnm[#1]{\unskip,\space#1}\fi
\bibitem[{Ali et~al.(2024)Ali, Ullah, ur~Rahman, Shah and Wang}]{ali2024novel}
\bibinfo{author}{Ali, W.}, \bibinfo{author}{Ullah, R.},
  \bibinfo{author}{ur~Rahman, W.}, \bibinfo{author}{Shah, S.A.},
  \bibinfo{author}{Wang, W.}, \bibinfo{year}{2024}.
\newblock \bibinfo{title}{A novel application of neural time series for dynamic
  characteristic analysis in underwater markov chain passive target tracking}.
\newblock \bibinfo{journal}{Ocean Engineering} \bibinfo{volume}{313},
  \bibinfo{pages}{119607}.
\bibitem[{Alvarez and Lawrence(2008)}]{alvarez2008sparse}
\bibinfo{author}{Alvarez, M.}, \bibinfo{author}{Lawrence, N.},
  \bibinfo{year}{2008}.
\newblock \bibinfo{title}{Sparse convolved gaussian processes for multi-output
  regression}.
\newblock \bibinfo{journal}{Advances in neural information processing systems}
  \bibinfo{volume}{21}.
\bibitem[{Beckers et~al.(2021)Beckers, Colombo, Hirche and
  Pappas}]{beckers2021online}
\bibinfo{author}{Beckers, T.}, \bibinfo{author}{Colombo, L.J.},
  \bibinfo{author}{Hirche, S.}, \bibinfo{author}{Pappas, G.J.},
  \bibinfo{year}{2021}.
\newblock \bibinfo{title}{Online learning-based trajectory tracking for
  underactuated vehicles with uncertain dynamics}.
\newblock \bibinfo{journal}{IEEE Control Systems Letters} \bibinfo{volume}{6},
  \bibinfo{pages}{2090--2095}.
\bibitem[{Benedetto and Fickus(2003)}]{benedetto2003finite}
\bibinfo{author}{Benedetto, J.J.}, \bibinfo{author}{Fickus, M.},
  \bibinfo{year}{2003}.
\newblock \bibinfo{title}{Finite normalized tight frames}.
\newblock \bibinfo{journal}{Advances in Computational Mathematics}
  \bibinfo{volume}{18}, \bibinfo{pages}{357--385}.
\bibitem[{Clapp and Etienne-Cummings(2006)}]{clapp2006single}
\bibinfo{author}{Clapp, M.A.}, \bibinfo{author}{Etienne-Cummings, R.},
  \bibinfo{year}{2006}.
\newblock \bibinfo{title}{Single ping-multiple measurements: Sonar bearing
  angle estimation using spatiotemporal frequency filters}.
\newblock \bibinfo{journal}{IEEE Transactions on Circuits and Systems I:
  Regular Papers} \bibinfo{volume}{53}, \bibinfo{pages}{769--783}.
\bibitem[{Du et~al.(2022)Du, Wang, Chai, Xiang, Zhang and
  Huang}]{du2022configuration}
\bibinfo{author}{Du, Z.}, \bibinfo{author}{Wang, W.}, \bibinfo{author}{Chai,
  H.}, \bibinfo{author}{Xiang, M.}, \bibinfo{author}{Zhang, F.},
  \bibinfo{author}{Huang, Z.}, \bibinfo{year}{2022}.
\newblock \bibinfo{title}{Configuration analysis method and geometric
  interpretation of {UUV}s cooperative localization based on error ellipse}.
\newblock \bibinfo{journal}{Ocean Engineering} \bibinfo{volume}{244},
  \bibinfo{pages}{110299}.
\bibitem[{Ebrahimi et~al.(2022)Ebrahimi, Ardeshiri and
  Khanghah}]{ebrahimi2022bearing}
\bibinfo{author}{Ebrahimi, M.}, \bibinfo{author}{Ardeshiri, M.},
  \bibinfo{author}{Khanghah, S.A.}, \bibinfo{year}{2022}.
\newblock \bibinfo{title}{Bearing-only 2d maneuvering target tracking using
  smart interacting multiple model filter}.
\newblock \bibinfo{journal}{Digital Signal Processing} \bibinfo{volume}{126},
  \bibinfo{pages}{103497}.
\bibitem[{Ferri et~al.(2018)Ferri, Munafo and LePage}]{ferri2018autonomous}
\bibinfo{author}{Ferri, G.}, \bibinfo{author}{Munafo, A.},
  \bibinfo{author}{LePage, K.D.}, \bibinfo{year}{2018}.
\newblock \bibinfo{title}{An autonomous underwater vehicle data-driven control
  strategy for target tracking}.
\newblock \bibinfo{journal}{IEEE Journal of Oceanic Engineering}
  \bibinfo{volume}{43}, \bibinfo{pages}{323--343}.
\bibitem[{Fossen(2011)}]{fossen2011handbook}
\bibinfo{author}{Fossen, T.I.}, \bibinfo{year}{2011}.
\newblock \bibinfo{title}{Handbook of {M}arine {C}raft {H}ydrodynamics and
  {M}otion {C}ontrol}.
\newblock \bibinfo{publisher}{John Wiley \& Sons}.
\bibitem[{Fu et~al.(2022)Fu, Yang, Zhu and Chen}]{fu2022leader}
\bibinfo{author}{Fu, Y.}, \bibinfo{author}{Yang, Z.}, \bibinfo{author}{Zhu,
  S.}, \bibinfo{author}{Chen, C.}, \bibinfo{year}{2022}.
\newblock \bibinfo{title}{Leader-following entrapping control of {AUV}s using
  bearing measurements in local coordinate frames}.
\newblock \bibinfo{journal}{Guidance, Navigation and Control}
  \bibinfo{volume}{2}, \bibinfo{pages}{2250025}.
\bibitem[{Hao et~al.(2021)Hao, Wang, Wei, Chen and Wang}]{hao2021imm}
\bibinfo{author}{Hao, L.}, \bibinfo{author}{Wang, Y.}, \bibinfo{author}{Wei,
  C.H.}, \bibinfo{author}{Chen, Z.W.}, \bibinfo{author}{Wang, X.N.},
  \bibinfo{year}{2021}.
\newblock \bibinfo{title}{An imm-pkf method for helicopter tracking using
  acoustic arrays}, in: \bibinfo{booktitle}{2021 4th International Conference
  on Information Communication and Signal Processing (ICICSP)},
  \bibinfo{organization}{IEEE}. pp. \bibinfo{pages}{224--228}.
\bibitem[{Haykin(2004)}]{haykin2004kalman}
\bibinfo{author}{Haykin, S.}, \bibinfo{year}{2004}.
\newblock \bibinfo{title}{Kalman {F}iltering and {N}eural {N}etworks}.
\newblock \bibinfo{publisher}{John Wiley \& Sons}.
\bibitem[{He et~al.(2019)He, Shin and Tsourdos}]{he2019trajectory}
\bibinfo{author}{He, S.}, \bibinfo{author}{Shin, H.S.},
  \bibinfo{author}{Tsourdos, A.}, \bibinfo{year}{2019}.
\newblock \bibinfo{title}{Trajectory optimization for target localization with
  bearing-only measurement}.
\newblock \bibinfo{journal}{IEEE Transactions on Robotics}
  \bibinfo{volume}{35}, \bibinfo{pages}{653--668}.
\bibitem[{He et~al.(2020)He, Shin and Tsourdos}]{he2020trajectory}
\bibinfo{author}{He, S.}, \bibinfo{author}{Shin, H.S.},
  \bibinfo{author}{Tsourdos, A.}, \bibinfo{year}{2020}.
\newblock \bibinfo{title}{Trajectory optimization for multitarget tracking
  using joint probabilistic data association filter}.
\newblock \bibinfo{journal}{Journal of Guidance, Control, and Dynamics}
  \bibinfo{volume}{43}, \bibinfo{pages}{170--178}.
\bibitem[{Hu and Zhang(2022)}]{hu2021bearing}
\bibinfo{author}{Hu, B.B.}, \bibinfo{author}{Zhang, H.T.},
  \bibinfo{year}{2022}.
\newblock \bibinfo{title}{Bearing-only motional target-surrounding control for
  multiple unmanned surface vessels}.
\newblock \bibinfo{journal}{IEEE Transactions on Industrial Electronics}
  \bibinfo{volume}{69}, \bibinfo{pages}{3988--3997}.
\bibitem[{Kanagawa et~al.(2018)Kanagawa, Hennig, Sejdinovic and
  Sriperumbudur}]{kanagawa2018gaussian}
\bibinfo{author}{Kanagawa, M.}, \bibinfo{author}{Hennig, P.},
  \bibinfo{author}{Sejdinovic, D.}, \bibinfo{author}{Sriperumbudur, B.K.},
  \bibinfo{year}{2018}.
\newblock \bibinfo{title}{Gaussian processes and kernel methods: A review on
  connections and equivalences}.
\newblock \bibinfo{journal}{arXiv preprint arXiv:1807.02582} .
\bibitem[{Kleeman(1999)}]{kleeman1999fast}
\bibinfo{author}{Kleeman, L.}, \bibinfo{year}{1999}.
\newblock \bibinfo{title}{Fast and accurate sonar trackers using double pulse
  coding}, in: \bibinfo{booktitle}{Proceedings 1999 IEEE/RSJ International
  Conference on Intelligent Robots and Systems. Human and Environment Friendly
  Robots with High Intelligence and Emotional Quotients (Cat. No. 99CH36289)},
  \bibinfo{organization}{IEEE}. pp. \bibinfo{pages}{1185--1190}.
\bibitem[{Lederer(2023)}]{lederer2023gaussian}
\bibinfo{author}{Lederer, A.}, \bibinfo{year}{2023}.
\newblock \bibinfo{title}{Gaussian Processes in Control: Performance Guarantees
  through Efficient Learning}.
\newblock Ph.D. thesis. Technische Universit{\"a}t M{\"u}nchen.
\bibitem[{Lederer et~al.(2022)Lederer, Yang, Jiao and
  Hirche}]{lederer2022cooperative}
\bibinfo{author}{Lederer, A.}, \bibinfo{author}{Yang, Z.},
  \bibinfo{author}{Jiao, J.}, \bibinfo{author}{Hirche, S.},
  \bibinfo{year}{2022}.
\newblock \bibinfo{title}{Cooperative control of uncertain multiagent systems
  via distributed gaussian processes}.
\newblock \bibinfo{journal}{IEEE Transactions on Automatic Control}
  \bibinfo{volume}{68}, \bibinfo{pages}{3091--3098}.
\bibitem[{Leonard and Bahr(2016)}]{leonard2016autonomous}
\bibinfo{author}{Leonard, J.J.}, \bibinfo{author}{Bahr, A.},
  \bibinfo{year}{2016}.
\newblock \bibinfo{title}{Autonomous underwater vehicle navigation}.
\newblock \bibinfo{journal}{Springer {H}andbook of {O}cean {E}ngineering} ,
  \bibinfo{pages}{341--358}.
\bibitem[{Li et~al.(2021)Li, Wang, Zhang and Duan}]{li2021constrained}
\bibinfo{author}{Li, B.}, \bibinfo{author}{Wang, Y.}, \bibinfo{author}{Zhang,
  K.}, \bibinfo{author}{Duan, G.R.}, \bibinfo{year}{2021}.
\newblock \bibinfo{title}{Constrained feedback control for spacecraft
  reorientation with an optimal gain}.
\newblock \bibinfo{journal}{IEEE Transactions on Aerospace and Electronic
  Systems} \bibinfo{volume}{57}, \bibinfo{pages}{3916--3926}.
\bibitem[{Li et~al.(2023a)Li, Ning, He, Lee and Zhao}]{li2022three}
\bibinfo{author}{Li, J.}, \bibinfo{author}{Ning, Z.}, \bibinfo{author}{He, S.},
  \bibinfo{author}{Lee, C.H.}, \bibinfo{author}{Zhao, S.},
  \bibinfo{year}{2023}a.
\newblock \bibinfo{title}{Three-dimensional bearing-only target following via
  observability-enhanced helical guidance}.
\newblock \bibinfo{journal}{IEEE Transactions on Robotics}
  \bibinfo{volume}{39}, \bibinfo{pages}{1509--1526}.
\bibitem[{Li et~al.(2023b)Li, Lu, Li, Lu and Jin}]{li2023adaptive}
\bibinfo{author}{Li, X.}, \bibinfo{author}{Lu, B.}, \bibinfo{author}{Li, Y.},
  \bibinfo{author}{Lu, X.}, \bibinfo{author}{Jin, H.}, \bibinfo{year}{2023}b.
\newblock \bibinfo{title}{Adaptive interacting multiple model for underwater
  maneuvering target tracking with one-step randomly delayed measurements}.
\newblock \bibinfo{journal}{Ocean engineering} \bibinfo{volume}{280},
  \bibinfo{pages}{114933}.
\bibitem[{Li et~al.(2023c)Li, Wang, Qi, Hao and Li}]{li2023optimal}
\bibinfo{author}{Li, X.}, \bibinfo{author}{Wang, Y.}, \bibinfo{author}{Qi, B.},
  \bibinfo{author}{Hao, Y.}, \bibinfo{author}{Li, S.}, \bibinfo{year}{2023}c.
\newblock \bibinfo{title}{Optimal maneuver strategy for an autonomous
  underwater vehicle with bearing-only measurements}.
\newblock \bibinfo{journal}{Ocean Engineering} \bibinfo{volume}{278},
  \bibinfo{pages}{114350}.
\bibitem[{Liu et~al.(2018)Liu, Cai and Ong}]{liu2018remarks}
\bibinfo{author}{Liu, H.}, \bibinfo{author}{Cai, J.}, \bibinfo{author}{Ong,
  Y.S.}, \bibinfo{year}{2018}.
\newblock \bibinfo{title}{Remarks on multi-output gaussian process regression}.
\newblock \bibinfo{journal}{Knowledge-Based Systems} \bibinfo{volume}{144},
  \bibinfo{pages}{102--121}.
\bibitem[{Mattingley and Boyd(2012)}]{mattingley2012cvxgen}
\bibinfo{author}{Mattingley, J.}, \bibinfo{author}{Boyd, S.},
  \bibinfo{year}{2012}.
\newblock \bibinfo{title}{Cvxgen: A code generator for embedded convex
  optimization}.
\newblock \bibinfo{journal}{Optimization and Engineering} \bibinfo{volume}{13},
  \bibinfo{pages}{1--27}.
\bibitem[{Ning et~al.(2024)Ning, Zhang, Li, Chen and Zhao}]{ning2024bearing}
\bibinfo{author}{Ning, Z.}, \bibinfo{author}{Zhang, Y.}, \bibinfo{author}{Li,
  J.}, \bibinfo{author}{Chen, Z.}, \bibinfo{author}{Zhao, S.},
  \bibinfo{year}{2024}.
\newblock \bibinfo{title}{A bearing-angle approach for unknown target motion
  analysis based on visual measurements}.
\newblock \bibinfo{journal}{The International Journal of Robotics Research}
  \bibinfo{volume}{43}, \bibinfo{pages}{1228--1249}.
\bibitem[{Nusrat et~al.(2022)Nusrat, Li, Cheng, Qazi and
  Xu}]{nusrat2022underwater}
\bibinfo{author}{Nusrat, A.}, \bibinfo{author}{Li, Y.}, \bibinfo{author}{Cheng,
  C.}, \bibinfo{author}{Qazi, H.}, \bibinfo{author}{Xu, L.},
  \bibinfo{year}{2022}.
\newblock \bibinfo{title}{Underwater bearing only tracking using optimal
  observer maneuver strategies}.
\newblock \bibinfo{journal}{Journal of Marine Science and Engineering}
  \bibinfo{volume}{10}, \bibinfo{pages}{576}.
\bibitem[{Omainska et~al.(2023)Omainska, Yamauchi, Lederer, Hirche and
  Fujita}]{omainska2023rigid}
\bibinfo{author}{Omainska, M.}, \bibinfo{author}{Yamauchi, J.},
  \bibinfo{author}{Lederer, A.}, \bibinfo{author}{Hirche, S.},
  \bibinfo{author}{Fujita, M.}, \bibinfo{year}{2023}.
\newblock \bibinfo{title}{Rigid motion gaussian processes with se (3) kernel
  and application to visual pursuit control}.
\newblock \bibinfo{journal}{IEEE Control Systems Letters} \bibinfo{volume}{7},
  \bibinfo{pages}{2665--2670}.
\bibitem[{Paull et~al.(2013)Paull, Saeedi, Seto and Li}]{paull2013auv}
\bibinfo{author}{Paull, L.}, \bibinfo{author}{Saeedi, S.},
  \bibinfo{author}{Seto, M.}, \bibinfo{author}{Li, H.}, \bibinfo{year}{2013}.
\newblock \bibinfo{title}{Auv navigation and localization: A review}.
\newblock \bibinfo{journal}{IEEE Journal of oceanic engineering}
  \bibinfo{volume}{39}, \bibinfo{pages}{131--149}.
\bibitem[{Qian et~al.(2024)Qian, Chen and Sun}]{qian2024maneuvering}
\bibinfo{author}{Qian, M.}, \bibinfo{author}{Chen, W.}, \bibinfo{author}{Sun,
  R.}, \bibinfo{year}{2024}.
\newblock \bibinfo{title}{A maneuvering target tracking algorithm based on
  cooperative localization of multi-uavs with bearing-only measurements}.
\newblock \bibinfo{journal}{IEEE Transactions on Instrumentation and
  Measurement} \bibinfo{volume}{73}, \bibinfo{pages}{1--11}.
\bibitem[{Shen et~al.(2016)Shen, Shi and Buckham}]{shen2016integrated}
\bibinfo{author}{Shen, C.}, \bibinfo{author}{Shi, Y.},
  \bibinfo{author}{Buckham, B.}, \bibinfo{year}{2016}.
\newblock \bibinfo{title}{Integrated path planning and tracking control of an
  {AUV}: A unified receding horizon optimization approach}.
\newblock \bibinfo{journal}{IEEE/ASME Transactions on Mechatronics}
  \bibinfo{volume}{22}, \bibinfo{pages}{1163--1173}.
\bibitem[{Snelson(2008)}]{snelson2008flexible}
\bibinfo{author}{Snelson, E.L.}, \bibinfo{year}{2008}.
\newblock \bibinfo{title}{Flexible and efficient Gaussian process models for
  machine learning}.
\newblock \bibinfo{publisher}{University of London, University College London
  (United Kingdom)}.
\bibitem[{Su et~al.(2022a)Su, Chen, Yang, Zhu and Guan}]{su2022bearing}
\bibinfo{author}{Su, H.}, \bibinfo{author}{Chen, C.}, \bibinfo{author}{Yang,
  Z.}, \bibinfo{author}{Zhu, S.}, \bibinfo{author}{Guan, X.},
  \bibinfo{year}{2022}a.
\newblock \bibinfo{title}{Bearing-based formation tracking control with
  time-varying velocity estimation}.
\newblock \bibinfo{journal}{IEEE Transactions on Cybernetics}
  \bibinfo{volume}{53}, \bibinfo{pages}{3961--3973}.
\bibitem[{Su et~al.(2022b)Su, Zhu, Yang, Chen and Guan}]{su2022bearingOE}
\bibinfo{author}{Su, H.}, \bibinfo{author}{Zhu, S.}, \bibinfo{author}{Yang,
  Z.}, \bibinfo{author}{Chen, C.}, \bibinfo{author}{Guan, X.},
  \bibinfo{year}{2022}b.
\newblock \bibinfo{title}{Bearing-based formation tracking control of {AUV}s
  with optimal gains tuning}.
\newblock \bibinfo{journal}{Ocean Engineering} \bibinfo{volume}{258},
  \bibinfo{pages}{111672}.
\bibitem[{Sui et~al.(2024)Sui, Deghat, Sun and Eskandari}]{sui2024adaptive}
\bibinfo{author}{Sui, D.}, \bibinfo{author}{Deghat, M.}, \bibinfo{author}{Sun,
  Z.}, \bibinfo{author}{Eskandari, M.}, \bibinfo{year}{2024}.
\newblock \bibinfo{title}{Adaptive bearing-only target localization and
  circumnavigation under unknown wind disturbance: Theory and experiments}.
\newblock \bibinfo{journal}{IEEE Robotics and Automation Letters} .
\bibitem[{Umlauft and Hirche(2019)}]{umlauft2019feedback}
\bibinfo{author}{Umlauft, J.}, \bibinfo{author}{Hirche, S.},
  \bibinfo{year}{2019}.
\newblock \bibinfo{title}{Feedback linearization based on {G}aussian processes
  with event-triggered online learning}.
\newblock \bibinfo{journal}{IEEE Transactions on Automatic Control}
  \bibinfo{volume}{65}, \bibinfo{pages}{4154--4169}.
\bibitem[{Van Der~Vaart and Van~Zanten(2011)}]{van2011information}
\bibinfo{author}{Van Der~Vaart, A.}, \bibinfo{author}{Van~Zanten, H.},
  \bibinfo{year}{2011}.
\newblock \bibinfo{title}{Information rates of nonparametric gaussian process
  methods.}
\newblock \bibinfo{journal}{Journal of Machine Learning Research}
  \bibinfo{volume}{12}, \bibinfo{pages}{2095--2119}.
\bibitem[{Van~Tran and Kim(2022)}]{van2022bearing}
\bibinfo{author}{Van~Tran, Q.}, \bibinfo{author}{Kim, J.},
  \bibinfo{year}{2022}.
\newblock \bibinfo{title}{Bearing-constrained formation tracking control of
  nonholonomic agents without inter-agent communication}.
\newblock \bibinfo{journal}{IEEE Control Systems Letters} \bibinfo{volume}{6},
  \bibinfo{pages}{2401--2406}.
\bibitem[{Wang et~al.(2024)Wang, Zhang, Zhao and Xu}]{wang2024target}
\bibinfo{author}{Wang, D.}, \bibinfo{author}{Zhang, Z.}, \bibinfo{author}{Zhao,
  Y.}, \bibinfo{author}{Xu, C.}, \bibinfo{year}{2024}.
\newblock \bibinfo{title}{Target tracking with circumnavigation scheme using
  discrete bearing and control input}.
\newblock \bibinfo{journal}{IEEE Transactions on Aerospace and Electronic
  Systems} \bibinfo{volume}{60}, \bibinfo{pages}{4699--4714}.
\bibitem[{Wang et~al.(2022)Wang, Zhou, Xu and Gao}]{wang2022geometrically}
\bibinfo{author}{Wang, Z.}, \bibinfo{author}{Zhou, X.}, \bibinfo{author}{Xu,
  C.}, \bibinfo{author}{Gao, F.}, \bibinfo{year}{2022}.
\newblock \bibinfo{title}{Geometrically constrained trajectory optimization for
  multicopters}.
\newblock \bibinfo{journal}{IEEE Transactions on Robotics}
  \bibinfo{volume}{38}, \bibinfo{pages}{3259--3278}.
\bibitem[{Williams and Rasmussen(2006)}]{williams2006gaussian}
\bibinfo{author}{Williams, C.K.}, \bibinfo{author}{Rasmussen, C.E.},
  \bibinfo{year}{2006}.
\newblock \bibinfo{title}{Gaussian {P}rocesses for {M}achine {L}earning}.
\newblock \bibinfo{publisher}{MIT press Cambridge, MA}.
\bibitem[{Wynne et~al.(2021)Wynne, Briol and Girolami}]{wynne2021convergence}
\bibinfo{author}{Wynne, G.}, \bibinfo{author}{Briol, F.X.},
  \bibinfo{author}{Girolami, M.}, \bibinfo{year}{2021}.
\newblock \bibinfo{title}{Convergence guarantees for gaussian process means
  with misspecified likelihoods and smoothness}.
\newblock \bibinfo{journal}{Journal of Machine Learning Research}
  \bibinfo{volume}{22}, \bibinfo{pages}{1--40}.
\bibitem[{Yang et~al.(2020)Yang, Zhu, Chen, Feng and Guan}]{yang2020entrapping}
\bibinfo{author}{Yang, Z.}, \bibinfo{author}{Zhu, S.}, \bibinfo{author}{Chen,
  C.}, \bibinfo{author}{Feng, G.}, \bibinfo{author}{Guan, X.},
  \bibinfo{year}{2020}.
\newblock \bibinfo{title}{Entrapping a target in an arbitrarily shaped orbit by
  a single robot using bearing measurements}.
\newblock \bibinfo{journal}{Automatica} \bibinfo{volume}{113},
  \bibinfo{pages}{108805}.
\bibitem[{Zhang et~al.(2017)Zhang, Dufour, Anselmi, Laneuville and
  N{\`e}gre}]{zhang2017piecewise}
\bibinfo{author}{Zhang, H.}, \bibinfo{author}{Dufour, F.},
  \bibinfo{author}{Anselmi, J.}, \bibinfo{author}{Laneuville, D.},
  \bibinfo{author}{N{\`e}gre, A.}, \bibinfo{year}{2017}.
\newblock \bibinfo{title}{Piecewise optimal trajectories of observer for
  bearings-only tracking by quantization}, in: \bibinfo{booktitle}{2017 20th
  International Conference on Information Fusion (Fusion)},
  \bibinfo{organization}{IEEE}. pp. \bibinfo{pages}{1--7}.
\bibitem[{Zhao et~al.(2013)Zhao, Chen and Lee}]{zhao2013optimal}
\bibinfo{author}{Zhao, S.}, \bibinfo{author}{Chen, B.M.}, \bibinfo{author}{Lee,
  T.H.}, \bibinfo{year}{2013}.
\newblock \bibinfo{title}{Optimal sensor placement for target localisation and
  tracking in 2d and 3d}.
\newblock \bibinfo{journal}{International Journal of Control}
  \bibinfo{volume}{86}, \bibinfo{pages}{1687--1704}.
\bibitem[{Zhou and Roumeliotis(2011)}]{zhou2011multirobot}
\bibinfo{author}{Zhou, K.}, \bibinfo{author}{Roumeliotis, S.I.},
  \bibinfo{year}{2011}.
\newblock \bibinfo{title}{Multirobot active target tracking with combinations
  of relative observations}.
\newblock \bibinfo{journal}{IEEE Transactions on Robotics}
  \bibinfo{volume}{27}, \bibinfo{pages}{678--695}.

\end{thebibliography}

\end{document}